\pdfoutput=1
\documentclass[cernpreprint,UKenglish,texlive=2016,texmf]{atlasdoc}
\usepackage[subfig]{atlaspackage}
\usepackage{atlasbiblatex}
 
\usepackage{atlasphysics}
 
\usepackage{todonotes}
\usepackage{multirow}
\usepackage{upgreek}
 
\graphicspath{{logos/}{figures/}}
 
\usepackage{ANA-TOPQ-2018-18-PAPER-defs}

\AtlasTitle{Measurements of top-quark pair single- and double-differential cross-sections in the all-hadronic channel in $pp$ collisions at $\sqrt{s}=13\,\ensuremath{\text{Te\kern -0.1em V}}$ using the ATLAS detector}

\AtlasAbstract{
 
Differential cross-sections are measured for top-quark pair production in the all-hadronic decay mode, using proton--proton collision events collected by the ATLAS experiment in which all six decay jets are separately resolved.
Absolute and normalised single- and double-differential cross-sections are measured at particle and parton level as a function of various kinematic variables.
Emphasis is placed on well-measured observables in fully reconstructed final states, as well as on the study of correlations between the top-quark pair system and additional jet radiation identified in the event.
The study is performed using data from proton--proton collisions at $\sqrt{s}=13\,\ensuremath{\text{Te\kern -0.1em V}}$ collected by the ATLAS detector at CERN's Large Hadron Collider in 2015 and 2016, corresponding to an integrated luminosity of \mbox{36.1\,fb$^{-1}$}.
The rapidities of the individual top quarks and of the top-quark pair are well modelled by several independent event generators. Significant mismodelling is observed in the transverse momenta of the leading three jet emissions, while the leading top-quark transverse momentum and top-quark pair transverse momentum are both found to be incompatible with several theoretical predictions.
}
 
\author{The ATLAS Collaboration}
 
\AtlasRefCode{ANA-TOPQ-2018-18}
 
\PreprintIdNumber{CERN-EP-2020-063}

\AtlasJournalRef{\JHEP 01 (2021) 033}
\AtlasDOI{10.1007/JHEP01(2021)033}

\hypersetup{pdftitle={ATLAS document},pdfauthor={The ATLAS Collaboration}}

\begin{document}
 
\maketitle
 
\tableofcontents

\section{Introduction}
\label{sec:intro}
 
As the heaviest particle of the Standard Model (SM), the top quark and its properties provide insights into a wide range of topics, including proton structure and precision electroweak physics.
Top-quark pair production is also the most significant background to many searches for physics beyond the Standard Model (BSM).
Therefore, improving the accuracy of theoretical models for this production process is of central importance to the collider physics programme.
 
The Large Hadron Collider~\cite{LHC} (LHC) is the first top-quark factory and thus provides an unprecedented opportunity to study the physics of the top quark.
This paper reports the results of measurements of differential cross-sections for the production of top-quark pairs in the final state with the largest branching ratio, namely the decay of each top quark into a bottom quark and two additional quarks. The measurements are performed in the six-jet topology, using data collected by the ATLAS detector~\cite{PERF-2007-01} at a centre-of-mass energy $\sqrt{s}$ of $13~\TeV$ in 2015 and 2016 and corresponding to 36.1~\invfb\ of proton--proton ($pp$) collisions.
 
Single- and double-differential distributions of the kinematic properties of the top-quark--top-antiquark ($\ttbar$) system are presented. They can be used to strengthen constraints on parton distribution functions (PDFs) and tuning of precision cross-section computations.
Correlations between the $\ttbar$ system and associated jet production are also measured, and are compared with predictions of multi-leg matrix element calculations.
Both the absolute and normalised differential cross-sections are presented.
 
Previous measurements of the differential cross-sections of top-quark pair production, particularly in association with additional jets, mainly used the lepton-plus-jets (\ljets) and dileptonic decay modes~\cite{TOPQ-2018-15, CMS-TOP-17-002,CMS-TOP-16-014,TOPQ-2017-01,TOPQ-2016-08,CMS-TOP-16-007,TOPQ-2015-02,CMS-TOP-14-013,TOPQ-2016-04,CMS-TOP-16-008,TOPQ-2015-06,CMS-TOP-14-012}, while the all-hadronic decay mode was studied at lower $\sqrt{s}$ by the CMS Collaboration~\cite{CMS-TOP-14-018,CMS-TOP-11-007} and in the highly boosted regime, at high transverse momentum (\pt)~\cite{TOPQ-2016-09}, by the ATLAS Collaboration.
This analysis considers events in which all three quarks from each top-quark decay are resolved into distinct jets, leading to at least six jets in the final state.
This complements the measurements made in this channel using large-radius jets~\cite{TOPQ-2016-09}, which are limited to the region of top-quark transverse momentum above 350~\GeV.
The resolved all-hadronic final state is admittedly subject to a larger background contamination from multi-jet production.
However, this final state avoids kinematic ambiguities due to the presence of neutrinos accompanying the leptonic decays.
This allows a full reconstruction of the top-quark pair without recourse to the missing transverse momentum, which has relatively poor experimental resolution~\cite{PERF-2016-07} and provides no information about longitudinal momentum.
The good momentum resolution for both top quarks enables characterisation of the kinematic properties of additional jet radiation accompanying the $\ttbar$ system in relation to the top-quark pair kinematics.
 
Differential distributions measured in data are presented with corrections both to the stable-particle level in a fiducial phase space and to the parton level in the full phase space.
The paper presents a set of measurements of the $\ttbar$ production cross-section as a function of properties of the reconstructed top quark (transverse momentum and rapidity) and of the $\ttbar$ system (transverse momentum, rapidity and invariant mass) as well as additional variables.
Taking various reference objects such as the leading top quark, the leading jet and the leading extra jet, angular separations and transverse momentum ratios between the additional jet radiation and these reference objects are measured.
The measured differential cross-sections are compared with predictions from a variety of Monte Carlo (MC) event generators at next-to-leading order (NLO): \POWHEGBOXV{v2}~\cite{POWHEG:1,POWHEG:2,POWHEG:Box,POWHEG:topq,Frixione_2007} and \MGMCatNLO~\cite{aMCatNLO}, interfaced with \PYTHIAEIGHT~\cite{Pythia8} and \HERWIGSEVEN~\cite{Herwig7}, and \SHERPAV{2.2}~\cite{Sherpa2}.
 
The paper is structured as follows. The ATLAS detector is described in Section~\ref{sec:detector}. Next, in Section~\ref{sec:datamc}, a description is given of the data and MC samples used in the paper. The event reconstruction and the selection criteria applied are defined respectively in Sections~\ref{sec:objdef} and~\ref{sec:evtsel}. Section~\ref{sec:bkgest} explains the procedure used to evaluate the multi-jet background, while the list of observables measured is presented in Section~\ref{sec:observables}. The detector and reconstruction effects are corrected by unfolding the data, via a procedure described in Section~\ref{sec:unfolding}. The systematic uncertainties and the results are presented in Sections~\ref{sec:syst} and~\ref{sec:results}. Finally, the conclusions of the analysis are summarised in Section~\ref{sec:conclusion}.
 
\section{ATLAS detector}
\label{sec:detector}
 
The ATLAS detector~\cite{PERF-2007-01} is a multipurpose particle physics detector with a forward--backward symmetric cylindrical geometry and nearly $4\pi$ coverage in solid angle, up to $|\eta|=4.9.$\footnote{ATLAS uses a right-handed Cartesian coordinate system with its origin at the nominal interaction point (IP) in the centre of the detector. The
$z$-axis is along the beam pipe, and the $x$-axis points from the IP to the centre of the LHC ring. Cylindrical coordinates ($r$, $\phi$) are used in the transverse plane, $\phi$ being the azimuthal angle around the beam pipe. The rapidity is defined as $y=(1/2)\ln[(E+p_z)/(E-p_z)]$, while the pseudorapidity is defined in terms of the polar angle $\theta$ as $\eta = -\ln \tan(\theta/2)$.}
The layout of the detector is based on four superconducting magnet systems, comprising a thin solenoid surrounding the inner tracking detectors (ID) plus a barrel and two endcap toroids generating the magnetic field for a large muon spectrometer.
The ID includes two silicon sub-detectors, namely an inner pixel detector and an outer microstrip tracker, inside a transition radiation tracker (TRT) based on gas-filled drift tubes. The innermost pixel layer, the insertable B-layer~\cite{ATLAS-TDR-2010-19, PIX-2018-001}, was added before the start of 13 TeV LHC operation at an average radius of 33 $mm$ around a new, thinner beam pipe.
The calorimeters are located between the ID and the muon system. The lead/liquid-argon (LAr) electromagnetic (EM) calorimeter is split into two regions: the barrel ($|\eta | < 1.475$) and the endcaps ($1.375 < |\eta| < 3.2$). The hadronic calorimeter is divided into four regions: the barrel ($|\eta| < 1.0$) and the extended barrel ($0.8 < |\eta| < 1.7$) made of scintillator/steel, the endcaps ($1.5 < |\eta| < 3.2$) with LAr/copper modules, and the forward calorimeter ($3.1 < |\eta| < 4.9$) composed of LAr/copper and LAr/tungsten modules.
 
A two-level trigger system~\cite{TRIG-2016-01} selects events for further analysis. The first level of the trigger reduces the event rate to about 100~kHz using hardware-based trigger algorithms acting on a subset of detector information. The second level, the high-level trigger, further reduces the average event rate to about 1000~Hz by using a combination of fast online algorithms and reconstruction software with algorithms similar to the offline versions.
 
\section{Collision data and simulated event samples}
\label{sec:datamc}
 
The data for this analysis were recorded with the ATLAS detector from $pp$ collisions at $\sqrt{s} = 13$~\TeV\ in 2015 and 2016 with an average number of $pp$ interactions per bunch crossing $\langle\mu\rangle$ of around 23.\footnote{A reference inelastic cross-section of 80~mb is assumed when converting between instantaneous luminosity and $\mu$.} The selected data sample corresponds to an integrated luminosity of 36.1~\invfb\ with an uncertainty of 2.1\%~\cite{ATLAS-CONF-2019-021}, obtained using the LUCID-2 detector~\cite{LUCID2} for the primary luminosity measurements. Only the data collected while all sub-detectors were operational are considered.
 
The events for this analysis were collected using a multi-jet trigger. This trigger selects events containing six jets with a minimum $\pt$ of 45~\GeV\ in the central $\eta$ region of the detector; the $\eta$ acceptance of all six jets changed from $|\eta|<3.2$ in 2015 to $|\eta|<2.4$ in 2016 to reduce triggered event rates. In the high-level trigger, jets are reconstructed with the \antikt{} jet algorithm~\cite{Cacciari:2008gp} using a radius parameter $R$ of 0.4 and are calibrated as described in Section~\ref{sec:obj_detector}. This trigger was chosen as it provides a high efficiency ($>98\%$) for signal events and does not require jets originating from $b$-quarks, which is crucial for evaluating background contributions in data.
 
The physics processes of interest in this analysis are \ttbar{} events with both $W$ bosons decaying hadronically (all-hadronic signal), \ttbar{} events with at least one $W$ boson decaying leptonically (non-all-hadronic background) and multi-jet production from pure strong-interaction processes (multi-jet background).
 
\subsection{Top-quark pair simulation samples}
\label{sec:samples}
 
The MC generators listed in Table~\ref{tab:MC} were used to simulate $\ttbar$ event samples for unfolding corrections (\Sect{\ref{sec:unfolding}}), systematic uncertainty estimates and comparison with results at the pre- and post-unfolding levels. The top-quark mass $\mt$ and width were set to 172.5~\GeV\ and 1.32~\GeV, respectively, in all MC event generators; these values are compatible with the most recent measurements~\cite{TOPQ-2017-03,TOPQ-2017-02}.
\begin{table*}[htbp]
\caption{Summary of \ttbar MC samples used in the analysis, showing the generator for the hard-scattering process, cross-section $\sigma$ normalisation precision, PDF choices for the hard-process matrix element (ME) and parton shower (PS), as well as the parton shower and hadronisation generator and the corresponding tune sets and scales.}
\label{tab:MC}
\renewcommand\arraystretch{1.2}
\footnotesize
\centering
\aboverulesep=0ex
\belowrulesep=0ex
\begin{tabular}{|l|c|c|c|c|c|c|} \hline
Application & \ttbar signal & \multicolumn{2}{c|}{\ttbar radiation syst.} & \ttbar PS syst. & \ttbar ME syst. & \ttbar comparison \\
\hline
\multirow{2}{*}{Generator} & \multicolumn{4}{c|}{\multirow{2}{*}{\POWHEGBOXV{v2}}} & \MADGRAPHFIVE & \multirow{2}{*}{\SHERPAV{2.2.1}}  \\
& \multicolumn{4}{c|}{} & \AMCatNLO 2.6.0 & \\
\hline
$\sigma$ precision & \multicolumn{6}{c|}{$\NNLO$ + $\NNLL$} \\
\hline
PDF for ME & \multicolumn{6}{c|}{NNPDF3.0NLO} \\
\hline
Parton shower & \multicolumn{3}{c|}{\PYTHIAEIGHT} & \HERWIGSEVEN & \PYTHIAEIGHT & ME+PS@NLO \\
\hline
PDF for PS & \multicolumn{3}{c|}{NNPDF2.3LO} & MMHT2014 & \multicolumn{2}{c|}{NNPDF2.3LO} \\
\hline
Tune set & A14 & Var3cUp & Var3cDown & H7UEMMHT & A14 & --  \\
\hline
\multirow{2}{*}{Scales} & \multirow{2}{*}{$\hdamp\!=\!1.5\mt$} & $\hdamp\!=\!3\mt$ & $\hdamp\!=\!1.5\mt$ & \multirow{2}{*}{$\hdamp\!=\!1.5\mt$} & \multirow{2}{*}{$\upmu_q=\HT/2$} & \multirow{2}{*}{--} \\
& & $\upmu_\mathrm{R,F} = 0.5$ & $\upmu_\mathrm{R,F} = 2.0$ & & & \\
\hline
\end{tabular}
\end{table*}

The \textsc{EvtGen} v1.2.0 program~\cite{EvtGen} was used to simulate the decay of bottom and charm hadrons, except for samples generated with \SHERPA~\cite{Sherpa2}.
Multiple overlaid $pp$ collisions (pile-up) were simulated with the low-\pt QCD processes of \PYTHIA~8.186~\cite{Pythia8} using a set of tuned parameters called the A3 tune set~\cite{ATL-PHYS-PUB-2016-017} and the NNPDF2.3LO~\cite{Ball:2012cx} set of parton distribution functions (PDFs).
 
The detector response was simulated using the $\GEANT4$ framework~\cite{Agostinelli:2002hh,SOFT-2010-01}. The data and MC events were reconstructed with the same software algorithms.
 
For the generation of \ttbar\ events, matrix elements (ME) were calculated at NLO in QCD using the \POWHEGBOXV{v2}~\cite{POWHEG:2,POWHEG:Box} event generator with the \textsc{NNPDF3.0NLO} PDF set~\cite{Ball:2014uwa}.
\PYTHIA~8.210~\cite{Sj_strand_2015} with the \textsc{NNPDF2.3LO}~\cite{Ball:2012cx} PDF set and the A14~\cite{ATL-PHYS-PUB-2014-021} tune set was used to simulate the parton shower, fragmentation and underlying event.
The $\hdamp$ parameter, which controls the $\pt$ of the first gluon or quark emission beyond the Born configuration in \POWHEGBOXV{v2}, was set to $1.5\mt$.
The main effect of this parameter is to regulate the high-$\pt$ emission against which the $\ttbar$ system recoils.
A dynamic value $\sqrt{\mt^2 + p_{\mathrm{T},t}^2}$ was used for the factorisation and renormalisation scales ($\upmu_\mathrm{F}$ and $\upmu_\mathrm{R}$ respectively).
Signal $\ttbar$ events generated with those settings are referred to as the nominal signal sample.
 
The effects of different levels of initial-state radiation (ISR) were evaluated using two samples with different factorisation and renormalisation scales relative to the nominal signal sample, as well as a different $\hdamp$ parameter value. Specifically, two settings for \PHPYEIGHT were compared~\cite{ATL-PHYS-PUB-2016-004}:
\begin{itemize}
\item In one sample, $\upmu_\mathrm{F,R}$ were reduced by a factor of 0.5, the $\hdamp$ parameter was increased to $3\mt$, and the Var3cUp A14 tune variation was used. In all the following figures and tables, the predictions based on this MC sample are referred to as `\textsc{PWG+PY8} Up'.
\item In the other sample, $\upmu_\mathrm{F,R}$ were increased by a factor of 2, the $\hdamp$ parameter was set to $1.5\mt$ as in the nominal sample, and the Var3cDown A14 tune variation was used. In all the following figures and tables, the predictions based on this MC sample are referred to as `\textsc{PWG+PY8} Down'.
\end{itemize}
 
To estimate the effect of choosing different parton shower and hadronisation algorithms, a \PHHSEVEN\ sample was generated with \POWHEG set up in the same way as for the nominal sample. The parton shower, hadronisation and underlying-event simulation were produced with \HERWIGSEVEN~\cite{Herwig7} (version 7.0.4) using the \textsc{MMHT2014lo68cl} PDF set and \textsc{H7-UE-MMHT} tune set~\cite{MMHT}. Detector simulation was performed using a fast simulation based on a parameterisation of the performance of the ATLAS electromagnetic and hadronic calorimeters \cite{ATL-SOFT-PUB-2014-001} (\ATLFAST) and
using $\GEANT4$ elsewhere.
 
The impact of the choice of matrix element generator was evaluated using events generated with \MCNPYEIGHT\ at NLO accuracy. The events were generated with version 2.6.0 of \MGMCatNLO~\cite{aMCatNLO} and $\upmu_q = H_\mathrm{T}/2$ (with $H_\mathrm{T}$ the scalar sum of the $p_\mathrm{T}$ of all outgoing partons) for the shower starting-scale functional form~\cite{ATL-PHYS-PUB-2017-007}. As in the \PHPYEIGHT\ samples, the \textsc{NNPDF3.0NLO} PDF set was used for the matrix element and the \textsc{NNPDF2.3LO} set for the parton shower.
Calorimeter simulation was performed using \ATLFAST.
 
An additional sample of \ttbar\ events was generated with \SHERPAV{2.2.1} to provide an extra point of comparison~\cite{Sherpa2}.
This sample was produced at NLO in QCD for up to one additional parton emission and at LO for up to four additional partons using the MEPS\@NLO merging scheme~\cite{Hoeche:2012yf} with the CKKW merging scale fixed at 30~\GeV~\cite{ATL-PHYS-PUB-2017-007}.
Loop integrals were calculated with OpenLoops~\cite{OpenLoops2}.
The shower, factorisation and renormalisation scales were set to $\upmu_\mathrm{F,R} = \sqrt{\mt^2 + 0.5(p_{\mathrm{T},t}^2 + p_{\mathrm{T},\bar{t}}^2)}$, and the \textsc{NNPDF2.3LO} PDF set was used.
 
The cross-section used to normalise the $t\bar{t}$ samples was $\sigma_{t\bar{t}} = 832^{+20}_{-29}(\mathrm{scale}) \pm 35~(\mathrm{PDF}, \alphas)$~pb, as calculated with the Top++2.0 program at NNLO in perturbative QCD including soft-gluon resummation to next-to-next-to-leading-log order (NNLL)~\cite{Toppp,Toppp:EW,Baernreuther:2012ws,Czakon:2012zr,Czakon:2012pz,Czakon:2013goa} and assuming $\mt$ = 172.5~\GeV{}. The first uncertainty comes from the independent variation of the factorisation and renormalisation scales, $\upmu_{\mathrm{F}}$ and $\upmu_{\mathrm{R}}$, while the second one is associated with variations in the PDF and $\alphas$, following the PDF4LHC prescription~\cite{PDF4LHC} with the MSTW2008 68\% CL NNLO~\cite{Martin_2009}, CT10 NNLO~\cite{CT10} and NNPDF2.3 5f FFN~\cite{Ball:2012cx} PDF sets.

Top-quark pair events in which at least one of the $W$ bosons decays into a lepton and a neutrino are a source of background contamination if the leptons are not identified.
Simulated \ttbar\ events with one or two leptonic decays were produced with the same settings used for the nominal signal sample.
 
 
\section{Object reconstruction}
\label{sec:objdef}
The following sections describe the detector-, particle- and parton-level objects used to characterise the final-state event topology and to define the fiducial and full phase-space regions for the measurements. The final state of interest in this measurement includes jets, some of which may be tagged as originating from $b$-quarks, but contains no isolated electrons, muons or $\tau$-leptons.
 
\subsection{Detector-level object reconstruction}
\label{sec:obj_detector}
Primary vertices are formed from reconstructed tracks which are spatially compatible with the interaction region~\cite{PERF-2015-01}. The hard-scatter primary vertex is chosen to be the one with at least two associated tracks and the highest $\sum$ $p^2_\mathrm{T}$, where the sum extends over all tracks with $\pt > 0.4$~\GeV\ matched to the vertex.
 
Jets are reconstructed from topological clusters of calorimeter cells that are noise-suppressed and calibrated to the electromagnetic scale~\cite{PERF-2014-07} using the \antikt{} algorithm with a radius parameter $R = 0.4$ as implemented in FastJet~\cite{Cacciari:2011ma}. The jets are corrected using a subtraction procedure that accounts for the jet area to estimate and remove the average energy contributed by pile-up interactions~\cite{Cacciari:2007fd}; these corrections can change the jet four-momentum. This procedure is followed by a jet-energy-scale (JES) calibration that restores the jet energy to the mean response in a particle-level simulation, refined by applying a series of additional calibrations that correct finer variations due to jet flavour and detector geometry and \textit{in situ} corrections that match the data to the simulation energy scale~\cite{PERF-2016-04}.
 
Jets must satisfy $\pT > 25$~\GeV\ and $|\eta|<2.5$, and survive the removal of overlaps with leptons, as described below. To reduce the number of jets that originate from pile-up, the jet vertex tagger (JVT)~\cite{PERF-2014-03} is used to identify jets associated with the hard-scatter vertex. Every jet with $\pT<60$~\GeV\ and $|\eta|<2.4$ must satisfy the criterion JVT > 0.59.
The JVT discriminant is based on the degree of association between the hard-scatter vertex and tracks matched to the jet by a ghost-association technique described in Ref.~\cite{Cacciari:2008gn}.
 
Jets containing $b$-hadrons are tagged as `$b$-jets' using a multivariate discriminant (MV2c10)~\cite{ATL-PHYS-PUB-2016-012}. It combines information from the impact parameters of displaced tracks and from the location and topological properties of secondary and tertiary decay vertices reconstructed within the jet. The jets are considered $b$-tagged if the value of the discriminant is larger than a threshold applied to the discriminant output value, chosen to provide a specific $b$-jet tagging efficiency in the nominal $t\bar{t}$ sample. In this analysis, a threshold corresponding to 70\% $b$-jet tagging efficiency is chosen. The corresponding rejection factors for jets initiated by charm quarks or lighter quark flavours are approximately 12 and 380, respectively~\cite{PERF-2012-04}.
 
Electron candidates are reconstructed from clusters of energy in the calorimeter combined with an inner detector (ID) track that is re-fitted using Gaussian sum filters and calibrated using a multivariate regression~\cite{PERF-2017-03,PERF-2017-01}.
They must satisfy $\pt > 15$~\GeV\ and $|\eta_\mathrm{clus}|<1.37$ or $1.52 < |\eta_\mathrm{clus}| < 2.47 $ and satisfy the `tight' likelihood-based identification criteria based on shower shapes in the EM calorimeter, track quality and detection of transition radiation produced in the TRT~\cite{ATL-PHYS-PUB-2011-006}. Isolation requirements based on calorimeter and tracking quantities are used to reduce the background from jets misidentified as prompt electrons (fake electrons) or due to semileptonic decays of heavy-flavour hadrons (non-prompt real electrons)~\cite{PERF-2017-01}. The isolation criteria are $\pt$- and $\eta$-dependent and ensure efficiencies of 90\% for electrons with $\pt$ > 25~\GeV\ and 99\% for electrons with $\pt$ > 60~\GeV.
 
Muon candidates are reconstructed using high-quality ID tracks combined with tracks reconstructed in the muon spectrometer~\cite{PERF-2015-10}. They must satisfy $\pt > 15$~\GeV\ and $|\eta| < 2.5$. To reduce the background from muons originating from heavy-flavour decays inside jets, muons are required to be isolated using track quality and isolation criteria similar to those applied to electrons.
 
Hadronically decaying $\tau$-lepton ($\tau_{\mathrm{had}}$) candidates are reconstructed from hadronic jets associated with either one or three ID tracks with a total charge of $\pm 1$~\cite{ATL-PHYS-PUB-2015-045,ATLAS-CONF-2017-029}. Candidate $\tau$-leptons with $\pT > 25$~\GeV\ and $|\eta| < 2.5$ are considered. A boosted decision tree (BDT) discriminant is used to distinguish $\tau_{\mathrm{had}}$ candidates from quark- or gluon-initiated jets, for which the `medium' working point is used. A second BDT is used to eliminate electrons misidentified as $\tau$-leptons, using the `loose' working point.
 
For objects satisfying both the jet and lepton selection criteria, a procedure called `overlap removal' is applied to assign objects a unique particle hypothesis, favouring well-identified and isolated particles. If an electron candidate shares a track with a muon candidate, the electron is removed, as it is likely to result from final-state radiation (FSR). If a jet and an electron are within $\Delta R=\sqrt{(\Delta \eta)^2 + (\Delta\phi)^2}$ < 0.2 the jet is discarded. If the distance in $\Delta R$ between a surviving jet and an electron is smaller than 0.4, then the electron is discarded. If a muon track is matched to a jet by ghost-association, or a jet and a muon are within $\Delta R < 0.2$, then the jet is removed if its $\pT$, total track $\pT$ and number of tracks are consistent with muon FSR or energy loss. If the distance in $\Delta R$ between a jet and a muon candidate is $\Delta R < 0.4$, the muon is discarded. Finally, if the distance in $\Delta R$ between a jet and a $\tau$-lepton jet is $\Delta R < 0.2$, then the jet is discarded.
 
\subsection{Particle- and parton-level object and phase-space definitions}
\label{sec:obj_particle}
 
Particle-level objects in simulated events are defined using only stable particles, i.e.\ particles with a mean lifetime $\tau > 30$~ps. The fiducial phase space for the measurements presented in this paper is defined using a series of requirements applied to particle-level objects, analogous to those used in the selection of the detector-level objects described above.
 
Electrons and muons are required not to originate from a hadron in the MC generator's `truth' record, whether directly or through a $\tau$-lepton decay. This ensures that the lepton is from an electroweak decay without requiring a direct $W$-boson match.  The four-momenta of the bare leptons are then modified (`dressed') by adding the four-momenta of all photons within a cone of size $\Delta R = 0.1$ to take into account final-state photon radiation. Dressed electrons are then required to have $\pt > 15$~\GeV\ and $|\eta|<1.37$ or $1.52 < |\eta| < 2.47$. Dressed muons are required to have $\pt > 15$~\GeV\ and $|\eta| <2.5$.
 
Particle-level jets are reconstructed using the same \antikt{} algorithm used at detector level. The jet-reconstruction procedure takes as input all stable particles, except for charged leptons not from hadron decays as described above, inside a radius $R=0.4$. Particle-level jets are required to have $\pt > 25$~\GeV\ and $|\eta| < 2.5$. A jet is identified as a $b$-jet if a hadron containing a $b$-quark is matched to the jet using the ghost-association procedure; the hadron must have $\pt > 5$~\GeV.
 
The simulated top-quark four-momenta are recorded after parton showering, but before decays are simulated, and correspond to the parton-level description of the event. The full phase space is defined by the set of $\ttbar$ pairs in which both top quarks decay hadronically. The measurements presented in this paper cover the entire phase space.
 
\section{Event selection and reconstruction}
\label{sec:evtsel}
 
A series of selection criteria are applied to define the signal region (SR) containing a pure sample of resolved all-hadronic top-quark pair events.
Events are removed if detector defects or data corruption are identified or if the events do not contain a primary vertex.
Events must contain at least six jets with $\pt>55~\GeV$ and $|\eta|<2.4$ to be in a regime where the trigger is highly efficient. Additional jets must pass the selection requirement described in Section~\ref{sec:obj_detector}.
Exactly two $b$-tagged jets must be found among all jets.
A veto is applied to events containing at least one electron or muon with $\pt>15~\GeV$ or a $\tau$-lepton with $\pT>25~\GeV$.
 
Subsequently, a $\ttbar$ reconstruction procedure is implemented to suppress backgrounds from multi-jet production and to calculate the observables to be measured (\Sect{\ref{sec:observables}}).
 
\subsection{Kinematic reconstruction of the $t\bar{t}$ system}
\label{sec:evtsel:topreco}
 
The identification of two top-quark candidates from the many jets in the event is a combinatorially complex problem.
Each \bjet\ is assigned to one top-quark candidate, and permutations are formed for each set of four jets selected from the remaining jets in the event.
These four `light' jets are paired to form $W$-boson candidates, and each $W$-boson candidate is, in turn, matched with one of the \bjets\ to form a top-quark candidate.
For the $W$-boson pairings and $b$--$W$ pairings, all unique permutations are considered.
A chi-square discriminant $\chi^2$ is computed for each permutation to judge whether the considered permutation is compatible with the hypothesis of a top-quark pair; the permutation with the smallest chi-square $\chi_\mathrm{min}^2$ is chosen as the one best describing the event as the product of a top-quark pair decay.
 
The $\chi^2$ discriminant is
\begin{equation*}
\chi^2 = \frac{ (m_{b_1 j_1 j_2 }-m_{b_2 j_3 j_4})^2}{2 \sigma_{t}^2}
+ \frac{ (m_{j_1 j_2 }-m_{W})^2}{\sigma_{W}^2}
+ \frac{ (m_{j_3 j_4 }-m_{W})^2}{\sigma_{W}^2},
\end{equation*}
where $m_{t,1}=m_{b_1 j_1 j_2}$ and $m_{t,2}=m_{b_2 j_3 j_4}$ are the invariant masses of the jets associated with the decay products of the leading and sub-leading top quark, sorted in \pt, respectively.\footnote{In this paper, `leading', `sub-leading' etc.\ are always taken to refer to $\pt$-ordering, for brevity.}
Similarly, $m_{j_1 j_2}$ and $m_{j_3 j_4}$ are the invariant masses of the jets associated with the decay products of the $W$ bosons from the top quarks.
The $W$-boson mass is taken to be $m_{W}=80.4~\GeV$~\cite{PDG}.
Finally, $\sigma_{t}$ and $\sigma_{W}$ respectively represent the detector resolutions for the top-quark and $W$-boson masses assuming the correct jet matching, as determined from simulated \ttbar\ events in which the jet assignments were fixed unambiguously by matching jets to decay partons.
These values are fixed to $\sigma_{t}=17.6~\GeV$ and $\sigma_{W}=9.3~\GeV$ for reconstruction of detector-level events, and $\sigma_{t}=10.7~\GeV$ and $\sigma_{W}=5.9~\GeV$ for particle-level reconstruction.
The permutation selected using $\chi_\mathrm{min}^2$ successfully matches all jets to top decay partons in approximately 75\% of \ttbar events with exactly six jets, while combinatorial confusion degrades the matching by 10\% in events with one additional jet and up to 30\% in events with three additional jets.
At particle level, the accuracy is higher at 85\% in events with exactly six jets, and 60--75\% in events with seven to nine jets.
 
\subsection{Multi-jet background rejection}
\label{sec:evtsel:multijetrej}
 
The $\chi_\mathrm{min}^2$ is used as a first discriminant to reject background events; multi-jet events produce larger $\chi_\mathrm{min}^2$ values, hence events are rejected if they have $\chi_\mathrm{min}^2>10$.
In addition, the masses of the two reconstructed top quarks are required to be in the range $130<(m_{t,1},m_{t,2})<200~\GeV$.
 
The top--antitop quarks are normally produced back-to-back in the transverse plane, hence the two $b$-tagged jets are produced at large angles. In contrast, the dominant mechanism for producing \bjets in background multi-jet events is gluon splitting $g\rightarrow b\bar{b}$, which typically results in nearly collinear \bjets. Therefore, the $\Delta R$ distance between the two \bjets, $\dRbb$, is required to be larger than 2.
Similarly, the larger of the two angles between a $b$-tagged jet and its associated $W$ boson, $\dRbWmax$, has good discriminating power due to the tendency for the top-quark decay products to be slightly collimated, and thus the requirement $\dRbWmax<2.2$ is imposed.
 
\Tab{\ref{tab:EventSelection:particleLevel}} summarises the selection criteria defining the signal region at reconstruction level.
The fiducial phase space used for unfolding to particle level is defined by the same selections, with two exceptions.
First, no trigger selection need be applied, as the six-jet selection ensures full efficiency.
Second, in place of the $b$-tagging requirements, the `truth' $b$-hadron labelling is used, as described in \Sect{\ref{sec:obj_particle}}.
 
\begin{table}[htbp]
\caption{Summary of selection requirements.}
\label{tab:EventSelection:particleLevel}
\centering
\aboverulesep=0ex
\belowrulesep=0ex
\renewcommand{\arraystretch}{1.4}
\sisetup{retain-explicit-plus}
\sisetup{round-mode = places, round-precision = 2}
\begin{tabular}{
|c|c|
}
\toprule
Requirement & Event selection \\
\midrule
Multi-jet trigger & 6 jets, $\pt > 45~\GeV$ \\
\midrule
Exactly 0 vertex-matched isolated leptons & $\mu$: $\pt > 15~\GeV$, $|\eta|<2.5$ \\
& $e$: $\pt > 15~\GeV$, $|\eta|<2.47$, excluding $1.37<|\eta|<1.52$ \\
& $\tau$: $\pt > 25~\GeV$, $|\eta|<2.5$ \\
\midrule
At least 6 jets & 6 leading jets: $\pt > 55~\GeV$ \\
& Sub-leading jets: $\pt > 25~\GeV$ \\
\midrule
Exactly 2 $b$-jets & $b$-tagging at 70\% efficiency \\
\midrule
Top mass & $130~\GeV < (\mtl, \mtsl) < 200~\GeV$ \\
\midrule
Reconstructed $\chi^2_\mathrm{min}$ & $\chi^2_\mathrm{min} < 10.0$ \\
\midrule
$\Delta R$ between $b$-jets & $\dRbb > 2.0$ \\
\midrule
Maximum $\Delta R$ between $b$-jet and $W$ & $\dRbWmax <2.2$ \\
\bottomrule
\end{tabular}
\end{table}
 
In the data, 44\,621 events pass the full event selection while the signal purity is predicted to be 68\% for the nominal all-hadronic \ttbar\ sample.
 
\section{Background estimation}
\label{sec:bkgest}
 
The signal region of the resolved all-hadronic topology is contaminated by two major sources of background.
The contribution of top-quark pairs decaying into non-hadronic final states is expected to be 5\% of the predicted number of selected all-hadronic events and 3\% of the total data yield. The non-hadronic contribution is estimated using the same MC simulated samples as for the signal but filtering instead for at least one leptonic $W$-boson decay.
The total single-top-quark contribution is estimated to be below 2\% of the selected data and well within both the MC and data statistical uncertainties. For this reason it is not considered further.
 
Multi-jet production forms the most significant source of background contamination, at about a third of the total number of selected events.
This is estimated using a data-driven procedure, as described below.
 
\subsection{Data-driven estimate of multi-jet background}
 
The estimate of the multi-jet background component uses the `ABCD method', which can be applied whenever there are two variables that each provide good signal--background discrimination, while their distributions in the background process are uncorrelated.
A similar method was used in previous measurements~\cite{TOPQ-2012-15,TOPQ-2016-09}.
The best performing pair of discriminating variables are the \bjet\ multiplicity ($\nbjets$) and a combination of the two top-quark-candidate masses. The masses of the two top-quark candidates are used to define two different mass regions as described in Table~\ref{tab:2d_mass_reg}.
 
\begin{table}[htbp]
\caption{Definition of the mass region based on the $\mt$ of the two top-quark candidates.}
\label{tab:2d_mass_reg}
\centering
\aboverulesep=0ex
\belowrulesep=0ex
\begin{tabular}{|l|l|}
\toprule
Mass region & Condition \\
\midrule
Tail & At least one top quark with $\mt < 120~\GeV$ or $\mt > 250~\GeV$ \\
Peak & Both top quarks have $130~\GeV < \mt < 200~\GeV$ \\
\bottomrule
\end{tabular}
\end{table}
 
The two variables identify six different regions as shown in Table~\ref{tab:RefinedABCD}.
The signal region is region $D$, defined by $\nbjets = 2$ and $130~\GeV < \mt < 200~\GeV$ for both top-quark candidates, together with the other criteria in \Tab{\ref{tab:EventSelection:particleLevel}}.
Background control regions are defined by a lower \bjet\ multiplicity and/or in the sidebands of the top-quark-candidate mass distribution.
In the control regions, at least one top-quark candidate must satisfy $\mt<120~\GeV$ or $\mt>250~\GeV$.
Excluding events where one top-quark candidate is in the signal region mass window and the other falls in either of the intermediate ranges $120~\GeV<\mt<130~\GeV$ or $200~\GeV<\mt<250~\GeV$ strongly reduces the signal contamination in the control regions with a negligible increase in the total statistical uncertainty, improving the overall robustness of the estimate.
 
\begin{table}[htbp]
\caption{Division into orthogonal regions according to the $\nbjets$ variable and a combination of the two top-quark masses as defined in Table~\ref{tab:2d_mass_reg}.}
\label{tab:RefinedABCD}
\centering
\aboverulesep=0ex
\belowrulesep=0ex
\begin{tabular}{|c|c|c|}
\toprule
& Tail & Peak \\
\midrule
$\nbjets = 0$ & $A_0$  & $B_0$  \\
$\nbjets = 1$ & $A_1$  & $B_1$  \\
$\nbjets = 2$ & $C$    & $D$  \\
\bottomrule
\end{tabular}
\end{table}
 
The background is estimated independently for each bin in the measured distributions, while the total expected multi-jet yield is estimated from the inclusive yields in the control regions.
The differential background estimate $D$ in one bin of a generic observable $X$ is defined as:
\begin{equation*}
D(X) = \frac{B_1(X) \cdot C(X)}{A_1(X)},
\label{eq:simpleABCD}
\end{equation*}
where the control region background yields $\{A_1,B_1,C\}$ are determined by subtracting the MC $\ttbar$ predictions (of all decay modes) from the data yields in each region.
 
A parallel estimate $D'$ is made using regions $A_0$ and $B_0$ to assess the systematic uncertainty of the method, which accounts for potential differences between the kinematic properties of the various flavour components of the multi-jet background.
This is defined as:
\begin{equation}
D'(X) = \frac{B_0(X) \cdot C(X)}{A_0(X)},
\label{eq:qcdsyst}
\end{equation}
such that $\Delta D = D' - D$ gives the systematic uncertainty of the nominal prediction $D$.
 
Table~\ref{tab:CRContamination} shows the fraction of signal and background \ttbar events estimated from MC simulation in the various regions.
More signal contamination is observed in regions with $b$-tagged jets, but sufficient multi-jet background purity is observed in all regions such that signal mismodelling should not substantially bias the multi-jet background prediction.
 
\begin{table*}[htbp]
\caption{Fractional yields from top-quark pair production processes in the different regions, defined by the values assumed by $\nbjets$ and the two top-quark masses $\mt$ as defined in Table~\ref{tab:2d_mass_reg}.}
\label{tab:CRContamination}
\centering
\aboverulesep=0ex
\belowrulesep=0ex
\begin{tabular}{|c|l|c|c|} \toprule
Region & Definition & All-hadronic \ttbar/Data & Non all-hadronic \ttbar/Data\\
\midrule
$A_0$ & $\nbjets = 0$ tail & ~1.87\% & 0.19\% \\
$B_0$ & $\nbjets = 0$ peak & ~0.96\% & 0.08\% \\
$A_1$ & $\nbjets = 1$ tail & ~3.35\% & 0.69\% \\
$B_1$ & $\nbjets = 1$ peak & 16.1\% & 1.16\% \\
$C$   & $\nbjets = 2$ tail & 16.1\% & 2.90\% \\
$D$   & $\nbjets = 2$ peak & 66.1\% & 3.35\% \\
\bottomrule
\end{tabular}
\end{table*}
 
\begin{figure}[htb]
\begin{center}
\subfloat[]{\includegraphics[width=0.32\textwidth]{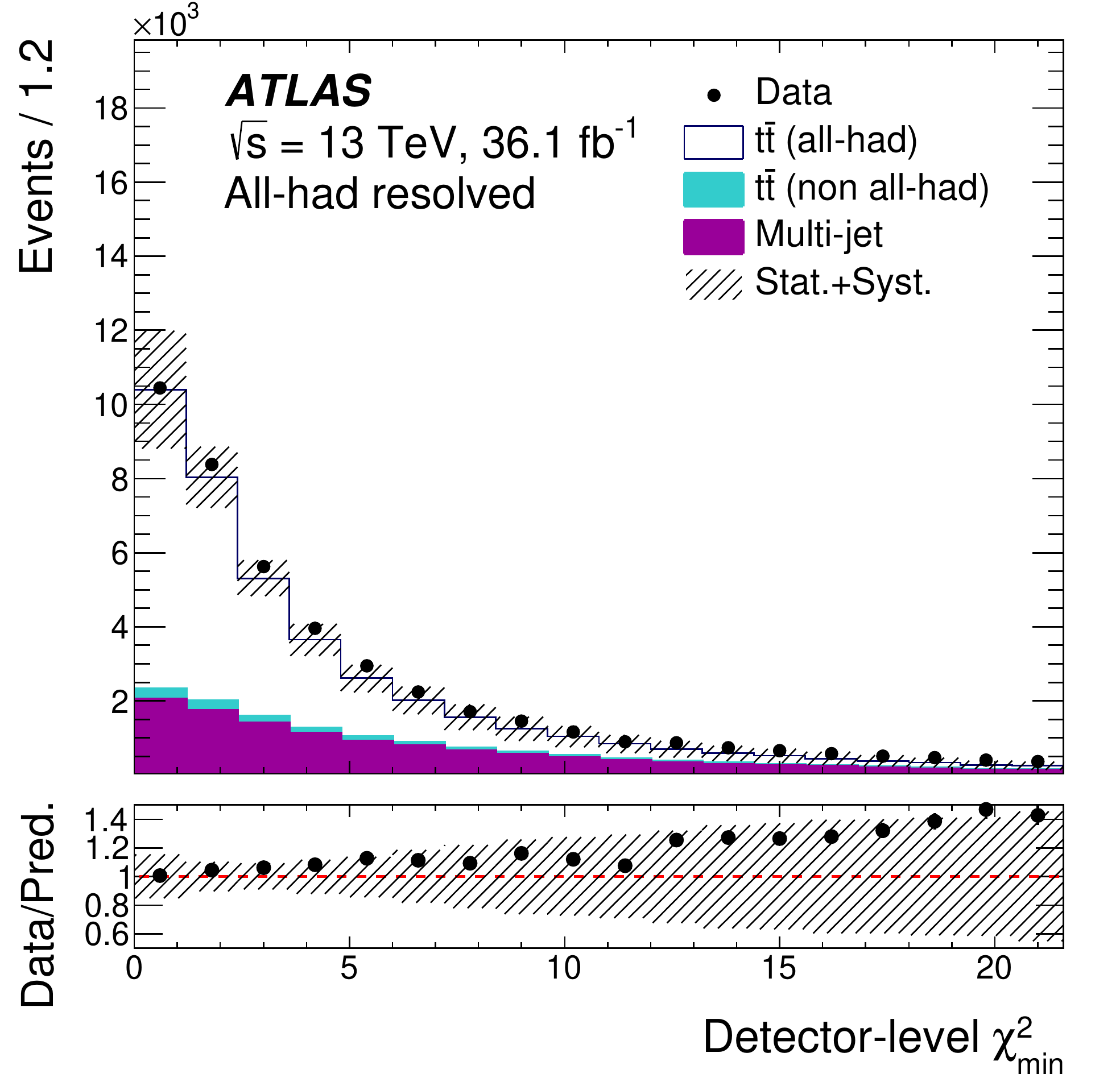}}
\subfloat[]{\includegraphics[width=0.32\textwidth]{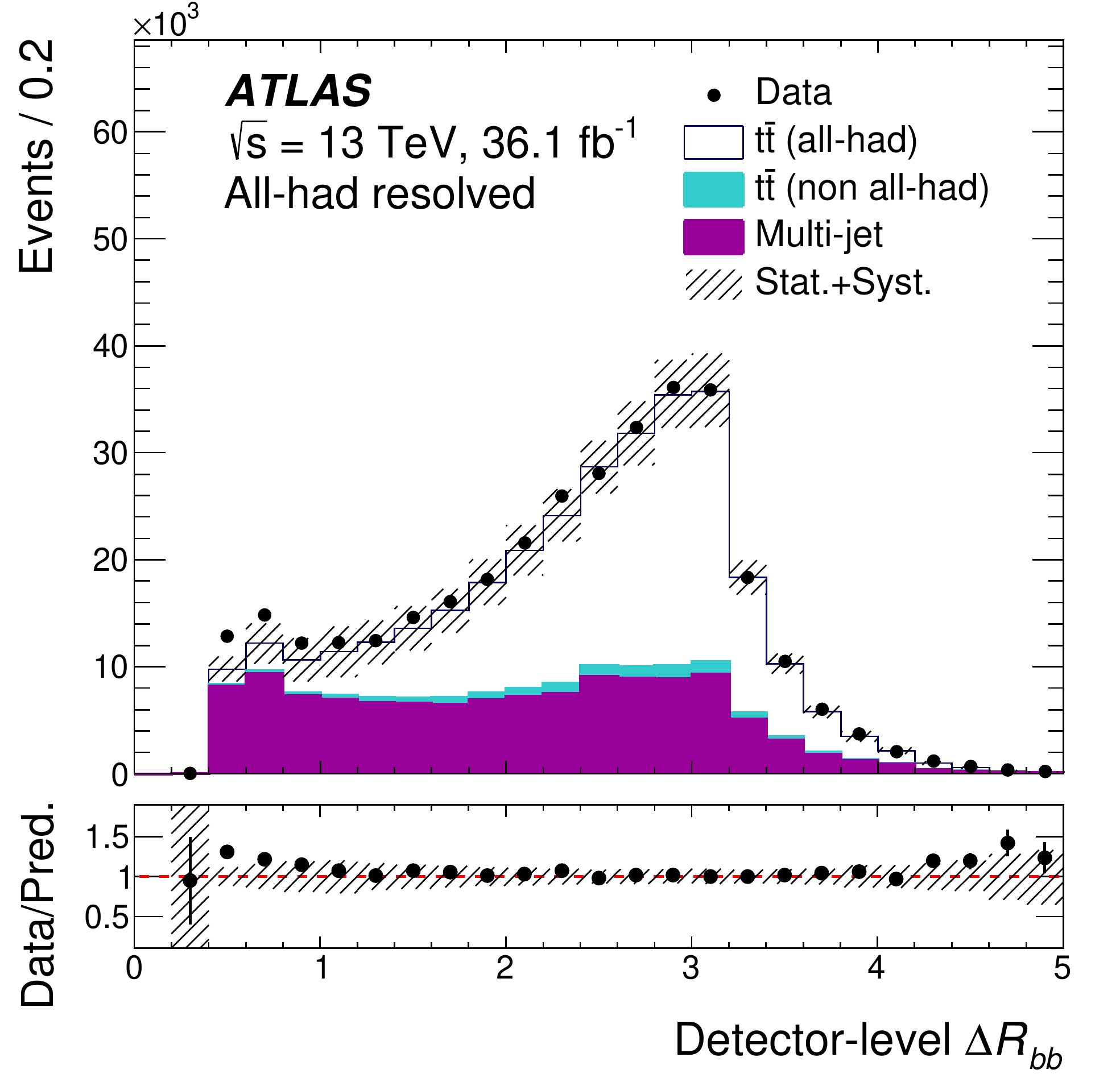}}
\subfloat[]{\includegraphics[width=0.32\textwidth]{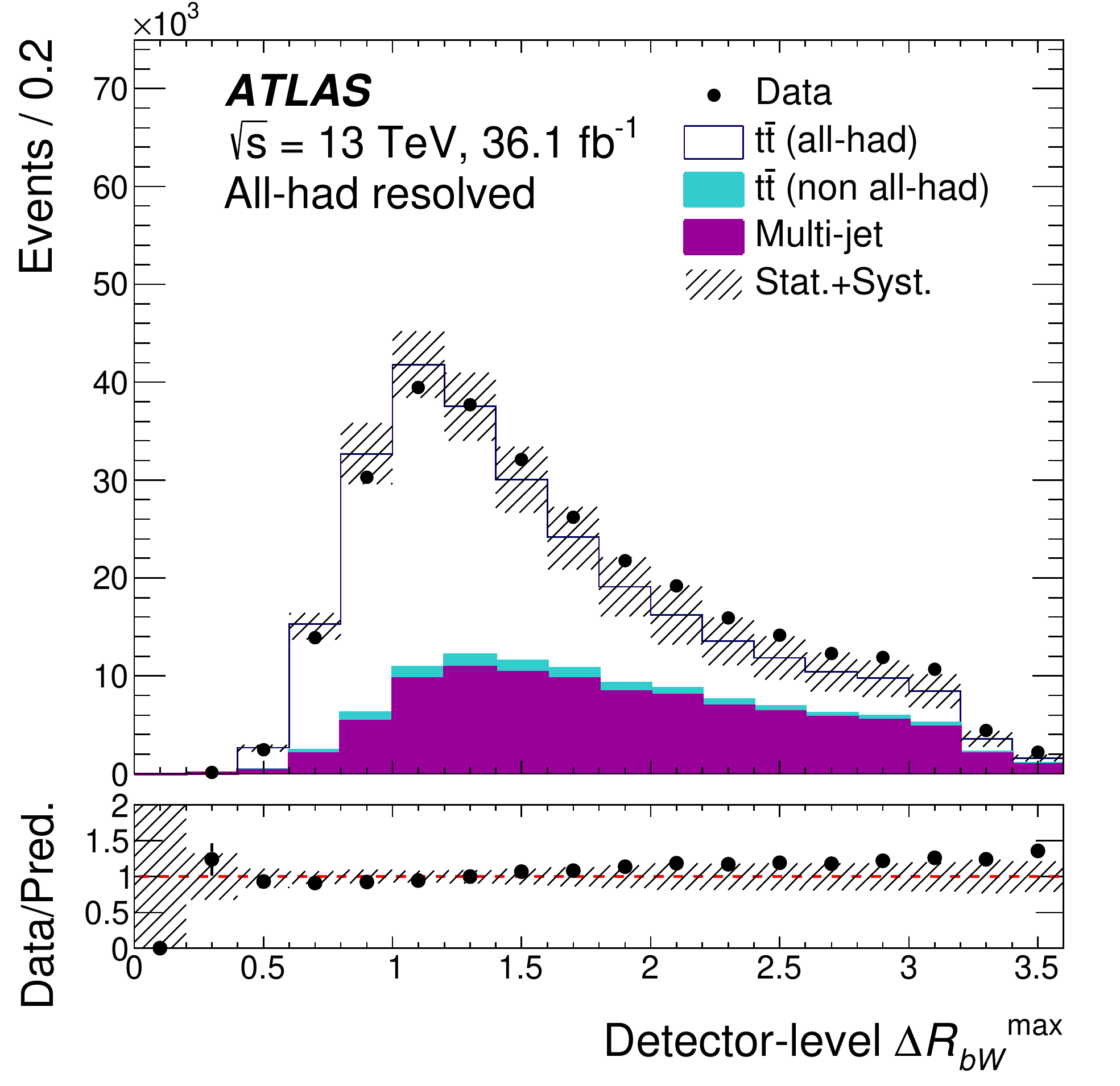}}
 
\subfloat[]{\includegraphics[width=0.32\textwidth]{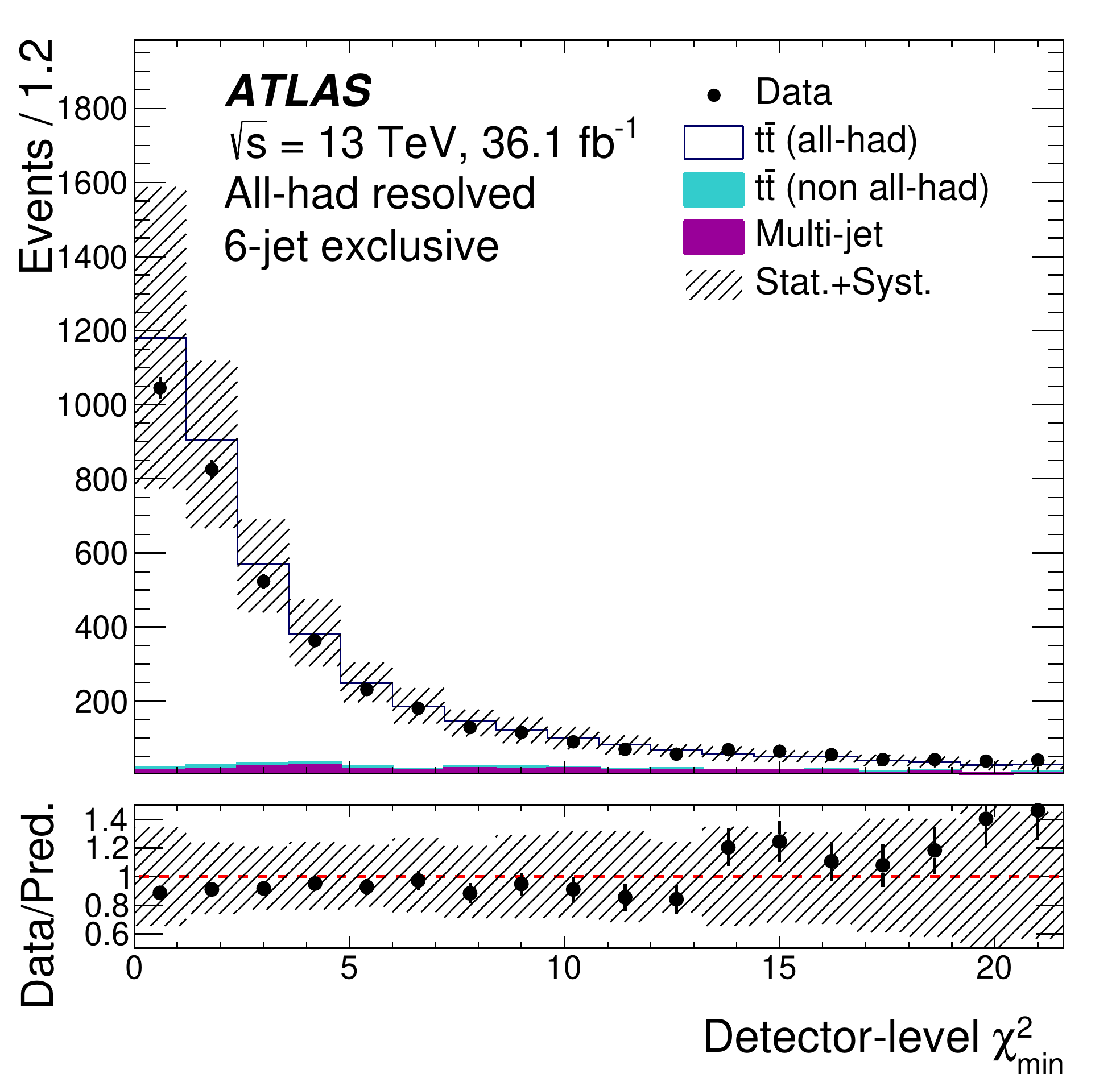}}
\subfloat[]{\includegraphics[width=0.32\textwidth]{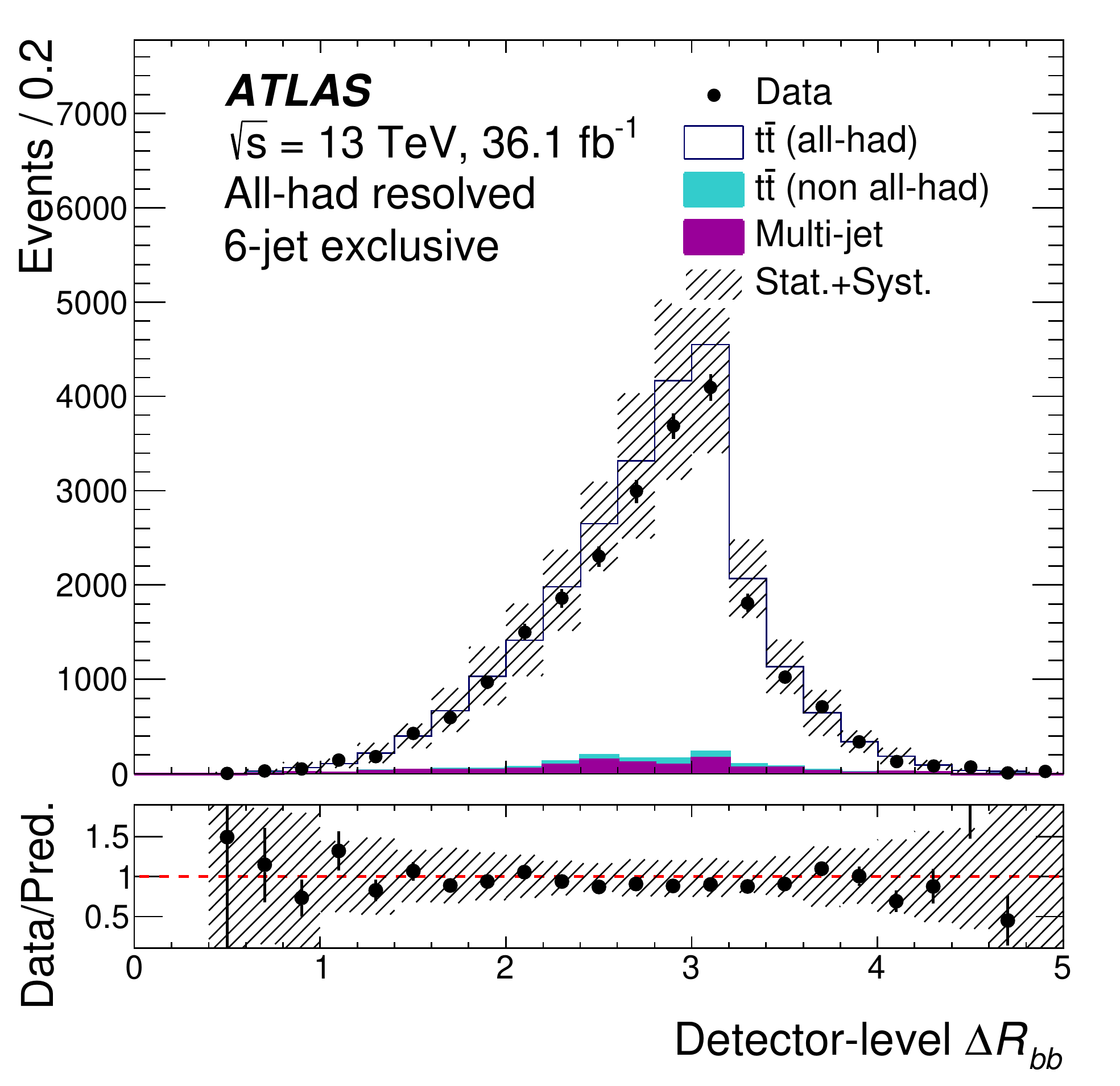}}
\subfloat[]{\includegraphics[width=0.32\textwidth]{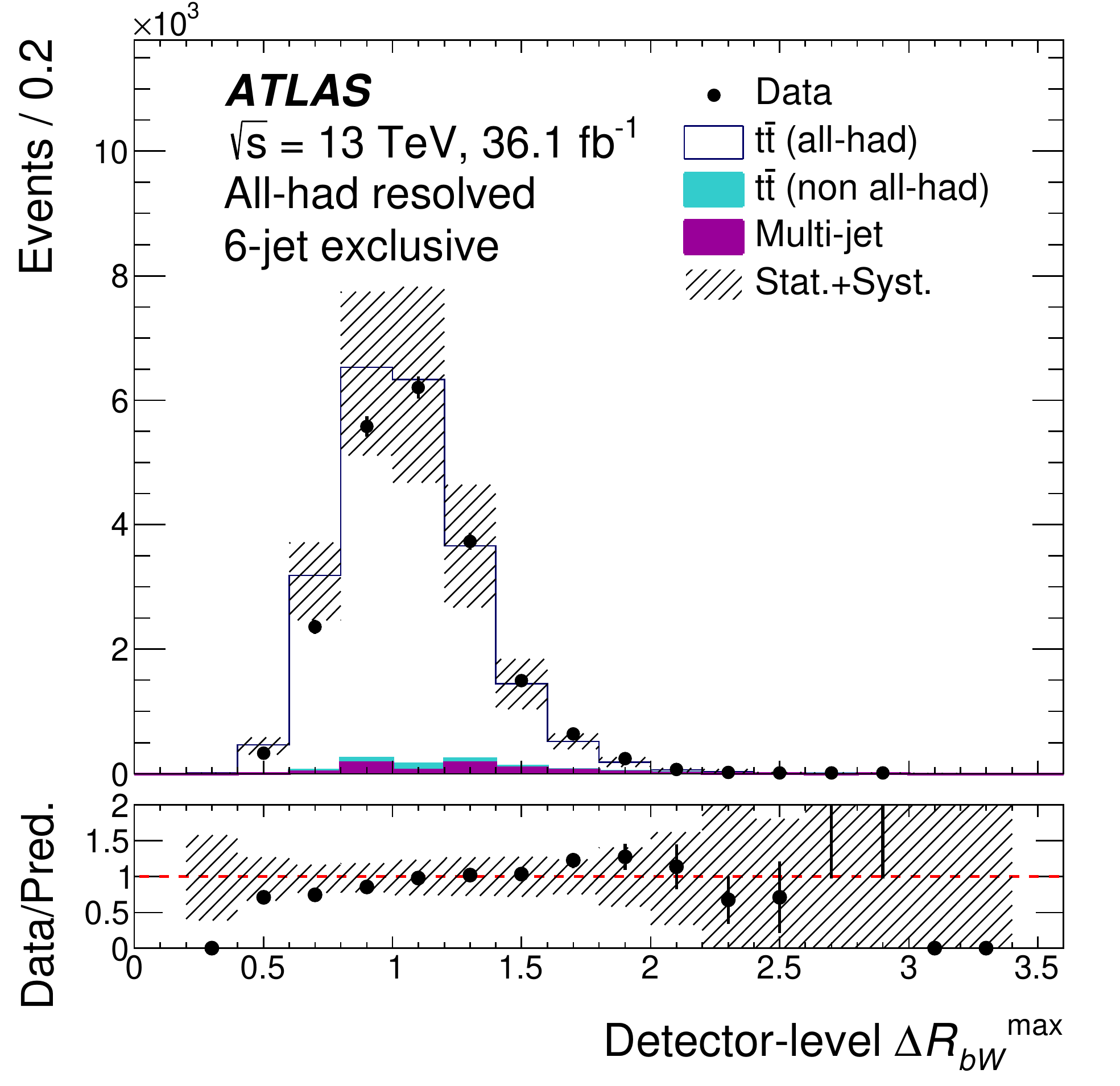}}
\caption{\small{Detector-level distributions in the signal regions as a function of the (left) $\chi^2_\mathrm{min}$, (middle) $\dRbb$ and (right) $\dRbWmax$, for (top) all selected events and (bottom) exclusive six-jet events.
The signal prediction is based on the \PHPYEIGHT\ generator.
The background is the sum of the data-driven multi-jet estimate and the MC-based expectation for the contributions of non-all-hadronic \ttbar\ production processes.
Statistical uncertainties combined with the combined systematic uncertainties for the applied selection are shown in hatching.
Data points are placed at the centre of each bin and overflow events are included in the last bins.}}
\label{fig:discriminatingvariables}
\end{center}
\end{figure}

The spectra of observables used to define the signal region, namely $\chi^2_\mathrm{min}$, $\dRbb$ and $\dRbWmax$ are presented in Figure~\ref{fig:discriminatingvariables}.
These plots are done in an `$N-1$' requirement configuration: the plot for a particular variable is made after applying all signal region requirements except that on the variable being displayed.
The $m_{t,1}$ and $m_{t,2}$ spectra are not shown since those observables are used to define the control regions in the multi-jet estimation.
 
Although the total predicted event yields do not perfectly reproduce the data distributions everywhere, they are compatible with data within the sum in quadrature of the statistical and systematic uncertainties.
The dominant source of uncertainties in the six-jet case is the \ttbar theoretical modelling (parton shower and initial-state radiation), whereas the systematic uncertainty of the multi-jet estimate dominates the inclusive jet distributions.
Together, the comparisons indicate an adequate description of the signal and background processes.
 
The event yields after this selection are shown in \Tab{\ref{tab:EventYields:particleLevel}} for data and the simulated MC signal and background.
 
\begin{table}[htbp]
\caption{
Event yields for data, signal and background processes after the signal region selection.
Uncertainties are quoted as the sum in quadrature of statistical and detector-level systematic uncertainties.
The composition of the selected events is also given in terms of the fractional contribution of the signal and background processes to the total yield.
}
\label{tab:EventYields:particleLevel}
\centering
\renewcommand{\arraystretch}{1.4}
\aboverulesep=0ex
\belowrulesep=0ex
\newcommand{\MCsigfig}{6}
\newcommand{\MCerrsigfig}{4}
\newcommand*{\numRF}[2]{\num[round-mode=figures,round-precision=#2]{#1}}
\newcommand*{\numRP}[2]{\num[round-mode=places, round-precision=#2]{#1}}
\newcommand{\dataprecisionhighstat}{5}
\newcommand{\dataprecision}{0}
\sisetup{retain-explicit-plus}
\sisetup{round-mode = places, round-precision = 2}
\begin{tabular}{
|c|r@{\,}l |c|}
\toprule
Process & \multicolumn{2}{c|}{Event yield} & Fraction \\
\midrule
$t\bar{t}$ (all-hadronic) &
\numRF{29500}{\MCsigfig} & ${}^{+\numRF{2000}{\MCerrsigfig}}_{-\numRF{2500}{\MCerrsigfig}}$ & 68\%\\
\midrule
$t\bar{t}$ (non-all-hadronic) &
\numRF{1490}{\MCsigfig} & ${}^{+\numRF{140}{\MCerrsigfig}}_{-\numRF{120}{\MCerrsigfig}}$ & ~~3\% \\
\midrule
Multi-jet background &
\numRF{12600}{\MCsigfig} & ${}^{+\numRF{1900}{\MCerrsigfig}}_{-\numRF{1900}{\MCerrsigfig}}$ & 29\% \\
\midrule
Total prediction &
\numRF{43500}{\MCsigfig} & ${}^{+\numRF{2800}{\MCerrsigfig}}_{-\numRF{3000}{\MCerrsigfig}}$ &\\
\midrule
Data &
\multicolumn{2}{c|}{\numRP{44621}{\dataprecision}} & \\
\bottomrule
\end{tabular}
\end{table}
 
\begin{figure}[htbp]
\centering
\includegraphics[width=0.5\textwidth]{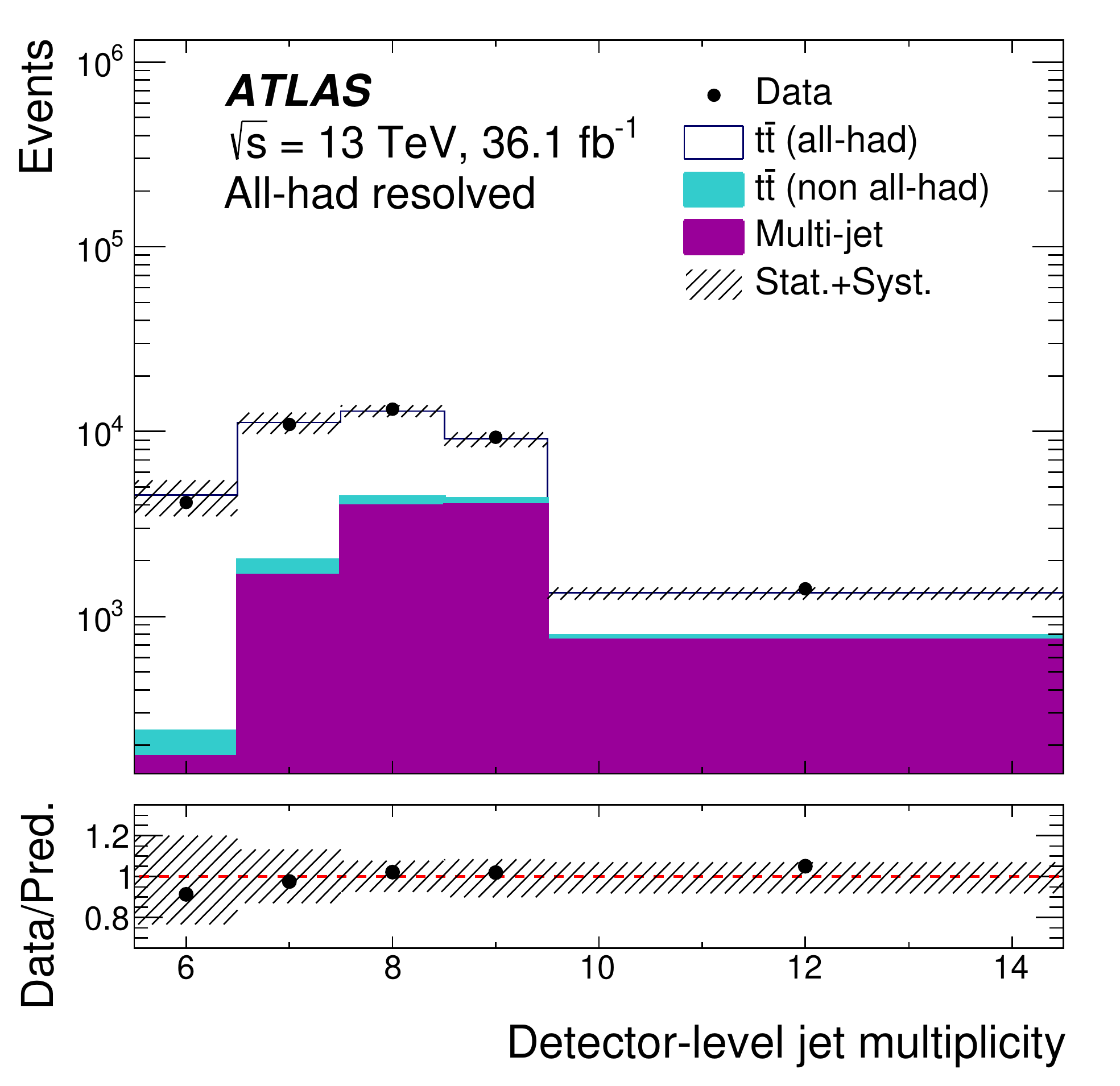}
\caption{
Jet multiplicity in the SR.
The signal prediction is based on the \PHPYEIGHT\ generator. The background is the sum of the data-driven multi-jet estimate  and the MC-based expectation for the contributions of non-all-hadronic \ttbar\ production processes.
Statistical uncertainties combined with systematic uncertainties are shown in hatching.
Data points are placed at the centre of each bin and overflow events are included in the last bin.
\label{fig:bkgest:Njets}
}
\end{figure}

 
Figure~\ref{fig:bkgest:Njets} shows the jet multiplicity distribution for selected events in data compared with the total SM prediction.
This demonstrates that the six-jet bin is essentially pure \ttbar, with negligible multi-jet background contamination, and in fact the nominal MC signal yield slightly exceeds the data yield.
In higher jet multiplicity bins the combinatorial difficulty in correctly identifying the jets from the top-quark decays increases, resulting in a growing multi-jet background contribution.
 
\FloatBarrier
\section{Observables}
\label{sec:observables}
The differential cross-sections are measured as functions of a variety of observables sensitive to the kinematics of top-quark pair production and accompanying radiation.
The all-hadronic final state makes each of the top-quark decay products directly measurable in the detector, thus this final state is especially suited to determining the kinematics of the individual top quarks and of the $\ttbar$ system.
These variables rely on the reconstruction of the $\ttbar$ system, which is described in \Sect{\ref{sec:evtsel:topreco}}.
 
\subsection{Single-differential cross-section measurements}
In the following subsections, the observables used to measure the single-differential cross-sections are described.
 
\subsubsection{Kinematic observables of the top quarks and $\ttbar$ system}
A set of baseline observables is presented. These variables describe the characteristic features of the four-momenta of the individual top quarks and the $\ttbar$ system. The cross-section is measured at both the particle and parton levels as a function of the transverse momentum (\pttl and \pttsl) and absolute value of rapidity (|$\ytl$| and |$\ytsl$|) of the leading and sub-leading top quarks.
For the $\ttbar$ system, the transverse momentum \ptttbar, the absolute value of the rapidity |$\yttbar$| and the mass \mttbar\ are measured.
 
In addition, differential cross-sections as functions of the observables listed below are measured. The following variables provide further information about the properties of the $\ttbar$ system and are sensitive to more than one aspect of $\ttbar$ production:
\begin{itemize}
\item the scalar sum of the \pt of the two top quarks, denoted \HTttbar;
\item the absolute value of the average rapidity of the two top quarks, $|\ytl + \ytsl|/2$, denoted |\boostttbar|;
\item $\exp(|\ytl - \ytsl|)$, denoted \chittbar;
\item the angular distance in $\phi$ between top quarks, denoted \deltaPhittbar.
\end{itemize}
The $|\boostttbar|$ observable is expected to be sensitive to the PDF description, while the $\chittbar$ variable has sensitivity to small rapidity differences between the top quarks and is of particular interest since many processes not included in the SM are predicted to peak at low values of $\chittbar$~\cite{EXOT-2014-15,TOPQ-2016-09}.
 
Differential cross-sections as functions of another set of observables are measured at particle level, such that they may be used to constrain the modelling of the direction and the $\pt$-sharing of the top quarks and their decay products by various matrix element and parton shower MC generators.
These observables comprise directional observables and transverse momentum ratios, as listed below:
\begin{itemize}
\item the cross products of the jet directions $\left|[\hat{b_1}\times(\hat{j_1}\times \hat{j_2})]\times[\hat{b_2}\times(\hat{j_3}\times \hat{j_4})]\right|$, denoted $|\Pcross|$. The ($b$-)jets are those obtained from the top pair reconstruction described in Section~\ref{sec:evtsel:topreco};
\item the out-of-plane momentum defined as the projection of the top-quark three-momentum onto the direction perpendicular to the plane defined by the other top quark and the beam axis ($\hat{z}$) in the laboratory frame, $\left | \vec{p}^{t,1} \cdot (\vec{p}^{t,2} \times \hat{z})/|\vec{p}^{t,2}\times \hat{z}| \right |$~\cite{Pout}, denoted \Pout;
\item the ratio of the \pt of the sub-leading top quark to the \pt of the leading top quark, denoted \zttbar;
\item the ratio of the $W$-boson \pt to the associated top quark's \pt (leading or sub-leading), denoted \RWtttbar;
\item the ratio of the $W$-boson \pt to the associated $b$-quark's \pt (leading or sub-leading), denoted \RWbttbar.
\end{itemize}
These observables were first studied in the 8~\TeV\ \ljets differential cross-section measurement~\cite{TOPQ-2015-06} and were also included in the measurement of boosted top quark pairs in the hadronic signature at 13~\TeV~\cite{TOPQ-2016-09}. By repeating these measurements in the resolved channel, it is possible to complement the results of the latter publication.
Furthermore, the channel used in the analysis described in this paper does not have neutrinos in the final state, avoiding the dependency on the \met, whose resolution is affected by all measured jets in the event. Hence, the resolution is expected to be better for all directional observables such as \Pout, \chittbar\ and \deltaPhittbar. Given that four-momenta are available for all visible decay products, a new variable \Pcross\ is introduced using only the direction of the jets, for which the absolute value is measured.

\subsubsection{Jet observables}
 
A set of jet-related observables is presented. These variables are unfolded at the particle level in the fiducial phase space. The differential cross-section is measured as a function of the number of reconstructed jets ($\njets$). In addition, a set of observables sensitive to the angular and energy correlations between the additional jets and the top quarks is listed below. The additional jets are those jets that are not associated with either top quark by the reconstruction procedure. The closest top quark refers to the top candidate with the smaller \dR\ separation from the jet in question:
\begin{itemize}
\item \dR\ between the leading, sub-leading, sub-subleading extra jet and the closest top quark, denoted \dRxonetc, \dRxtwotc, \dRxthreetc;
\item ratio of the leading, sub-leading, sub-subleading extra jet's \pt\ to the leading top quark's \pt, denoted \Rptxonetl, \Rptxtwotl, \Rptxthreetl;
\item ratio of the leading, sub-leading, sub-subleading extra jet's \pt\ to the leading jet's \pt, denoted \RptxoneJ, \RptxtwoJ, \RptxthreeJ;
\item ratio of \ptttbar{} to the \pt of the leading extra jet, denoted \Rptttbarxone.
\end{itemize}
 
The \dR\ separation is measured relative to the closest top quark, as collinear emissions are favoured, and furthermore the sub-leading top quark is more likely to have lost momentum via a hard emission. The first \pt\ ratio uses the leading top quark as a reference for the hard scale in the event, while the second is sensitive to emissions beyond the first, in particular soft gluons that may not be resolved as jets, allowing a test of resummation effects.
 
Further constraints can be placed on correlations between the angles and between the transverse momenta of additional jets themselves, which are of particular interest for multi-leg matrix element calculations, by measuring differential cross-sections as a function of the following observables:
\begin{itemize}
\item \dR\ between the leading extra jet and the leading jet, denoted \dRxoneJ;
\item \dR\ between the sub-leading, sub-subleading extra jet and the leading extra jet, denoted \dRxtwoX, \dRxthreeX;
\item ratio of the sub-leading, sub-subleading extra jet's \pt\ to the leading extra jet's \pt, denoted \RptxtwoX, \RptxthreeX.
\end{itemize}
Since ISR scales with the partonic centre-of-mass energy, when the leading extra jet is the hardest object in the event its transverse momentum serves well as a reference for the energy scale of the interaction.
 
\subsection{Double-differential measurements}
 
The observables described below are used for double-differential measurements at both the particle and parton levels. The measurements of these observables allow better understanding of correlations between different aspects of the \ttbar\ system kinematics. These combinations are useful for extracting information about PDFs and measuring the top-quark pole mass from the differential cross-section measurements~\cite{CMS-TOP-17-001,TOPQ-2018-17}. The combinations considered are:
\begin{itemize}
\item $\pttl$, $\pttsl$, |$\ytl$|, |$\ytsl$|, $\ptttbar$ and $\absyttbar$ in bins of \mttbar;
\item $\pttl$ in bins of $\pttsl$;
\item |$\ytl$| in bins of |$\ytsl$|.
 
\end{itemize}
Additional observables are measured differentially at the particle level only, as functions of the jet multiplicity, and can be used to tune and constrain the parameters of MC generators. The combinations considered are:
\begin{itemize}
\item $\pttl$, $\pttsl$, $\ptttbar$, $\Pout$, $\dPhittbar$ and $|\Pcross|$ in bins of $\njets$.
\end{itemize}

 
\section{Unfolding strategy}
\label{sec:unfolding}
 
The measured differential cross-sections are obtained from the reconstruction-level distributions using an unfolding technique which corrects for detector and reconstruction effects such as efficiency, acceptance and resolution. The iterative Bayesian unfolding method~\cite{unfold:bayes} as implemented in RooUnfold~\cite{Adye:2011gm} is used.
 
For each observable, the unfolding procedure starts from the number of events observed in data at reconstruction level in bin $j$ of the distribution $N_\mathrm{obs}^j$, from which the background event yield $N_\mathrm{bkg}^j$, estimated as described in Section~\ref{sec:bkgest}, is subtracted.
Then the corrections are applied. All corrections are evaluated using the nominal MC $\ttbar$ simulation (Table~\ref{tab:MC}).
 
\subsection{Unfolding at particle level}
\label{sec:unfolding_particle}
 
As the first step, an acceptance correction is applied. The acceptance correction in bin $j$ is defined as the fraction of signal events reconstructed in this bin that also pass the particle-level selection:
\begin{equation*}
f^j_\mathrm{acc}  \equiv \frac{ N_{ \mathrm{reco} \land \mathrm{part} }^j }{N_\mathrm{reco}^j } {.}
\end{equation*}
 
This correction is a bin-by-bin factor which corrects for events that are generated outside the fiducial phase-space region but pass the reconstruction-level selection.
The resulting distribution is then unfolded to the particle level, defined in \Sect{\ref{sec:evtsel}}.
 
The unfolding step uses as input a migration matrix $\mathcal{M}$ derived from simulated \ttbar{} samples which maps the particle-level bin $i$ in which an event falls to the bin $j$ in which it is reconstructed.
The probability for particle-level events to be reconstructed in the same bin is represented by the elements on the diagonal, while the off-diagonal elements describe the fraction of particle-level events that migrate into other bins.
Therefore, the elements of each row sum to unity (within rounding).
For each observable, the number of bins is based on the resolution of the ATLAS detector and reconstruction algorithms and optimised to perform under stable unfolding conditions.
 
The unfolding is performed using four iterations to balance the unfolding stability relative to the previous iteration and the growth of the statistical uncertainty, which is limited to be below 0.1\%.
 
Finally, an efficiency correction $\epsilon$ is applied to the unfolded spectrum, correcting the result by a bin-by-bin factor to the fiducial phase space.
The efficiency correction in bin $i$ is defined as the fraction of the events generated in a particle-level bin $i$ that pass the inclusive reconstruction-level selection:
\begin{equation*}
\epsilon^i \equiv \frac{ N_{\mathrm{part} \land \mathrm{reco}}^i }{N_\mathrm{part}^i } {.}
\end{equation*}
This factor corrects for the inefficiency of the event selection and reconstruction.
 
\begin{figure}[htb]
\begin{center}
\subfloat[]{\includegraphics[width=0.45\textwidth]{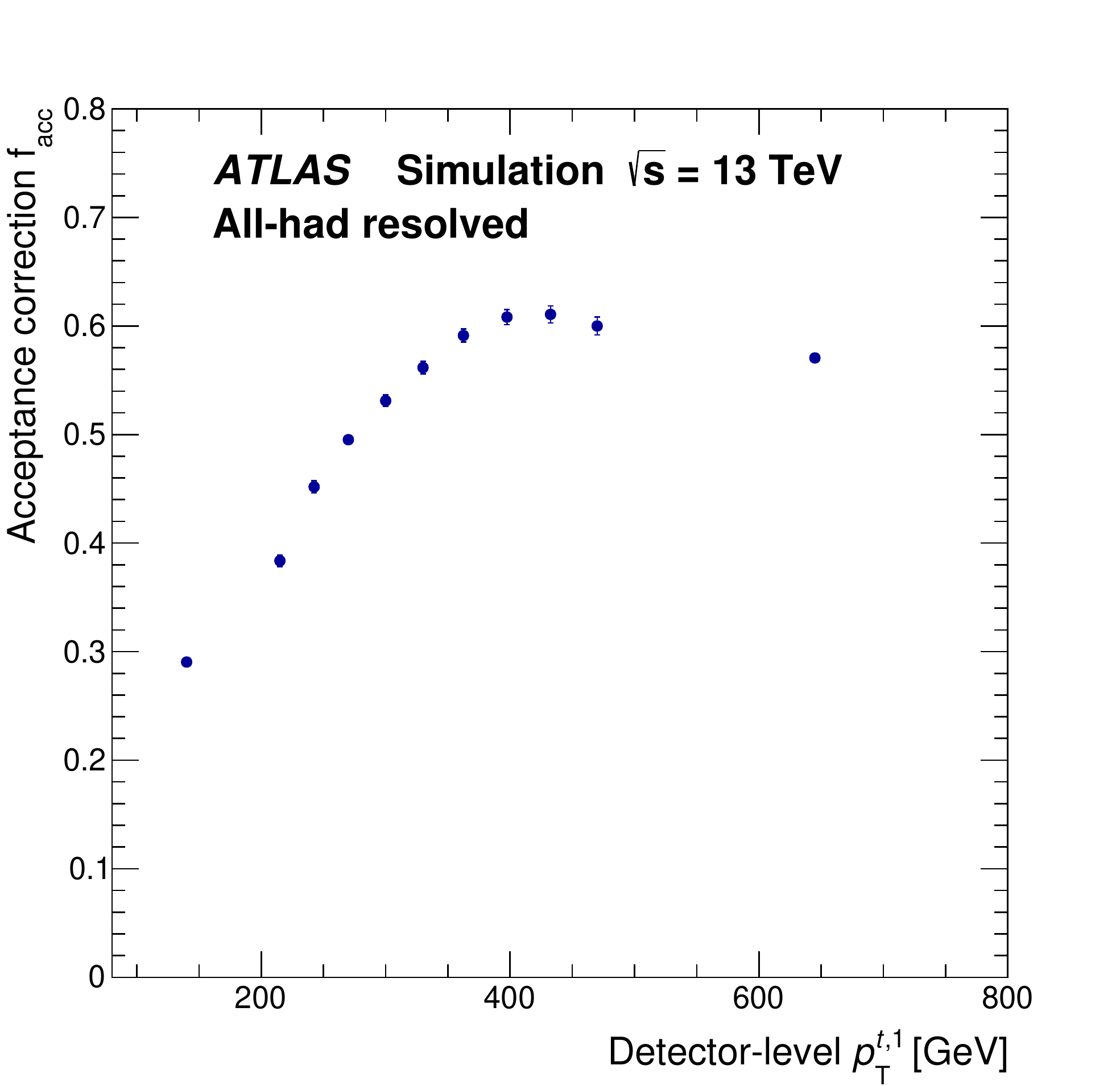}\label{acc:particle_t1_pt}}
\subfloat[]{\includegraphics[width=0.45\textwidth]{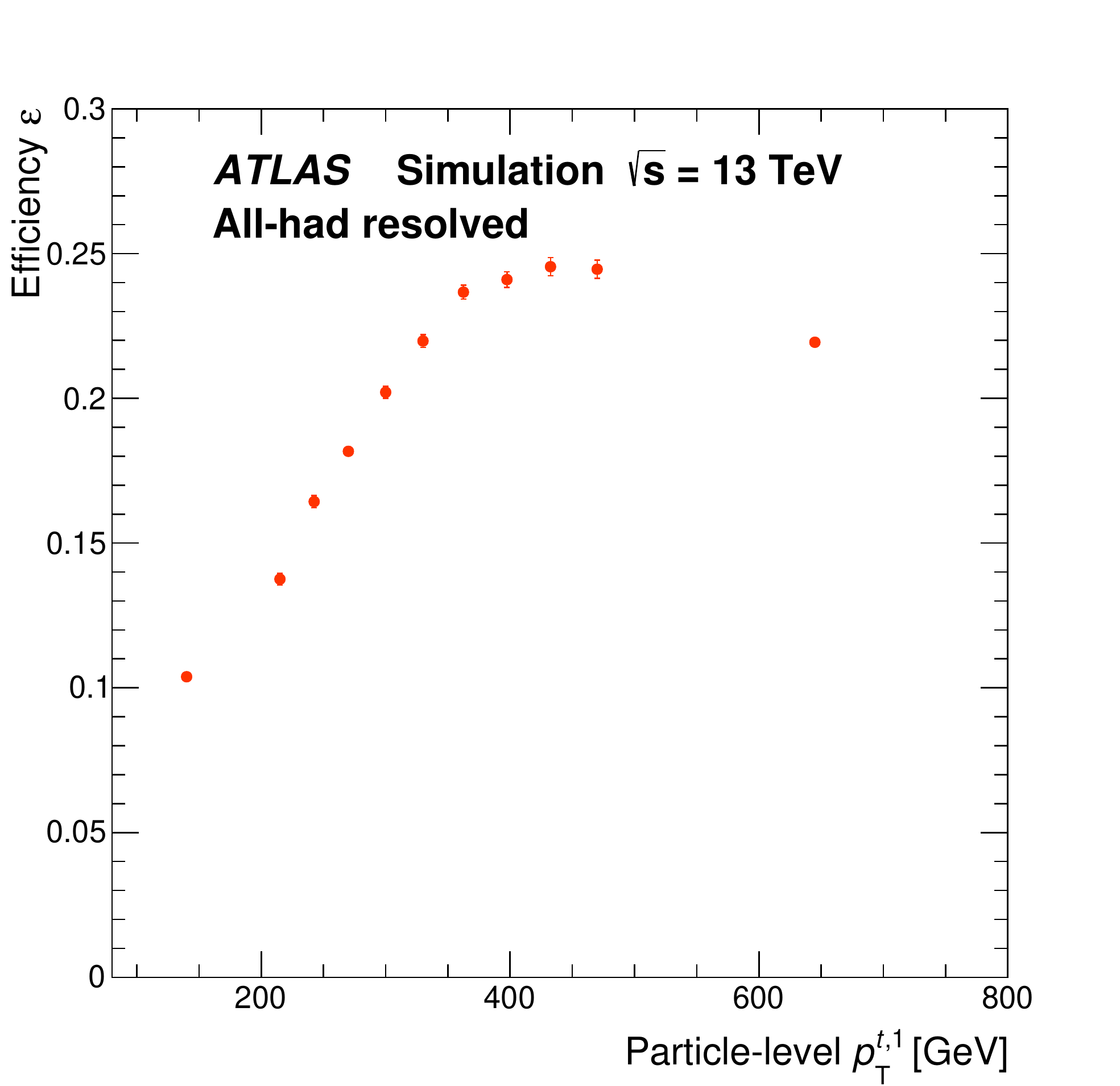}\label{eff:particle_t1_pt}}
 
\subfloat[]{\includegraphics[width=0.45\textwidth]{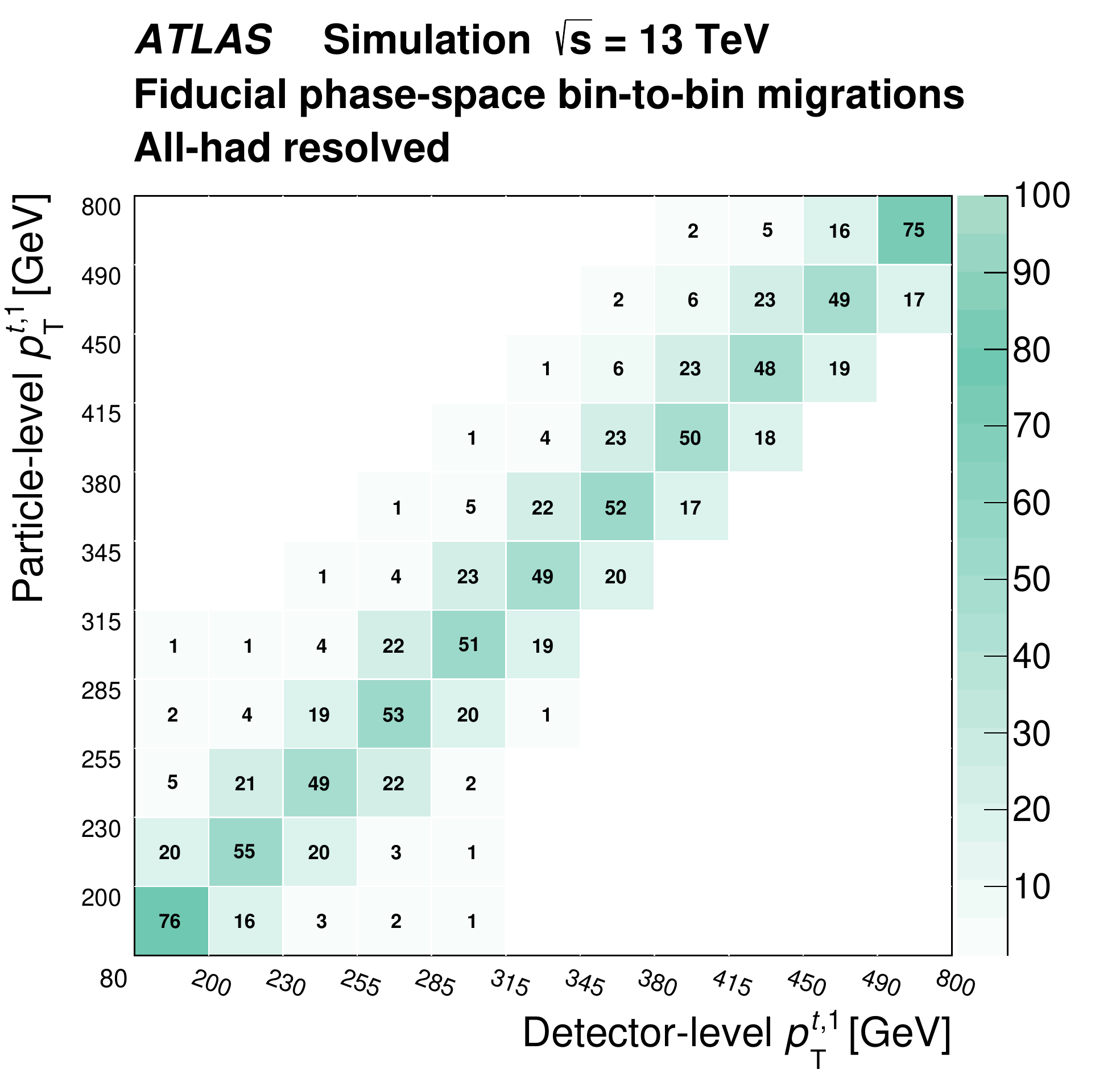}\label{migration:particle_t1_pt}}
\caption{\small{The~(a) acceptance $f_\mathrm{acc}$} and~(b) efficiency $\epsilon$ corrections in bins of detector- and particle-level $\pttl$, respectively, and~(c) the particle-to-detector-level migration matrix (evaluated with the MC $\ttbar$ signal sample) for the transverse momentum of the leading top quark.}
\label{fig:corrections_particle}
\end{center}
\end{figure}

As an example, Figure~\ref{fig:corrections_particle} shows the corrections and the migration matrix for the case of the $\pt$ of the leading top quark.
 
The extraction of the absolute differential cross-section for an observable $X$ at particle level is then summarised by the following expression:
\begin{equation*}
\frac{\mathrm{d}\sigma^\mathrm{fid}}{\mathrm{d}X^i} \equiv \frac{1}{\mathcal{L} \cdot \Delta X^i} \cdot  \frac{1}{\epsilon^i} \cdot \sum_j \mathcal{M}^{-1} \cdot  f_\mathrm{acc}^j \cdot \left(N_\mathrm{obs}^j - N_\mathrm{bkg}^j\right)\hbox{,}
\end{equation*}
where the index $j$ iterates over bins of observable $X$ at reconstruction level while the index $i$ labels bins at particle level, $\Delta X^i$ is the bin width, $\mathcal{L}$ is the integrated luminosity, and the inverted migration matrix as obtained with the iterative unfolding procedure is symbolised by $\mathcal{M}^{-1}$.
The integrated fiducial cross-section $\sigma^\mathrm{fid}$ is obtained by integrating the unfolded cross-section over all bins, and its value is used to compute the normalised differential cross-sections:
\begin{equation*}
\frac{1}{\sigma^\mathrm{fid}}\cdot\frac{\mathrm{d}\sigma^\mathrm{fid}}{\mathrm{d}X^i}\hbox{.}
\end{equation*}
\FloatBarrier
 
\subsection{Unfolding at parton level}
\label{sec:unfolding_parton}
 
The measurements are extrapolated to the full phase space of the $\ttbar$ system using the same procedure as extrapolation to the fiducial phase space.
The binning is re-optimised because of the different resolution; this leads to similar migration matrices.
Since in this case the measurements are unfolded to the full phase space, the acceptance correction is irrelevant, but large efficiency corrections are needed due to the larger extrapolation.
 
\begin{figure}[htb]
\begin{center}
\subfloat[]{\includegraphics[width=0.45\textwidth]{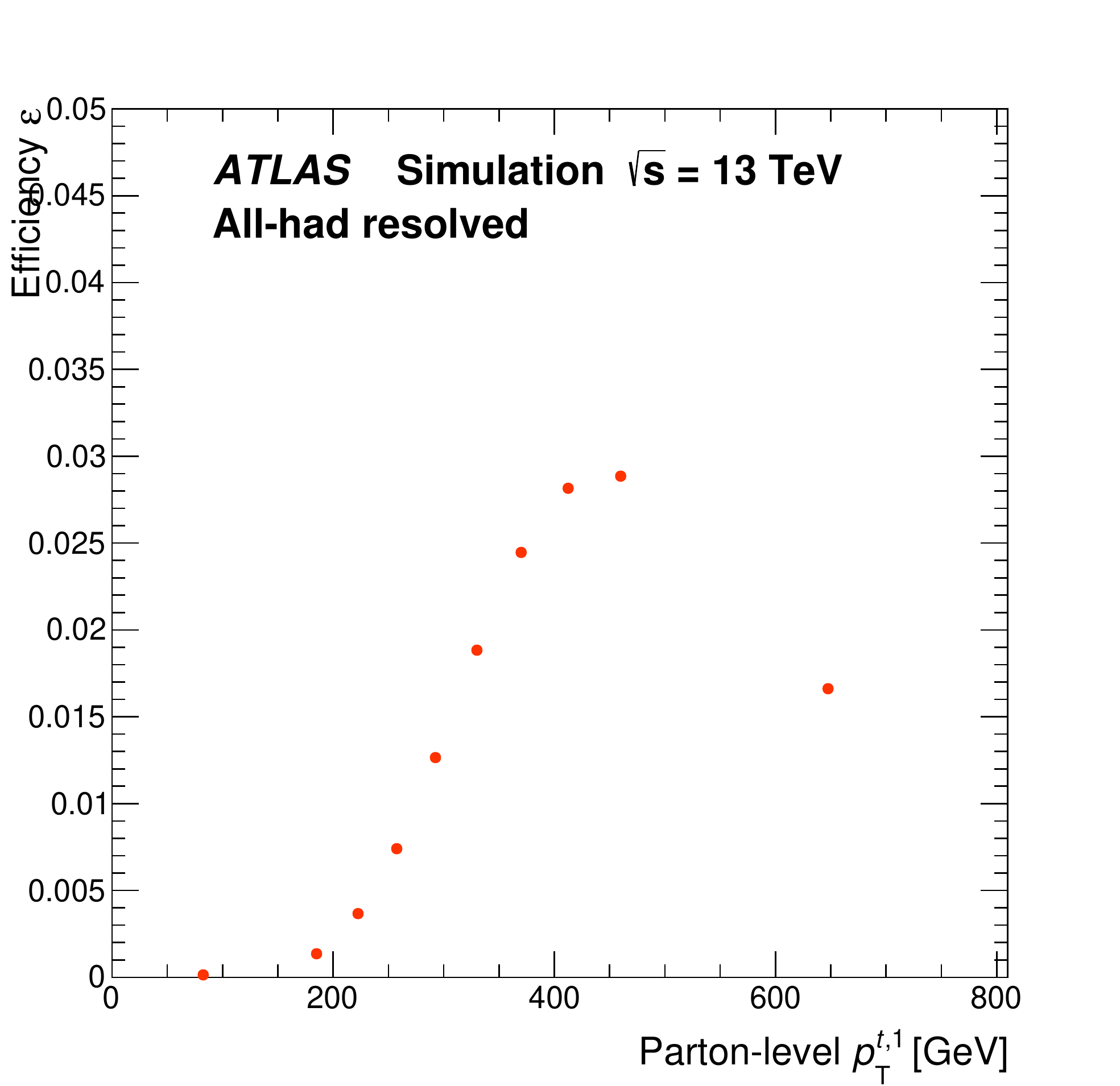}\label{eff:parton_t1_pt}}
\subfloat[]{\includegraphics[width=0.45\textwidth]{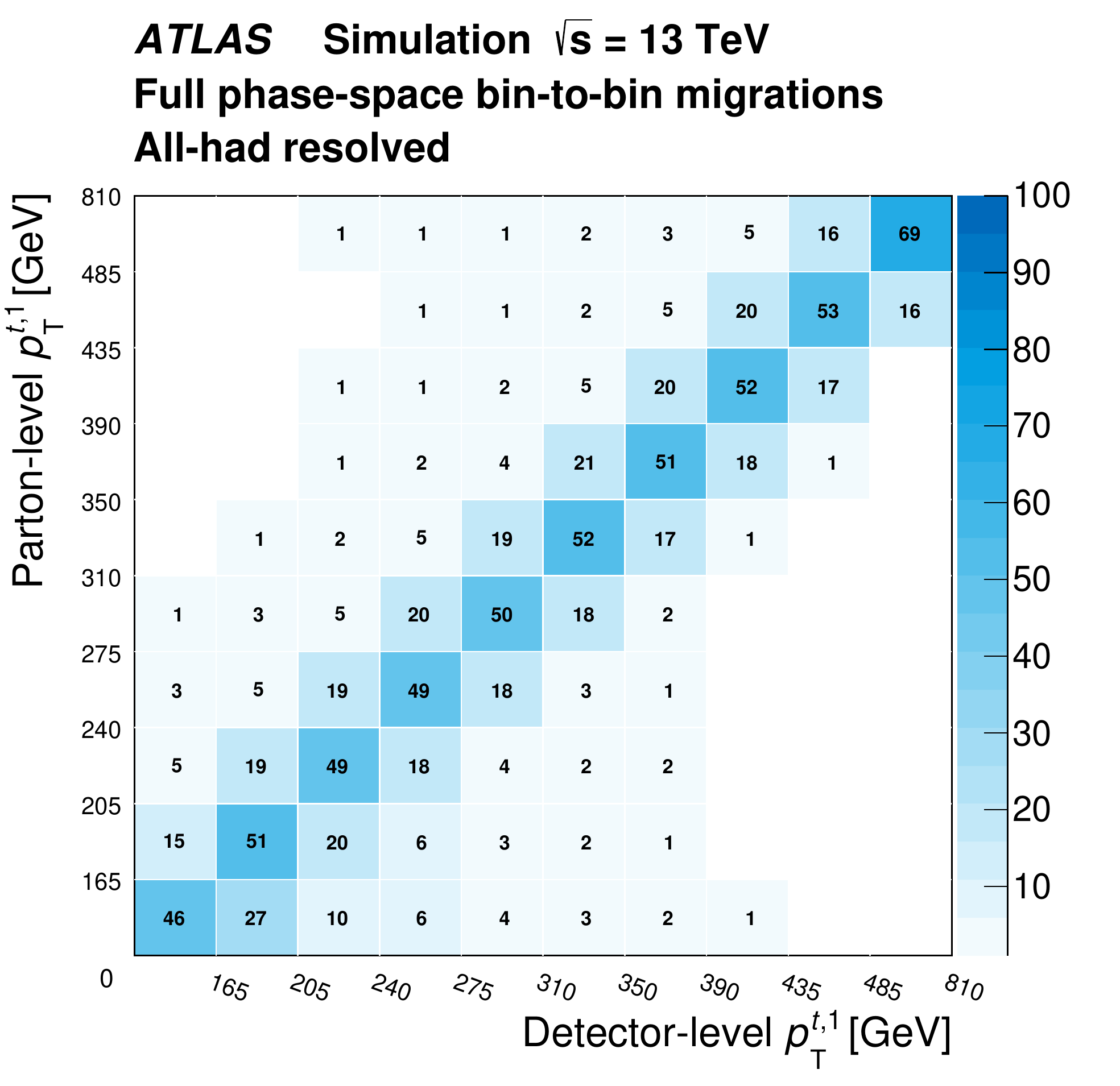}\label{migration:parton_t1_pt}} \\
\caption{\small{The~(a) efficiency $\epsilon$ corrections in bins of the parton-level $\pttl$ and (b) parton-to-detector-level migration matrix (evaluated with the MC $\ttbar$ signal sample) for the transverse momentum of the leading top quark. The acceptance correction $f_\mathrm{acc}$ is identically 1 and is not displayed.}}
\label{fig:corrections_parton}
\end{center}
\end{figure}

As an example, Figure~\ref{fig:corrections_parton} shows the efficiency corrections and the migration matrix for the case of the $\pt$ of the leading top quark.
 
The unfolding procedure is summarised by:
\begin{equation*}
\frac{\mathrm{d}\sigma^\mathrm{full}}{\mathrm{d}X^i} \equiv \frac{1}{\mathcal{L} \cdot \mathcal{B} \cdot \Delta X^i} \cdot  \frac{1}{\epsilon^i} \cdot \sum_j \mathcal{M}^{-1} \cdot \left(N_\mathrm{obs}^j - N_\mathrm{bkg}^j\right)\hbox{,}
\end{equation*}
where the index $j$ iterates over bins of observable $X$ at reconstruction level while the index $i$ labels bins at the parton level, $\Delta X^i$ is the bin width, $\mathcal{B}=0.456$ is the all-hadronic branching ratio~\cite{PDG}, $\mathcal{L}$ is the integrated luminosity, and the inverted migration matrix as obtained with the iterative unfolding procedure is symbolised by $\mathcal{M}^{-1}$.
 
 
\FloatBarrier
\section{Systematic uncertainties}
\label{sec:syst}
 
Several sources of systematic uncertainty affect the measured differential cross-sections. The systematic uncertainties due to detector effects and the ones related to the modelling of the signal and background MC components are found to be more relevant than uncertainties from the unfolding procedure (described in Section~\ref{sec:unfolding}).
 
Each systematic uncertainty is evaluated before and after the unfolding procedure. Deviations from the nominal predictions were evaluated separately for the upward (+1 standard deviation) and downward (-1 standard deviation) variations for each bin of each observable; in the case of a single variation, the single deviation was symmetrised.
 
In the absence of backgrounds, the uncertainty in the predictions ${\Delta S}_\mathrm{syst}$ would be evaluated as the difference between the nominal and alternative MC signal samples using the formula ${\Delta S}_\mathrm{syst} = S_\mathrm{syst} - S_\mathrm{nominal}$.
To account for the effect of the uncertainties in the background yields, the total predictions $T$ need to be compared instead: ${\Delta S}_\mathrm{syst} = T_\mathrm{syst} - T_\mathrm{nominal}$.
The total predictions, for both nominal and systematically varied samples, are given by the sum of the all-hadronic signal sample, the non-all-hadronic contribution and by the multi-jet background estimated when using those $\ttbar$ samples.
Hence, for the estimate of the uncertainty in the signal modelling, the non-all-hadronic events and the multi-jet events are considered fully correlated with the all-hadronic signal sample.
 
The varied MC detector-level spectrum is then unfolded using the background subtraction and corrections evaluated with the nominal $\ttbar$ signal sample and the unfolded result is compared with the corresponding particle- or parton-level distribution.
All detector- and background-related systematic uncertainties are evaluated using the nominal MC generator, while alternative event generators are employed to assess the systematic uncertainties related to the $\ttbar$ system modelling as described in Section~\ref{sec:syst_signal_modelling_uncertainties}.
In the latter case, the corrections derived from the nominal event generator are used to unfold the detector-level spectra of the alternative event generator.
 
The detector-related uncertainties are described briefly in Section~\ref{sec:syst_experimental_uncertainties}, and the uncertainties in the \ttbar{} signal and background modelling are discussed in Sections~\ref{sec:syst_signal_modelling_uncertainties} and~\ref{sec:syst_background_modelling_uncertainties}, respectively.
 
\subsection{Experimental uncertainties}
\label{sec:syst_experimental_uncertainties}
 
The experimental uncertainties quantify the degree to which the simulated detector response is trusted to reproduce collision data for each of the reconstructed objects as well as other empirical uncertainties in object reconstruction and calibration.
For a given source of systematic uncertainty, its impact on the measurement is evaluated by replacing the nominal MC predictions for signal and non-all-hadronic $\ttbar$ background with their systematic variations, then rerunning the multi-jet background estimate and unfolding the data using the nominal correction factors.
Due to the selected final state, the main experimental systematic uncertainties arise from jet reconstruction and flavour tagging. As events with leptons are removed, the uncertainties associated to lepton reconstruction and identification are negligible.
 
The uncertainty in the combined 2015+2016 integrated luminosity is 2.1\%~\cite{ATLAS-CONF-2019-021}.
 
\subsubsection{Jet reconstruction}
The uncertainty in the JES was estimated by using a combination of simulation, test beam data and $in$ $situ$ measurements. Additional contributions from jet flavour composition, $\eta$-intercalibration, punch-through, single-particle response, calorimeter response to different jet flavours and pile-up are taken into account, resulting in 29 independent sub-components of the systematic uncertainty~\cite{PERF-2012-01,PERF-2016-04,PERF-2011-05}.
 
The uncertainty due to the difference in jet-energy resolution (JER) between the data and MC events was evaluated by smearing the MC jet transverse momentum according to the jet resolution as a function of the jet $\pT$ and $\eta$~\cite{ATL-PHYS-PUB-2015-015}.
Uncertainties in the efficiency of the JVT criterion were determined from efficiency measurements made on $Z\to ee/\mu\mu$ +jets events~\cite{ATLAS-CONF-2014-018} and are applied as variations of the jet-by-jet efficiency corrections.
 
Given the all-hadronic final state, the JES modelling is the most important source of experimental uncertainties, contributing at the 5--10\% level. The JER systematics are usually at the level of $1\%$, except where inflated by the statistical uncertainties.
 
\subsubsection{$b$-tagging}
Systematic uncertainties associated with tagging jets originating from $b$-quarks are separated into three categories:
the efficiency of the tagging algorithm for tagging $b$-initiated jets, the misidentification rates for jets initiated by $c$-quarks and finally the misidentification rates for jets originating from light-quark flavours.
These efficiencies were estimated from data and parameterised as a function of \pt\ and $\eta$~\cite{ATL-PHYS-PUB-2017-011}.
Uncertainties in the efficiencies arise from factors used to correct for the differences between the
simulation and data in each of the categories.
The uncertainties in the simulation modelling of the $b$-tagging performance are assessed by studying $b$-jets in dileptonic $\ttbar$ events.
While the systematic uncertainties of the $c$-jet and light-jet tagging efficiencies are generally at the sub-percent level, the uncertainty in the $b$-jet tagging efficiency can be as large as 5\%.

\subsection{Signal modelling}
\label{sec:syst_signal_modelling_uncertainties}
The choice of MC generator used in the signal modelling (Table~\ref{tab:MC}) affects the kinematic properties of simulated $\ttbar$ events, the reconstruction efficiencies and the estimate of the multi-jet background.
 
\subsubsection{MC generator: matrix element calculations plus parton shower and hadronisation models}
 
Signal and background $\ttbar$ events simulated with generator configurations other than the nominal one are used to assess the impact of using different NLO matrix element calculations, as well as the impact of different parton shower and hadronisation models.
Consistent detector simulation is used for both the nominal and systematic variations.
 
The uncertainty due to the choice of the generator is determined by unfolding a \MCNPYEIGHT sample using corrections and response matrices from the nominal sample. The unfolded result is then compared with the truth-level spectrum of the \MCNPYEIGHT sample and the relative difference is used as the systematic uncertainty from the ME generator.
 
The uncertainty due to the choice of the parton shower and hadronisation is determined by unfolding a \PHHSEVEN sample using corrections and response matrices from the nominal sample. The unfolded result is then compared with the truth-level spectrum of the \PHHSEVEN sample and the relative difference is used as the systematic uncertainty from the parton shower and hadronisation.
 
The resulting systematic uncertainties are found to depend strongly on the variable being evaluated.
The matrix-element and parton-shower variations are found to be the most significant sources of systematic uncertainty, among all the systematic uncertainties, and usually affect the tails of the distributions by no more than 20\%, although for most distributions the effect is at the percent level.
 
\subsubsection{Initial-state QCD radiation}
The amount of ISR changes the number of jets in the event as well as the transverse momentum of the \ttbar{} system.
To evaluate the uncertainty linked to the modelling of the ISR, $\ttbar$ MC samples with modified ISR modelling are used.
In particular, the unfolding was performed on samples generated similarly to the nominal sample but with the factorisation and renormalisation scales as well as the value of the \hdamp\ parameter co-varied as described in Section~\ref{sec:samples}.
 
In each case, the spectrum unfolded using the nominal sample is compared with the truth-level spectrum of the corresponding ISR sample.
Being at the level of a few percent for most bins, the ISR variations are at most comparable to the parton shower and matrix element uncertainties.
 
\subsubsection{Parton distribution functions}
The impact of the choice of different PDF sets was assessed using the 30-eigenvector set of the PDF4LHC15 prescription~\cite{PDF4LHC}.
The effect of a different PDF choice modifies the efficiency, acceptance and potentially also the response matrix, i.e.\ the corrections used to correct the spectrum at the detector level to the particle level. The PDF choice effect was evaluated by unfolding the nominal \textsc{Powheg+Pythia8} sample using differently PDF-reweighted corrections.
The intra-PDF variations were combined to define a relative uncertainty as
\begin{equation*}
\delta_{\mathrm{intra}} \equiv \frac{ \sqrt{  \sum\limits_{i \in {\mathrm{sets}} } \left(U_i \cdot R_0 - T_0 \right)^2}}{T_0} {,}
\end{equation*}
while the relative inter-PDF variation between NNPDF3.0NLO and the PDF4LHC15 central PDF sets is evaluated as
\begin{equation*}
\delta_{\mathrm{inter}} \equiv \frac{U_\mathrm{NNPDF3.0NLO} \cdot R_0 -T_0}{T_0} {,}
\end{equation*}
where the $0$ ($i$) subscripts denote the PDF4LHC15 central (varied) PDF sets, $R$ represents the distribution at the detector level while $T$ symbolises the distribution at the particle level, and the unfolding procedure is represented by the $U$ factor.
The resulting uncertainties are at the sub-percent level, except for a few variables studied, where uncertainties at the level of 1--2\% are seen in sparsely populated bins of their distributions.
 
\subsubsection{MC generator: sample size}
To account for the limited size of the signal MC sample, pseudo-experiments are used to evaluate the impact of sample size. The event yield in each bin is generated from a Gaussian distribution with mean equal to the yield of the bin and standard deviation equal to the Poisson uncertainty of the bin yield. This smeared spectrum is then unfolded. The procedure is repeated $10\,000$ times, and the final statistical uncertainty is evaluated as the difference between the nominal prediction and the average over the $10\,000$ pseudo-experiments.
The resulting systematic uncertainty was found to be typically below $0.5$\%, increasing to 1--2\% in the tails of some distributions.
 
\subsection{Background modelling}
\label{sec:syst_background_modelling_uncertainties}
 
Two sources of uncertainty in the background predictions are assessed in addition to the effects of the signal modelling uncertainties on the background subtraction in the control regions.
The first is related to the finite number of events used in the evaluation of the background. This uncertainty is treated in the same way as the MC sample size and is listed as `Multi-jet Stat.' in all plots.
The second component represents the intrinsic error of the ABCD method used to estimate the multi-jet background. An alternative background prediction is made, substituting the 0-$b$-tag control regions for the 1-$b$-tag control regions in accord with Eq.~(\ref{eq:qcdsyst}), and used in the background subtraction step of the unfolding. The background systematic error is given by the difference between the unfolded distributions in the two scenarios and symmetrised.
The first bins of \pttsl\ have a larger background contamination and therefore suffer more from this uncertainty.
 
The statistical uncertainty of the multi-jet background estimation is small, usually under 5\%. The systematic uncertainty is usually sub-dominant relative to uncertainties from modelling and JES/JER, occasionally reaching 10\%. In rare cases, the uncertainty can be larger due to a low signal purity in the specific bin; this amplifies the contribution of the background and the corresponding systematic uncertainty.
 
The impact of using a fixed total \ttbar\ cross-section when computing the background prediction is also assessed. In the background estimation, the normalisation is varied by the uncertainty of the inclusive cross-section, which in relative terms is 5.2\% (see Section~\ref{sec:samples}). The corresponding uncertainty in the measurement is very small, normally less than 1\%, given that the background control regions have little contamination from signal.
 
\subsection{Systematic uncertainty summary}
 
A general overview of the dominant systematic uncertainties that affect the measurement is reported in Table~\ref{tab:inclusive_unc}. In the table, the systematic uncertainties that affect the inclusive cross-section measurement at both the particle and parton levels, grouped per type, are shown. The total cross-section measured in the fiducial phase space and compared with several MC predictions is reported in Section~\ref{sec:TotalXsParticle}.
 
\begin{table}[hbpt]
\caption{Summary of the main relative uncertainties in the inclusive cross-section measured at the particle and parton levels. The uncertainties are symmetrised.}
\label{tab:inclusive_unc}
\aboverulesep=0ex
\belowrulesep=0ex
\renewcommand*{\arraystretch}{1.1}
\centering
\begin{tabular}{|l|c|c|c|}
\hline
Source &  \multicolumn{2}{c|}{ Uncertainty [\%]} \\
\hline
& Particle level & Parton level  \\
\hline
PS/hadronisation & 8.2 & 7.9 \\
Multi-jet syst. & 7.7 & 7.7\\
JES/JER & 6.7 &  6.7 \\
ISR, PDF & 3.3 & 3.5 \\
ME generator & 2.4 &  5.3 \\
Flavour tagging & 2.2 & 2.2  \\
Luminosity & 2.1 &  2.1 \\
Multi-jet stat. & 0.6  & 0.6 \\
MC signal stat. & 0.3 & 0.3  \\
 
\hline
Stat.\ unc. & 0.7 &  0.7 \\
\hline
Stat.+syst.\ unc. & 14~~~~~ & 15~~~~~  \\
\hline
\end{tabular}
\end{table}

 
The dominant source of uncertainty at both the particle and parton levels is the contribution of the hadronisation component, followed by the contribution of the multi-jet estimation and the JES and JER uncertainties.
 
\FloatBarrier
\section{Results}
\label{sec:results}
 
In this section, the measurements of the differential cross-sections are reported.
First, the overall level of agreement between the measurements and various theoretical predictions is shown in the form of $\chi^2$ tables (\Sect{\ref{sec:results:dataMC}}).
A more detailed discussion of the modelling of individual observables follows (\Sect{\ref{sec:results:obs}}).
Then the total cross-section is compared to several predictions (\Sect{\ref{sec:TotalXsParticle}}).
Finally, comparisons are made between the results of this analysis and other measurements of specific observables (\Sect{\ref{sec:results:other}}).
 
\subsection{Overall assessment of data--MC agreement}
\label{sec:results:dataMC}
 
The level of agreement between the measured differential cross-sections and the theoretical predictions is quantified by calculating $\chi^2$ values. These are evaluated using the total covariance matrices of the uncertainties in the measurement; the uncertainties in the theoretical predictions are not included in this evaluation. The $\chi^2$ is given by the following relation:
\begin{equation*}
\chi^2 = V_{N_{\textrm b}}^{\textrm T} \cdot {\mathrm{Cov}}_{N_{\textrm b}}^{-1} \cdot V_{N_{\textrm b}} \,,
\end{equation*}
where $N_{\textrm b}$ is the number of bins of the spectrum and $V_{N_{\textrm b}}$ is the vector of differences between the measured and predicted cross-sections. ${\mathrm{Cov}}_{N_{\textrm b}}$ represents the covariance matrix.
The $p$-values (probabilities of obtaining a test $\chi^2$ result as extreme as the observed value) are then evaluated from the observed $\chi^2$ and the number of degrees of freedom (NDF), which is $N_{\textrm b}$.
 
The covariance matrix incorporates the statistical uncertainty and the systematic uncertainties from detector, signal and background modelling. It is obtained by performing pseudo-experiments where, in each pseudo-experiment, each bin of the data distribution is varied according to a Poisson distribution. Gaussian-distributed shifts are coherently added for each detector-modelling systematic uncertainty by scaling each Poisson-fluctuated bin with the expected relative variation from the associated systematic uncertainty effect. The varied distribution is then unfolded with the nominal corrections and additional Gaussian-distributed shifts are coherently added for each signal-modelling systematic uncertainty. The signal-modelling shifts are derived by using the expected relative variations from the associated systematic uncertainty to scale each bin of the Poisson-fluctuated distribution unfolded with nominal corrections. Such relative variations are defined as the difference between the generated and the unfolded cross-section of a given alternative model, using nominal corrections in the unfolding. The resulting changes in the unfolded distributions are used to compute this covariance matrix. For the calculation of covariance matrices associated to normalised differential cross-sections, the varied distributions are normalised to unity after all effects are included.
 
If the number of events in a~given bin of a~pseudo-experiment becomes negative due to the effect of the combined systematic shifts, this value is set to zero before the unfolding stage.  Differential cross-sections are obtained by unfolding each varied reconstructed distribution with the nominal corrections, and the results are used to compute the covariance matrix.
 
To compare only the shapes of the measured cross-sections and the predictions, the results are also presented as normalised cross-sections. This treatment reduces the contribution of uncertainties common to all bins of the distributions, highlighting shape differences relative to the absolute case.
For normalised differential cross-sections, $V_{N_{\textrm b}}$ is replaced with $V_{N_{\textrm b}-1}$, which is the vector of differences between data and prediction obtained by discarding the last one of the $N_{\textrm b}$ elements and, consequently,  ${\mathrm{Cov}}_{N_{\textrm b}-1}$ is the $(N_{\textrm b}-1) \times (N_{\textrm b}-1)$ sub-matrix derived from the full covariance matrix of the normalised measurements by discarding the corresponding row and column. The sub-matrix obtained in this way is invertible and allows the $\chi^2$ to be computed. The $\chi^2$ value does not depend on the choice of element discarded for the vector $V_{N_{\textrm b}-1}$ and the corresponding sub-matrix ${\mathrm{Cov}}_{N_{\textrm b}-1}$. In this case, the NDF becomes $N_{\textrm b}-1$.
 
The $\chi^2$ values and their corresponding $p$-values are reported below for differential cross-sections measured at particle level in the fiducial phase space (\Sect{\ref{sec:results:chi2_particle}}) and at parton level in the full phase space (\Sect{\ref{sec:results:chi2_parton}}).
All observables introduced in Section~\ref{sec:observables} are included in these tables.
 
\subsubsection{Cross-sections in the fiducial phase space}
\label{sec:results:chi2_particle}
 
The quantitative comparisons among the single-differential particle-level results and theoretical predictions are shown in Tables~\ref{tab:chisquare:absolute:particle_1D} and \ref{tab:chisquare:relative:particle_1D}.
Overall, the MC generator that gives the best description of several single-differential distributions is \PHHSEVEN, followed by \PHPYEIGHT.
Other predictions are less accurate, with \MCNPYEIGHT and the \PHPYEIGHT Var3cDown variation giving the poorest agreement.
 
It is interesting to note, considering the shape of the distributions, that $\RWtl$ is not well described by any MC prediction while $\RptxthreeX$ is only described accurately by \MCNPYEIGHT. The two top-quark \pt observables are only described correctly by the \PHHSEVEN and \PHPYEIGHT Var3cUp predictions.
 
The results for the double-differential cross-sections are shown in Tables~\ref{tab:chisquare:absolute:particle_2D} and~\ref{tab:chisquare:relative:particle_2D} and demonstrate a larger difference between MC predictions. Again, \PHHSEVEN gives the best agreement overall. \MCNPYEIGHT and \PHPYEIGHT Var3cDown have the poorest agreement among all MC predictions.
 
\begin{table}[ht]
\caption{ Comparison of the measured particle-level absolute single-differential cross-sections with the predictions from several MC generators. For each prediction, a $\chi^2$ and a $p$-value are calculated using the covariance matrix of the measured spectrum. The number of degrees of freedom (NDF) is equal to the number of bins in the distribution.}
\label{tab:chisquare:absolute:particle_1D}
\footnotesize
\centering
\renewcommand\arraystretch{\arrstretch}
\resizebox*{0.7\textwidth}{!}{
\noindent\makebox[\textwidth]{
\begin{tabular}{|c|  r @{/} l c  | r @{/} l c  | r @{/} l c  | r @{/} l c  | r @{/} l c  | r @{/} l c |}
\hline
Observable  & \multicolumn{3}{c|}{\textsc{PWG+PY8}} & \multicolumn{3}{c|}{\textsc{PWG+PY8} Var. Up} & \multicolumn{3}{c|}{\textsc{PWG+PY8} Var. Down} & \multicolumn{3}{c|}{a\textsc{MC@NLO+PY8}} & \multicolumn{3}{c|}{\textsc{Sherpa}} & \multicolumn{3}{c|}{\textsc{PWG+H7}} \\
& \multicolumn{2}{c}{$\chi^{2}$/NDF} &  ~$p$-value  & \multicolumn{2}{c}{$\chi^{2}$/NDF} &  ~$p$-value  & \multicolumn{2}{c}{$\chi^{2}$/NDF} &  ~$p$-value  & \multicolumn{2}{c}{$\chi^{2}$/NDF} &  ~$p$-value  & \multicolumn{2}{c}{$\chi^{2}$/NDF} &  ~$p$-value  & \multicolumn{2}{c}{$\chi^{2}$/NDF} &  ~$p$-value  \\
\hline
$       \chi^{\ttbar}$ & {\ } 3.3 & 7 & 0.86  &  {\ } 4.3 & 7 & 0.74  &  {\ } 6.6 & 7 & 0.47  &  {\ } 3.8 & 7 & 0.80  &  {\ } 7.2 & 7 & 0.41  &  {\ } 5.4 & 7 & 0.61 \\
$ \Delta R^\mathrm{extra1}_\mathrm{jet1}$ & {\ } 11.6 & 12 & 0.48  &  {\ } 6.2 & 12 & 0.90  &  {\ } 17.4 & 12 & 0.14  &  {\ } 46.6 & 12 & $<$0.01  &  {\ } 23.2 & 12 & 0.03  &  {\ } 20.5 & 12 & 0.06 \\
$ \Delta R^\mathrm{extra1}_{t, \mathrm{close}}$ & {\ } 5.5 & 16 & 0.99  &  {\ } 19.2 & 16 & 0.26  &  {\ } 4.0 & 16 & 1.00  &  {\ } 8.3 & 16 & 0.94  &  {\ } 7.5 & 16 & 0.96  &  {\ } 6.1 & 16 & 0.99 \\
$ \Delta R^\mathrm{extra2}_\mathrm{extra1}$ & {\ } 6.2 & 15 & 0.98  &  {\ } 6.5 & 15 & 0.97  &  {\ } 7.3 & 15 & 0.95  &  {\ } 7.6 & 15 & 0.94  &  {\ } 9.3 & 15 & 0.86  &  {\ } 9.9 & 15 & 0.82 \\
$ \Delta R^\mathrm{extra2}_{t, \mathrm{close}}$ & {\ } 11.9 & 9 & 0.22  &  {\ } 21.0 & 9 & 0.01  &  {\ } 8.6 & 9 & 0.48  &  {\ } 7.0 & 9 & 0.64  &  {\ } 5.8 & 9 & 0.76  &  {\ } 7.4 & 9 & 0.59 \\
$ \Delta R^\mathrm{extra3}_\mathrm{extra1}$ & {\ } 2.9 & 6 & 0.82  &  {\ } 4.9 & 6 & 0.55  &  {\ } 2.8 & 6 & 0.83  &  {\ } 4.1 & 6 & 0.66  &  {\ } 3.7 & 6 & 0.72  &  {\ } 6.9 & 6 & 0.33 \\
$ \Delta R^\mathrm{extra3}_{t, \mathrm{close}}$ & {\ } 6.9 & 7 & 0.44  &  {\ } 11.5 & 7 & 0.12  &  {\ } 5.5 & 7 & 0.60  &  {\ } 8.8 & 7 & 0.27  &  {\ } 10.5 & 7 & 0.16  &  {\ } 7.1 & 7 & 0.42 \\
$ \Delta\phi^{\ttbar}$ & {\ } 4.4 & 6 & 0.62  &  {\ } 4.3 & 6 & 0.63  &  {\ } 11.5 & 6 & 0.08  &  {\ } 26.1 & 6 & $<$0.01  &  {\ } 3.9 & 6 & 0.69  &  {\ } 3.9 & 6 & 0.69 \\
$ H_\mathrm{T}^{\ttbar}$ & {\ } 23.4 & 11 & 0.02  &  {\ } 22.8 & 11 & 0.02  &  {\ } 32.8 & 11 & $<$0.01  &  {\ } 24.7 & 11 & 0.01  &  {\ } 13.7 & 11 & 0.25  &  {\ } 8.2 & 11 & 0.69 \\
$ R_\mathrm{Wb}^\mathrm{leading}$ & {\ } 4.8 & 6 & 0.57  &  {\ } 3.7 & 6 & 0.72  &  {\ } 5.3 & 6 & 0.51  &  {\ } 3.1 & 6 & 0.79  &  {\ } 8.0 & 6 & 0.24  &  {\ } 5.5 & 6 & 0.48 \\
$ R_\mathrm{Wb}^\mathrm{subleading}$ & {\ } 4.7 & 6 & 0.59  &  {\ } 4.0 & 6 & 0.68  &  {\ } 5.6 & 6 & 0.47  &  {\ } 2.2 & 6 & 0.90  &  {\ } 3.3 & 6 & 0.77  &  {\ } 4.1 & 6 & 0.67 \\
$ R_\mathrm{Wt}^\mathrm{leading}$ & {\ } 12.6 & 7 & 0.08  &  {\ } 15.1 & 7 & 0.03  &  {\ } 13.7 & 7 & 0.06  &  {\ } 12.1 & 7 & 0.10  &  {\ } 15.8 & 7 & 0.03  &  {\ } 12.5 & 7 & 0.08 \\
$ R_\mathrm{Wt}^\mathrm{subleading}$ & {\ } 2.3 & 6 & 0.89  &  {\ } 1.5 & 6 & 0.96  &  {\ } 3.8 & 6 & 0.71  &  {\ } 3.0 & 6 & 0.81  &  {\ } 4.5 & 6 & 0.61  &  {\ } 5.1 & 6 & 0.53 \\
$ R^\mathrm{pT, extra1}_\mathrm{jet1}$ & {\ } 9.2 & 5 & 0.10  &  {\ } 2.1 & 5 & 0.84  &  {\ } 17.0 & 5 & $<$0.01  &  {\ } 34.3 & 5 & $<$0.01  &  {\ } 3.6 & 5 & 0.60  &  {\ } 3.0 & 5 & 0.70 \\
$ R^\mathrm{pT, extra1}_{t,1}$ & {\ } 15.2 & 7 & 0.03  &  {\ } 3.8 & 7 & 0.80  &  {\ } 25.6 & 7 & $<$0.01  &  {\ } 19.8 & 7 & $<$0.01  &  {\ } 7.8 & 7 & 0.35  &  {\ } 6.4 & 7 & 0.49 \\
$ R^\mathrm{pT, extra2}_\mathrm{extra1}$ & {\ } 15.0 & 6 & 0.02  &  {\ } 19.2 & 6 & $<$0.01  &  {\ } 16.4 & 6 & 0.01  &  {\ } 9.2 & 6 & 0.16  &  {\ } 6.8 & 6 & 0.34  &  {\ } 9.8 & 6 & 0.13 \\
$ R^\mathrm{pT, extra2}_\mathrm{jet1}$ & {\ } 13.2 & 6 & 0.04  &  {\ } 13.2 & 6 & 0.04  &  {\ } 16.2 & 6 & 0.01  &  {\ } 9.0 & 6 & 0.18  &  {\ } 4.9 & 6 & 0.55  &  {\ } 10.9 & 6 & 0.09 \\
$ R^\mathrm{pT, extra2}_{t,1}$ & {\ } 6.9 & 5 & 0.23  &  {\ } 14.6 & 5 & 0.01  &  {\ } 5.9 & 5 & 0.32  &  {\ } 7.5 & 5 & 0.18  &  {\ } 11.3 & 5 & 0.05  &  {\ } 8.1 & 5 & 0.15 \\
$ R^\mathrm{pT, extra3}_\mathrm{extra1}$ & {\ } 10.5 & 5 & 0.06  &  {\ } 17.4 & 5 & $<$0.01  &  {\ } 9.5 & 5 & 0.09  &  {\ } 5.9 & 5 & 0.32  &  {\ } 13.1 & 5 & 0.02  &  {\ } 10.8 & 5 & 0.05 \\
$ R^\mathrm{pT, extra3}_\mathrm{jet1}$ & {\ } 5.6 & 4 & 0.23  &  {\ } 7.2 & 4 & 0.12  &  {\ } 5.9 & 4 & 0.21  &  {\ } 6.2 & 4 & 0.19  &  {\ } 5.2 & 4 & 0.26  &  {\ } 5.7 & 4 & 0.22 \\
$ R^\mathrm{pT, extra3}_{t,1}$ & {\ } 1.7 & 3 & 0.63  &  {\ } 2.3 & 3 & 0.51  &  {\ } 2.0 & 3 & 0.57  &  {\ } 3.8 & 3 & 0.29  &  {\ } 0.5 & 3 & 0.92  &  {\ } 2.1 & 3 & 0.54 \\
$ R^\mathrm{pT, \ttbar}_\mathrm{extra1}$ & {\ } 5.2 & 7 & 0.63  &  {\ } 3.9 & 7 & 0.80  &  {\ } 7.3 & 7 & 0.40  &  {\ } 5.7 & 7 & 0.58  &  {\ } 4.2 & 7 & 0.75  &  {\ } 9.4 & 7 & 0.23 \\
$          Z^{\ttbar}$ & {\ } 3.9 & 5 & 0.56  &  {\ } 11.0 & 5 & 0.05  &  {\ } 5.3 & 5 & 0.37  &  {\ } 10.3 & 5 & 0.07  &  {\ } 12.8 & 5 & 0.02  &  {\ } 5.5 & 5 & 0.36 \\
$  |P_\mathrm{cross}|$ & {\ } 4.5 & 10 & 0.92  &  {\ } 2.6 & 10 & 0.99  &  {\ } 6.8 & 10 & 0.74  &  {\ } 2.8 & 10 & 0.99  &  {\ } 2.9 & 10 & 0.98  &  {\ } 3.0 & 10 & 0.98 \\
$ |P_\mathrm{out}^{t,1}|$ & {\ } 2.7 & 7 & 0.91  &  {\ } 26.3 & 7 & $<$0.01  &  {\ } 5.5 & 7 & 0.60  &  {\ } 14.5 & 7 & 0.04  &  {\ } 6.5 & 7 & 0.48  &  {\ } 3.1 & 7 & 0.88 \\
$           |y^{t,1}|$ & {\ } 2.5 & 6 & 0.86  &  {\ } 3.8 & 6 & 0.71  &  {\ } 2.7 & 6 & 0.84  &  {\ } 4.4 & 6 & 0.62  &  {\ } 1.5 & 6 & 0.96  &  {\ } 4.1 & 6 & 0.66 \\
$           |y^{t,2}|$ & {\ } 3.2 & 6 & 0.78  &  {\ } 2.3 & 6 & 0.89  &  {\ } 3.5 & 6 & 0.75  &  {\ } 2.9 & 6 & 0.82  &  {\ } 4.9 & 6 & 0.55  &  {\ } 6.9 & 6 & 0.33 \\
$         |y^{\ttbar}|$ & {\ } 11.6 & 18 & 0.87  &  {\ } 11.9 & 18 & 0.85  &  {\ } 13.0 & 18 & 0.79  &  {\ } 14.0 & 18 & 0.73  &  {\ } 24.6 & 18 & 0.14  &  {\ } 11.5 & 18 & 0.87 \\
$ |y_\mathrm{boost}^{\ttbar}|$ & {\ } 11.2 & 15 & 0.73  &  {\ } 11.8 & 15 & 0.69  &  {\ } 12.4 & 15 & 0.65  &  {\ } 11.0 & 15 & 0.75  &  {\ } 16.5 & 15 & 0.35  &  {\ } 11.4 & 15 & 0.72 \\
$     N_\mathrm{jets}$ & {\ } 7.3 & 5 & 0.20  &  {\ } 0.9 & 5 & 0.97  &  {\ } 19.6 & 5 & $<$0.01  &  {\ } 24.1 & 5 & $<$0.01  &  {\ } 8.9 & 5 & 0.11  &  {\ } 6.5 & 5 & 0.26 \\
$ p_\mathrm{T}^{t,1} $ & {\ } 22.7 & 11 & 0.02  &  {\ } 19.5 & 11 & 0.05  &  {\ } 27.2 & 11 & $<$0.01  &  {\ } 14.6 & 11 & 0.20  &  {\ } 26.6 & 11 & $<$0.01  &  {\ } 8.2 & 11 & 0.69 \\
$ p_\mathrm{T}^{t,2} $ & {\ } 20.5 & 9 & 0.01  &  {\ } 11.0 & 9 & 0.27  &  {\ } 35.7 & 9 & $<$0.01  &  {\ } 34.7 & 9 & $<$0.01  &  {\ } 2.3 & 9 & 0.99  &  {\ } 7.7 & 9 & 0.56 \\
$         m^{\ttbar} $ & {\ } 17.0 & 9 & 0.05  &  {\ } 12.5 & 9 & 0.19  &  {\ } 22.4 & 9 & $<$0.01  &  {\ } 22.2 & 9 & $<$0.01  &  {\ } 7.6 & 9 & 0.57  &  {\ } 10.6 & 9 & 0.30 \\
$ p_\mathrm{T}^{\ttbar} $ & {\ } 4.8 & 8 & 0.78  &  {\ } 39.7 & 8 & $<$0.01  &  {\ } 7.2 & 8 & 0.51  &  {\ } 18.6 & 8 & 0.02  &  {\ } 15.7 & 8 & 0.05  &  {\ } 5.0 & 8 & 0.75 \\
\hline
\end{tabular}}}
\end{table}
 
\begin{table}[ht]
\caption{ Comparison of the measured particle-level normalised single-differential cross-sections with the predictions from several MC generators. For each prediction, a $\chi^2$ and a $p$-value are calculated using the covariance matrix of the measured spectrum. The number of degrees of freedom (NDF) is equal to the number of bins in the distribution minus one.}
\label{tab:chisquare:relative:particle_1D}
\footnotesize
\centering
\renewcommand\arraystretch{\arrstretch}
\resizebox*{0.7\textwidth}{!}{
\noindent\makebox[\textwidth]{
\begin{tabular}{|c|  r @{/} l c  | r @{/} l c  | r @{/} l c  | r @{/} l c  | r @{/} l c  | r @{/} l c |}
\hline
Observable  & \multicolumn{3}{c|}{\textsc{PWG+PY8}} & \multicolumn{3}{c|}{\textsc{PWG+PY8} Var. Up} & \multicolumn{3}{c|}{\textsc{PWG+PY8} Var. Down} & \multicolumn{3}{c|}{a\textsc{MC@NLO+PY8}} & \multicolumn{3}{c|}{\textsc{Sherpa}} & \multicolumn{3}{c|}{\textsc{PWG+H7}} \\
& \multicolumn{2}{c}{$\chi^{2}$/NDF} &  ~$p$-value  & \multicolumn{2}{c}{$\chi^{2}$/NDF} &  ~$p$-value  & \multicolumn{2}{c}{$\chi^{2}$/NDF} &  ~$p$-value  & \multicolumn{2}{c}{$\chi^{2}$/NDF} &  ~$p$-value  & \multicolumn{2}{c}{$\chi^{2}$/NDF} &  ~$p$-value  & \multicolumn{2}{c}{$\chi^{2}$/NDF} &  ~$p$-value  \\
\hline
$       \chi^{\ttbar}$ & {\ } 3.2 & 6 & 0.78  &  {\ } 3.4 & 6 & 0.75  &  {\ } 6.5 & 6 & 0.37  &  {\ } 4.0 & 6 & 0.67  &  {\ } 7.7 & 6 & 0.26  &  {\ } 3.3 & 6 & 0.77 \\
$ \Delta R^\mathrm{extra1}_\mathrm{jet1}$ & {\ } 10.0 & 11 & 0.53  &  {\ } 5.8 & 11 & 0.88  &  {\ } 16.0 & 11 & 0.14  &  {\ } 51.8 & 11 & $<$0.01  &  {\ } 18.7 & 11 & 0.07  &  {\ } 10.0 & 11 & 0.53 \\
$ \Delta R^\mathrm{extra1}_{t, \mathrm{close}}$ & {\ } 5.4 & 15 & 0.99  &  {\ } 18.6 & 15 & 0.23  &  {\ } 4.1 & 15 & 1.00  &  {\ } 9.7 & 15 & 0.84  &  {\ } 8.1 & 15 & 0.92  &  {\ } 6.1 & 15 & 0.98 \\
$ \Delta R^\mathrm{extra2}_\mathrm{extra1}$ & {\ } 6.3 & 14 & 0.96  &  {\ } 6.0 & 14 & 0.97  &  {\ } 7.6 & 14 & 0.91  &  {\ } 6.0 & 14 & 0.96  &  {\ } 9.0 & 14 & 0.83  &  {\ } 11.4 & 14 & 0.65 \\
$ \Delta R^\mathrm{extra2}_{t, \mathrm{close}}$ & {\ } 12.9 & 8 & 0.12  &  {\ } 18.3 & 8 & 0.02  &  {\ } 9.5 & 8 & 0.30  &  {\ } 6.0 & 8 & 0.65  &  {\ } 4.9 & 8 & 0.77  &  {\ } 4.1 & 8 & 0.84 \\
$ \Delta R^\mathrm{extra3}_\mathrm{extra1}$ & {\ } 3.1 & 5 & 0.68  &  {\ } 3.4 & 5 & 0.63  &  {\ } 3.2 & 5 & 0.67  &  {\ } 4.5 & 5 & 0.48  &  {\ } 2.7 & 5 & 0.75  &  {\ } 7.8 & 5 & 0.17 \\
$ \Delta R^\mathrm{extra3}_{t, \mathrm{close}}$ & {\ } 7.6 & 6 & 0.27  &  {\ } 9.3 & 6 & 0.16  &  {\ } 6.5 & 6 & 0.37  &  {\ } 11.3 & 6 & 0.08  &  {\ } 9.6 & 6 & 0.14  &  {\ } 7.8 & 6 & 0.25 \\
$ \Delta\phi^{\ttbar}$ & {\ } 4.5 & 5 & 0.48  &  {\ } 3.0 & 5 & 0.70  &  {\ } 10.4 & 5 & 0.07  &  {\ } 27.7 & 5 & $<$0.01  &  {\ } 3.3 & 5 & 0.66  &  {\ } 3.8 & 5 & 0.58 \\
$ H_\mathrm{T}^{\ttbar}$ & {\ } 18.8 & 10 & 0.04  &  {\ } 15.8 & 10 & 0.11  &  {\ } 26.7 & 10 & $<$0.01  &  {\ } 27.5 & 10 & $<$0.01  &  {\ } 11.7 & 10 & 0.31  &  {\ } 11.3 & 10 & 0.33 \\
$ R_\mathrm{Wb}^\mathrm{leading}$ & {\ } 5.1 & 5 & 0.41  &  {\ } 3.6 & 5 & 0.61  &  {\ } 5.3 & 5 & 0.38  &  {\ } 3.2 & 5 & 0.67  &  {\ } 7.7 & 5 & 0.17  &  {\ } 4.2 & 5 & 0.52 \\
$ R_\mathrm{Wb}^\mathrm{subleading}$ & {\ } 4.6 & 5 & 0.46  &  {\ } 3.8 & 5 & 0.58  &  {\ } 5.0 & 5 & 0.41  &  {\ } 2.5 & 5 & 0.77  &  {\ } 3.6 & 5 & 0.61  &  {\ } 2.1 & 5 & 0.83 \\
$ R_\mathrm{Wt}^\mathrm{leading}$ & {\ } 12.9 & 6 & 0.04  &  {\ } 14.6 & 6 & 0.02  &  {\ } 13.3 & 6 & 0.04  &  {\ } 14.8 & 6 & 0.02  &  {\ } 16.9 & 6 & $<$0.01  &  {\ } 14.2 & 6 & 0.03 \\
$ R_\mathrm{Wt}^\mathrm{subleading}$ & {\ } 2.1 & 5 & 0.83  &  {\ } 1.2 & 5 & 0.94  &  {\ } 3.1 & 5 & 0.68  &  {\ } 3.5 & 5 & 0.62  &  {\ } 4.9 & 5 & 0.43  &  {\ } 4.1 & 5 & 0.54 \\
$ R^\mathrm{pT, extra1}_\mathrm{jet1}$ & {\ } 9.0 & 4 & 0.06  &  {\ } 1.9 & 4 & 0.76  &  {\ } 15.9 & 4 & $<$0.01  &  {\ } 39.2 & 4 & $<$0.01  &  {\ } 3.1 & 4 & 0.54  &  {\ } 3.1 & 4 & 0.54 \\
$ R^\mathrm{pT, extra1}_{t,1}$ & {\ } 14.7 & 6 & 0.02  &  {\ } 3.5 & 6 & 0.75  &  {\ } 23.7 & 6 & $<$0.01  &  {\ } 22.8 & 6 & $<$0.01  &  {\ } 7.3 & 6 & 0.29  &  {\ } 7.6 & 6 & 0.27 \\
$ R^\mathrm{pT, extra2}_\mathrm{extra1}$ & {\ } 13.3 & 5 & 0.02  &  {\ } 16.5 & 5 & $<$0.01  &  {\ } 14.3 & 5 & 0.01  &  {\ } 9.6 & 5 & 0.09  &  {\ } 6.7 & 5 & 0.25  &  {\ } 7.8 & 5 & 0.17 \\
$ R^\mathrm{pT, extra2}_\mathrm{jet1}$ & {\ } 9.2 & 5 & 0.10  &  {\ } 9.7 & 5 & 0.08  &  {\ } 11.8 & 5 & 0.04  &  {\ } 4.4 & 5 & 0.50  &  {\ } 4.2 & 5 & 0.52  &  {\ } 6.4 & 5 & 0.27 \\
$ R^\mathrm{pT, extra2}_{t,1}$ & {\ } 6.5 & 4 & 0.16  &  {\ } 12.9 & 4 & 0.01  &  {\ } 5.0 & 4 & 0.29  &  {\ } 6.4 & 4 & 0.17  &  {\ } 10.9 & 4 & 0.03  &  {\ } 8.4 & 4 & 0.08 \\
$ R^\mathrm{pT, extra3}_\mathrm{extra1}$ & {\ } 11.3 & 4 & 0.02  &  {\ } 15.0 & 4 & $<$0.01  &  {\ } 10.7 & 4 & 0.03  &  {\ } 4.1 & 4 & 0.39  &  {\ } 11.8 & 4 & 0.02  &  {\ } 13.0 & 4 & 0.01 \\
$ R^\mathrm{pT, extra3}_\mathrm{jet1}$ & {\ } 5.7 & 3 & 0.13  &  {\ } 6.1 & 3 & 0.11  &  {\ } 6.0 & 3 & 0.11  &  {\ } 5.7 & 3 & 0.13  &  {\ } 4.4 & 3 & 0.22  &  {\ } 5.9 & 3 & 0.12 \\
$ R^\mathrm{pT, extra3}_{t,1}$ & {\ } 1.4 & 2 & 0.49  &  {\ } 1.9 & 2 & 0.39  &  {\ } 1.4 & 2 & 0.51  &  {\ } 2.0 & 2 & 0.37  &  {\ } 0.4 & 2 & 0.83  &  {\ } 1.2 & 2 & 0.55 \\
$ R^\mathrm{pT, \ttbar}_\mathrm{extra1}$ & {\ } 5.4 & 6 & 0.49  &  {\ } 3.5 & 6 & 0.75  &  {\ } 7.5 & 6 & 0.28  &  {\ } 6.2 & 6 & 0.40  &  {\ } 4.5 & 6 & 0.61  &  {\ } 10.7 & 6 & 0.10 \\
$          Z^{\ttbar}$ & {\ } 3.9 & 4 & 0.43  &  {\ } 10.7 & 4 & 0.03  &  {\ } 5.2 & 4 & 0.27  &  {\ } 12.1 & 4 & 0.02  &  {\ } 14.0 & 4 & $<$0.01  &  {\ } 4.6 & 4 & 0.33 \\
$  |P_\mathrm{cross}|$ & {\ } 4.1 & 9 & 0.91  &  {\ } 2.2 & 9 & 0.99  &  {\ } 5.8 & 9 & 0.76  &  {\ } 3.0 & 9 & 0.96  &  {\ } 2.9 & 9 & 0.97  &  {\ } 1.9 & 9 & 0.99 \\
$ |P_\mathrm{out}^{t,1}|$ & {\ } 2.8 & 6 & 0.84  &  {\ } 25.5 & 6 & $<$0.01  &  {\ } 5.3 & 6 & 0.50  &  {\ } 14.3 & 6 & 0.03  &  {\ } 7.3 & 6 & 0.30  &  {\ } 2.6 & 6 & 0.85 \\
$           |y^{t,1}|$ & {\ } 2.6 & 5 & 0.76  &  {\ } 3.8 & 5 & 0.58  &  {\ } 2.7 & 5 & 0.74  &  {\ } 4.6 & 5 & 0.46  &  {\ } 1.4 & 5 & 0.93  &  {\ } 3.4 & 5 & 0.63 \\
$           |y^{t,2}|$ & {\ } 3.0 & 5 & 0.70  &  {\ } 2.2 & 5 & 0.81  &  {\ } 3.2 & 5 & 0.67  &  {\ } 1.8 & 5 & 0.87  &  {\ } 3.7 & 5 & 0.59  &  {\ } 3.2 & 5 & 0.67 \\
$         |y^{\ttbar}|$ & {\ } 11.9 & 17 & 0.81  &  {\ } 11.8 & 17 & 0.81  &  {\ } 13.0 & 17 & 0.74  &  {\ } 15.4 & 17 & 0.57  &  {\ } 27.0 & 17 & 0.06  &  {\ } 11.5 & 17 & 0.83 \\
$ |y_\mathrm{boost}^{\ttbar}|$ & {\ } 11.5 & 14 & 0.64  &  {\ } 11.6 & 14 & 0.64  &  {\ } 12.2 & 14 & 0.59  &  {\ } 12.0 & 14 & 0.61  &  {\ } 18.2 & 14 & 0.20  &  {\ } 12.3 & 14 & 0.59 \\
$     N_\mathrm{jets}$ & {\ } 4.3 & 4 & 0.36  &  {\ } 0.8 & 4 & 0.94  &  {\ } 10.6 & 4 & 0.03  &  {\ } 27.3 & 4 & $<$0.01  &  {\ } 6.7 & 4 & 0.15  &  {\ } 2.5 & 4 & 0.64 \\
$ p_\mathrm{T}^{t,1} $ & {\ } 17.7 & 10 & 0.06  &  {\ } 14.1 & 10 & 0.17  &  {\ } 20.3 & 10 & 0.03  &  {\ } 14.7 & 10 & 0.14  &  {\ } 28.9 & 10 & $<$0.01  &  {\ } 11.4 & 10 & 0.33 \\
$ p_\mathrm{T}^{t,2} $ & {\ } 17.8 & 8 & 0.02  &  {\ } 7.7 & 8 & 0.47  &  {\ } 31.8 & 8 & $<$0.01  &  {\ } 38.7 & 8 & $<$0.01  &  {\ } 2.3 & 8 & 0.97  &  {\ } 10.0 & 8 & 0.27 \\
$         m^{\ttbar} $ & {\ } 14.9 & 8 & 0.06  &  {\ } 11.1 & 8 & 0.19  &  {\ } 18.7 & 8 & 0.02  &  {\ } 18.9 & 8 & 0.02  &  {\ } 6.6 & 8 & 0.58  &  {\ } 9.1 & 8 & 0.33 \\
$ p_\mathrm{T}^{\ttbar} $ & {\ } 5.0 & 7 & 0.67  &  {\ } 37.7 & 7 & $<$0.01  &  {\ } 6.7 & 7 & 0.47  &  {\ } 21.8 & 7 & $<$0.01  &  {\ } 16.5 & 7 & 0.02  &  {\ } 6.0 & 7 & 0.54 \\
\hline
\end{tabular}}}
\end{table}

\begin{table}[ht]
\caption{ Comparison of the measured particle-level absolute double-differential cross-sections with the predictions from several MC generators. For each prediction, a $\chi^2$ and a $p$-value are calculated using the covariance matrix of the measured spectrum. The number of degrees of freedom (NDF) is equal to the number of bins in the distribution.}
\label{tab:chisquare:absolute:particle_2D}
\footnotesize
\centering
\renewcommand\arraystretch{\arrstretch}
\resizebox*{0.7\textwidth}{!}{
\noindent\makebox[\textwidth]{
\begin{tabular}{|c|  r @{/} l c  | r @{/} l c  | r @{/} l c  | r @{/} l c  | r @{/} l c  | r @{/} l c |}
\hline
Observable  & \multicolumn{3}{c|}{\textsc{PWG+PY8}} & \multicolumn{3}{c|}{\textsc{PWG+PY8} Var. Up} & \multicolumn{3}{c|}{\textsc{PWG+PY8} Var. Down} & \multicolumn{3}{c|}{a\textsc{MC@NLO+PY8}} & \multicolumn{3}{c|}{\textsc{Sherpa}} & \multicolumn{3}{c|}{\textsc{PWG+H7}} \\
& \multicolumn{2}{c}{$\chi^{2}$/NDF} &  ~$p$-value  & \multicolumn{2}{c}{$\chi^{2}$/NDF} &  ~$p$-value  & \multicolumn{2}{c}{$\chi^{2}$/NDF} &  ~$p$-value  & \multicolumn{2}{c}{$\chi^{2}$/NDF} &  ~$p$-value  & \multicolumn{2}{c}{$\chi^{2}$/NDF} &  ~$p$-value  & \multicolumn{2}{c}{$\chi^{2}$/NDF} &  ~$p$-value  \\
\hline
$ \Delta\phi^{\ttbar} \textrm{ vs } N_\mathrm{jets}$ & {\ } 33.0 & 12 & $<$0.01  &  {\ } 26.1 & 12 & 0.01  &  {\ } 61.9 & 12 & $<$0.01  &  {\ } 73.6 & 12 & $<$0.01  &  {\ } 22.5 & 12 & 0.03  &  {\ } 13.0 & 12 & 0.37 \\
$ |P_\mathrm{cross}| \textrm{ vs } N_\mathrm{jets}$ & {\ } 20.0 & 13 & 0.10  &  {\ } 5.7 & 13 & 0.96  &  {\ } 43.5 & 13 & $<$0.01  &  {\ } 44.7 & 13 & $<$0.01  &  {\ } 23.4 & 13 & 0.04  &  {\ } 9.9 & 13 & 0.70 \\
$ |P_\mathrm{out}^{t,1}| \textrm{ vs } N_\mathrm{jets}$ & {\ } 49.2 & 14 & $<$0.01  &  {\ } 43.0 & 14 & $<$0.01  &  {\ } 78.7 & 14 & $<$0.01  &  {\ } 92.0 & 14 & $<$0.01  &  {\ } 33.2 & 14 & $<$0.01  &  {\ } 16.7 & 14 & 0.27 \\
$ |y^{\ttbar}| \textrm{ vs } m^{\ttbar} $ & {\ } 36.6 & 24 & 0.05  &  {\ } 28.0 & 24 & 0.26  &  {\ } 47.5 & 24 & $<$0.01  &  {\ } 39.6 & 24 & 0.02  &  {\ } 27.8 & 24 & 0.27  &  {\ } 22.7 & 24 & 0.54 \\
$ p_\mathrm{T}^{t,1} \textrm{ vs } N_\mathrm{jets}$ & {\ } 41.2 & 19 & $<$0.01  &  {\ } 27.2 & 19 & 0.10  &  {\ } 64.1 & 19 & $<$0.01  &  {\ } 61.0 & 19 & $<$0.01  &  {\ } 41.2 & 19 & $<$0.01  &  {\ } 29.5 & 19 & 0.06 \\
$ p_\mathrm{T}^{t,1} \textrm{ vs } m^{\ttbar}$ & {\ } 27.1 & 11 & $<$0.01  &  {\ } 18.3 & 11 & 0.08  &  {\ } 39.0 & 11 & $<$0.01  &  {\ } 37.1 & 11 & $<$0.01  &  {\ } 13.7 & 11 & 0.25  &  {\ } 13.4 & 11 & 0.27 \\
$ p_\mathrm{T}^{t,1} \textrm{ vs } p_\mathrm{T}^{t,2} $ & {\ } 21.6 & 12 & 0.04  &  {\ } 25.8 & 12 & 0.01  &  {\ } 30.9 & 12 & $<$0.01  &  {\ } 30.0 & 12 & $<$0.01  &  {\ } 30.0 & 12 & $<$0.01  &  {\ } 6.9 & 12 & 0.86 \\
$ p_\mathrm{T}^{t,2} \textrm{ vs } N_\mathrm{jets}$ & {\ } 25.9 & 14 & 0.03  &  {\ } 21.2 & 14 & 0.10  &  {\ } 45.1 & 14 & $<$0.01  &  {\ } 82.1 & 14 & $<$0.01  &  {\ } 26.2 & 14 & 0.02  &  {\ } 30.8 & 14 & $<$0.01 \\
$ p_\mathrm{T}^{t,2} \textrm{ vs } m^{\ttbar} $ & {\ } 15.1 & 12 & 0.24  &  {\ } 6.9 & 12 & 0.86  &  {\ } 31.1 & 12 & $<$0.01  &  {\ } 24.8 & 12 & 0.02  &  {\ } 8.2 & 12 & 0.77  &  {\ } 8.0 & 12 & 0.78 \\
$ p_\mathrm{T}^{\ttbar} \textrm{ vs } N_\mathrm{jets}$ & {\ } 34.4 & 11 & $<$0.01  &  {\ } 45.1 & 11 & $<$0.01  &  {\ } 59.6 & 11 & $<$0.01  &  {\ } 135.0 & 11 & $<$0.01  &  {\ } 27.1 & 11 & $<$0.01  &  {\ } 17.4 & 11 & 0.10 \\
$ p_\mathrm{T}^{\ttbar} \textrm{ vs } m^{\ttbar} $ & {\ } 19.0 & 11 & 0.06  &  {\ } 43.4 & 11 & $<$0.01  &  {\ } 23.6 & 11 & 0.01  &  {\ } 25.7 & 11 & $<$0.01  &  {\ } 23.7 & 11 & 0.01  &  {\ } 9.4 & 11 & 0.58 \\
\hline
\end{tabular}}}
\end{table}
 
\begin{table}[ht]
\caption{ Comparison of the measured particle-level normalised double-differential cross-sections with the predictions from several MC generators. For each prediction, a $\chi^2$ and a $p$-value are calculated using the covariance matrix of the measured spectrum. The number of degrees of freedom (NDF) is equal to the number of bins in the distribution minus one.}
\label{tab:chisquare:relative:particle_2D}
\footnotesize
\centering
\renewcommand\arraystretch{\arrstretch}
\resizebox*{0.7\textwidth}{!}{
\noindent\makebox[\textwidth]{
\begin{tabular}{|c|  r @{/} l c  | r @{/} l c  | r @{/} l c  | r @{/} l c  | r @{/} l c  | r @{/} l c |}
\hline
Observable  & \multicolumn{3}{c|}{\textsc{PWG+PY8}} & \multicolumn{3}{c|}{\textsc{PWG+PY8} Var. Up} & \multicolumn{3}{c|}{\textsc{PWG+PY8} Var. Down} & \multicolumn{3}{c|}{a\textsc{MC@NLO+PY8}} & \multicolumn{3}{c|}{\textsc{Sherpa}} & \multicolumn{3}{c|}{\textsc{PWG+H7}} \\
& \multicolumn{2}{c}{$\chi^{2}$/NDF} &  ~$p$-value  & \multicolumn{2}{c}{$\chi^{2}$/NDF} &  ~$p$-value  & \multicolumn{2}{c}{$\chi^{2}$/NDF} &  ~$p$-value  & \multicolumn{2}{c}{$\chi^{2}$/NDF} &  ~$p$-value  & \multicolumn{2}{c}{$\chi^{2}$/NDF} &  ~$p$-value  & \multicolumn{2}{c}{$\chi^{2}$/NDF} &  ~$p$-value  \\
\hline
$ \Delta\phi^{\ttbar} \textrm{ vs } N_\mathrm{jets}$ & {\ } 27.7 & 11 & $<$0.01  &  {\ } 24.3 & 11 & 0.01  &  {\ } 47.7 & 11 & $<$0.01  &  {\ } 82.5 & 11 & $<$0.01  &  {\ } 21.8 & 11 & 0.03  &  {\ } 14.5 & 11 & 0.21 \\
$ |P_\mathrm{cross}| \textrm{ vs } N_\mathrm{jets}$ & {\ } 12.6 & 12 & 0.40  &  {\ } 5.3 & 12 & 0.95  &  {\ } 22.8 & 12 & 0.03  &  {\ } 47.1 & 12 & $<$0.01  &  {\ } 16.2 & 12 & 0.18  &  {\ } 5.5 & 12 & 0.94 \\
$ |P_\mathrm{out}^{t,1}| \textrm{ vs } N_\mathrm{jets}$ & {\ } 47.4 & 13 & $<$0.01  &  {\ } 38.8 & 13 & $<$0.01  &  {\ } 68.3 & 13 & $<$0.01  &  {\ } 104.0 & 13 & $<$0.01  &  {\ } 29.2 & 13 & $<$0.01  &  {\ } 17.4 & 13 & 0.18 \\
$ |y^{\ttbar}| \textrm{ vs } m^{\ttbar} $ & {\ } 33.8 & 23 & 0.07  &  {\ } 25.2 & 23 & 0.34  &  {\ } 42.4 & 23 & $<$0.01  &  {\ } 38.8 & 23 & 0.02  &  {\ } 27.3 & 23 & 0.24  &  {\ } 23.9 & 23 & 0.41 \\
$ p_\mathrm{T}^{t,1} \textrm{ vs } N_\mathrm{jets}$ & {\ } 38.0 & 18 & $<$0.01  &  {\ } 25.2 & 18 & 0.12  &  {\ } 54.6 & 18 & $<$0.01  &  {\ } 70.0 & 18 & $<$0.01  &  {\ } 38.8 & 18 & $<$0.01  &  {\ } 29.6 & 18 & 0.04 \\
$ p_\mathrm{T}^{t,1} \textrm{ vs } m^{\ttbar}$ & {\ } 24.8 & 10 & $<$0.01  &  {\ } 16.4 & 10 & 0.09  &  {\ } 34.6 & 10 & $<$0.01  &  {\ } 38.2 & 10 & $<$0.01  &  {\ } 13.9 & 10 & 0.18  &  {\ } 14.0 & 10 & 0.17 \\
$ p_\mathrm{T}^{t,1} \textrm{ vs } p_\mathrm{T}^{t,2} $ & {\ } 15.1 & 11 & 0.18  &  {\ } 20.1 & 11 & 0.04  &  {\ } 19.8 & 11 & 0.05  &  {\ } 26.5 & 11 & $<$0.01  &  {\ } 31.4 & 11 & $<$0.01  &  {\ } 9.1 & 11 & 0.62 \\
$ p_\mathrm{T}^{t,2} \textrm{ vs } N_\mathrm{jets}$ & {\ } 23.8 & 13 & 0.03  &  {\ } 19.1 & 13 & 0.12  &  {\ } 37.9 & 13 & $<$0.01  &  {\ } 91.6 & 13 & $<$0.01  &  {\ } 21.9 & 13 & 0.06  &  {\ } 22.0 & 13 & 0.06 \\
$ p_\mathrm{T}^{t,2} \textrm{ vs } m^{\ttbar} $ & {\ } 13.9 & 11 & 0.24  &  {\ } 6.0 & 11 & 0.87  &  {\ } 27.9 & 11 & $<$0.01  &  {\ } 27.3 & 11 & $<$0.01  &  {\ } 7.9 & 11 & 0.72  &  {\ } 6.0 & 11 & 0.87 \\
$ p_\mathrm{T}^{\ttbar} \textrm{ vs } N_\mathrm{jets}$ & {\ } 28.7 & 10 & $<$0.01  &  {\ } 42.8 & 10 & $<$0.01  &  {\ } 42.2 & 10 & $<$0.01  &  {\ } 149.0 & 10 & $<$0.01  &  {\ } 26.7 & 10 & $<$0.01  &  {\ } 14.9 & 10 & 0.14 \\
$ p_\mathrm{T}^{\ttbar} \textrm{ vs } m^{\ttbar} $ & {\ } 17.2 & 10 & 0.07  &  {\ } 35.1 & 10 & $<$0.01  &  {\ } 21.2 & 10 & 0.02  &  {\ } 28.4 & 10 & $<$0.01  &  {\ } 20.4 & 10 & 0.03  &  {\ } 9.7 & 10 & 0.47 \\
\hline
\end{tabular}}}
\end{table}
 
\clearpage
 
\subsubsection{Cross-sections in the full phase space}
\label{sec:results:chi2_parton}
 
Tables~\ref{tab:chisquare:absolute:parton_1D} and \ref{tab:chisquare:relative:parton_1D} show the quantitative comparisons among the single-differential parton-level results and theoretical predictions, while Tables~\ref{tab:chisquare:absolute:parton_2D} and \ref{tab:chisquare:relative:parton_2D} show the results for the double-differential cross-sections.
Conclusions similar to those for the $\chi^2$ of the particle-level measurements can be drawn, although a few minor differences can be seen.
Once more, \PHHSEVEN and \PHPYEIGHT perform best in terms of reproducing the data, while \MCNPYEIGHT typically fails to reflect the data.
 
\begin{table}[ht]
\caption{ Comparison of the measured parton-level absolute single-differential cross-sections with the predictions from several MC generators. For each prediction, a $\chi^2$ and a $p$-value are calculated using the covariance matrix of the measured spectrum. The number of degrees of freedom (NDF) is equal to the number of bins in the distribution.}
\label{tab:chisquare:absolute:parton_1D}
\footnotesize
\centering
\renewcommand\arraystretch{\arrstretch}
\resizebox*{0.7\textwidth}{!}{
\noindent\makebox[\textwidth]{
\begin{tabular}{|c|  r @{/} l c  | r @{/} l c  | r @{/} l c  | r @{/} l c  | r @{/} l c  | r @{/} l c |}
\hline
Observable  & \multicolumn{3}{c|}{\textsc{PWG+PY8}} & \multicolumn{3}{c|}{\textsc{PWG+PY8} Var. Up} & \multicolumn{3}{c|}{\textsc{PWG+PY8} Var. Down} & \multicolumn{3}{c|}{a\textsc{MC@NLO+PY8}} & \multicolumn{3}{c|}{\textsc{Sherpa}} & \multicolumn{3}{c|}{\textsc{PWG+H7}} \\
& \multicolumn{2}{c}{$\chi^{2}$/NDF} &  ~$p$-value  & \multicolumn{2}{c}{$\chi^{2}$/NDF} &  ~$p$-value  & \multicolumn{2}{c}{$\chi^{2}$/NDF} &  ~$p$-value  & \multicolumn{2}{c}{$\chi^{2}$/NDF} &  ~$p$-value  & \multicolumn{2}{c}{$\chi^{2}$/NDF} &  ~$p$-value  & \multicolumn{2}{c}{$\chi^{2}$/NDF} &  ~$p$-value  \\
\hline
$       \chi^{\ttbar}$ & {\ } 2.7 & 7 & 0.91  &  {\ } 1.7 & 7 & 0.97  &  {\ } 4.7 & 7 & 0.69  &  {\ } 2.5 & 7 & 0.93  &  {\ } 2.7 & 7 & 0.92  &  {\ } 2.0 & 7 & 0.96 \\
$ \Delta\phi^{\ttbar}$ & {\ } 3.4 & 6 & 0.76  &  {\ } 2.7 & 6 & 0.85  &  {\ } 4.6 & 6 & 0.59  &  {\ } 14.7 & 6 & 0.02  &  {\ } 8.5 & 6 & 0.20  &  {\ } 4.6 & 6 & 0.59 \\
$ H_\mathrm{T}^{\ttbar}$ & {\ } 13.6 & 11 & 0.26  &  {\ } 11.0 & 11 & 0.44  &  {\ } 20.0 & 11 & 0.05  &  {\ } 14.1 & 11 & 0.23  &  {\ } 16.7 & 11 & 0.12  &  {\ } 11.0 & 11 & 0.45 \\
$           |y^{t,1}|$ & {\ } 0.8 & 7 & 1.00  &  {\ } 0.8 & 7 & 1.00  &  {\ } 0.7 & 7 & 1.00  &  {\ } 0.8 & 7 & 1.00  &  {\ } 0.7 & 7 & 1.00  &  {\ } 1.0 & 7 & 0.99 \\
$           |y^{t,2}|$ & {\ } 2.4 & 6 & 0.89  &  {\ } 2.3 & 6 & 0.89  &  {\ } 2.3 & 6 & 0.89  &  {\ } 2.4 & 6 & 0.88  &  {\ } 2.5 & 6 & 0.86  &  {\ } 2.3 & 6 & 0.89 \\
$        |y^{\ttbar}|$ & {\ } 7.8 & 12 & 0.80  &  {\ } 7.5 & 12 & 0.82  &  {\ } 7.8 & 12 & 0.80  &  {\ } 7.8 & 12 & 0.80  &  {\ } 8.7 & 12 & 0.73  &  {\ } 7.7 & 12 & 0.81 \\
$ |y_\mathrm{boost}^{\ttbar}|$ & {\ } 12.3 & 15 & 0.65  &  {\ } 12.0 & 15 & 0.68  &  {\ } 12.4 & 15 & 0.65  &  {\ } 12.3 & 15 & 0.66  &  {\ } 13.6 & 15 & 0.55  &  {\ } 12.2 & 15 & 0.66 \\
$ p_\mathrm{T}^{t,1} $ & {\ } 9.6 & 10 & 0.48  &  {\ } 13.3 & 10 & 0.21  &  {\ } 8.5 & 10 & 0.58  &  {\ } 7.3 & 10 & 0.70  &  {\ } 18.5 & 10 & 0.05  &  {\ } 8.0 & 10 & 0.63 \\
$ p_\mathrm{T}^{t,2} $ & {\ } 6.5 & 8 & 0.59  &  {\ } 3.3 & 8 & 0.92  &  {\ } 13.1 & 8 & 0.11  &  {\ } 8.0 & 8 & 0.44  &  {\ } 5.9 & 8 & 0.66  &  {\ } 4.5 & 8 & 0.81 \\
$         m^{\ttbar} $ & {\ } 9.3 & 9 & 0.41  &  {\ } 10.1 & 9 & 0.34  &  {\ } 8.8 & 9 & 0.46  &  {\ } 8.9 & 9 & 0.44  &  {\ } 9.1 & 9 & 0.43  &  {\ } 8.8 & 9 & 0.46 \\
$ p_\mathrm{T}^{\ttbar} $ & {\ } 1.5 & 5 & 0.91  &  {\ } 12.6 & 5 & 0.03  &  {\ } 2.2 & 5 & 0.83  &  {\ } 20.9 & 5 & $<$0.01  &  {\ } 9.6 & 5 & 0.09  &  {\ } 3.0 & 5 & 0.70 \\
\hline
\end{tabular}}}
\end{table}

\begin{table}[ht]
\caption{ Comparison of the measured parton-level normalised single-differential cross-sections with the predictions from several MC generators. For each prediction, a $\chi^2$ and a $p$-value are calculated using the covariance matrix of the measured spectrum. The number of degrees of freedom (NDF) is equal to the number of bins in the distribution minus one.}
\label{tab:chisquare:relative:parton_1D}
\footnotesize
\centering
\renewcommand\arraystretch{\arrstretch}
\resizebox*{0.7\textwidth}{!}{
\noindent\makebox[\textwidth]{
\begin{tabular}{|c|  r @{/} l c  | r @{/} l c  | r @{/} l c  | r @{/} l c  | r @{/} l c  | r @{/} l c |}
\hline
Observable  & \multicolumn{3}{c|}{\textsc{PWG+PY8}} & \multicolumn{3}{c|}{\textsc{PWG+PY8} Var. Up} & \multicolumn{3}{c|}{\textsc{PWG+PY8} Var. Down} & \multicolumn{3}{c|}{a\textsc{MC@NLO+PY8}} & \multicolumn{3}{c|}{\textsc{Sherpa}} & \multicolumn{3}{c|}{\textsc{PWG+H7}} \\
& \multicolumn{2}{c}{$\chi^{2}$/NDF} &  ~$p$-value  & \multicolumn{2}{c}{$\chi^{2}$/NDF} &  ~$p$-value  & \multicolumn{2}{c}{$\chi^{2}$/NDF} &  ~$p$-value  & \multicolumn{2}{c}{$\chi^{2}$/NDF} &  ~$p$-value  & \multicolumn{2}{c}{$\chi^{2}$/NDF} &  ~$p$-value  & \multicolumn{2}{c}{$\chi^{2}$/NDF} &  ~$p$-value  \\
\hline
$       \chi^{\ttbar}$ & {\ } 2.8 & 6 & 0.84  &  {\ } 1.8 & 6 & 0.94  &  {\ } 4.3 & 6 & 0.63  &  {\ } 2.7 & 6 & 0.85  &  {\ } 2.7 & 6 & 0.84  &  {\ } 2.1 & 6 & 0.91 \\
$ \Delta\phi^{\ttbar}$ & {\ } 3.2 & 5 & 0.68  &  {\ } 2.4 & 5 & 0.80  &  {\ } 4.4 & 5 & 0.49  &  {\ } 14.3 & 5 & 0.01  &  {\ } 7.3 & 5 & 0.20  &  {\ } 4.3 & 5 & 0.50 \\
$ H_\mathrm{T}^{\ttbar}$ & {\ } 18.8 & 10 & 0.04  &  {\ } 14.9 & 10 & 0.14  &  {\ } 27.3 & 10 & $<$0.01  &  {\ } 20.3 & 10 & 0.03  &  {\ } 22.4 & 10 & 0.01  &  {\ } 15.4 & 10 & 0.12 \\
$           |y^{t,1}|$ & {\ } 0.8 & 6 & 0.99  &  {\ } 0.8 & 6 & 0.99  &  {\ } 0.8 & 6 & 0.99  &  {\ } 0.9 & 6 & 0.99  &  {\ } 0.7 & 6 & 0.99  &  {\ } 1.0 & 6 & 0.98 \\
$           |y^{t,2}|$ & {\ } 2.5 & 5 & 0.77  &  {\ } 2.5 & 5 & 0.77  &  {\ } 2.5 & 5 & 0.78  &  {\ } 2.6 & 5 & 0.76  &  {\ } 2.8 & 5 & 0.73  &  {\ } 2.5 & 5 & 0.78 \\
$        |y^{\ttbar}|$ & {\ } 7.3 & 11 & 0.77  &  {\ } 7.0 & 11 & 0.80  &  {\ } 7.3 & 11 & 0.77  &  {\ } 7.2 & 11 & 0.78  &  {\ } 8.1 & 11 & 0.70  &  {\ } 7.2 & 11 & 0.78 \\
$ |y_\mathrm{boost}^{\ttbar}|$ & {\ } 11.6 & 14 & 0.64  &  {\ } 11.3 & 14 & 0.66  &  {\ } 11.6 & 14 & 0.64  &  {\ } 11.7 & 14 & 0.63  &  {\ } 12.8 & 14 & 0.54  &  {\ } 11.6 & 14 & 0.64 \\
$ p_\mathrm{T}^{t,1} $ & {\ } 12.1 & 9 & 0.21  &  {\ } 16.3 & 9 & 0.06  &  {\ } 10.5 & 9 & 0.31  &  {\ } 9.8 & 9 & 0.36  &  {\ } 23.5 & 9 & $<$0.01  &  {\ } 10.4 & 9 & 0.32 \\
$ p_\mathrm{T}^{t,2} $ & {\ } 7.2 & 7 & 0.41  &  {\ } 3.7 & 7 & 0.81  &  {\ } 14.4 & 7 & 0.04  &  {\ } 9.0 & 7 & 0.25  &  {\ } 5.6 & 7 & 0.59  &  {\ } 5.1 & 7 & 0.64 \\
$         m^{\ttbar} $ & {\ } 11.6 & 8 & 0.17  &  {\ } 12.6 & 8 & 0.13  &  {\ } 10.8 & 8 & 0.21  &  {\ } 11.5 & 8 & 0.17  &  {\ } 11.0 & 8 & 0.20  &  {\ } 11.0 & 8 & 0.20 \\
$ p_\mathrm{T}^{\ttbar} $ & {\ } 1.5 & 4 & 0.83  &  {\ } 9.8 & 4 & 0.04  &  {\ } 1.7 & 4 & 0.78  &  {\ } 12.8 & 4 & 0.01  &  {\ } 9.4 & 4 & 0.05  &  {\ } 2.2 & 4 & 0.70 \\
\hline
\end{tabular}}}
\end{table}
 
\begin{table}[ht]
\caption{ Comparison of the measured parton-level absolute double-differential cross-sections with the predictions from several MC generators. For each prediction, a $\chi^2$ and a $p$-value are calculated using the covariance matrix of the measured spectrum. The number of degrees of freedom (NDF) is equal to the number of bins in the distribution.}
\label{tab:chisquare:absolute:parton_2D}
\footnotesize
\centering
\renewcommand\arraystretch{\arrstretch}
\resizebox*{0.7\textwidth}{!}{
\noindent\makebox[\textwidth]{
\begin{tabular}{|c|  r @{/} l c  | r @{/} l c  | r @{/} l c  | r @{/} l c  | r @{/} l c  | r @{/} l c |}
\hline
Observable  & \multicolumn{3}{c|}{\textsc{PWG+PY8}} & \multicolumn{3}{c|}{\textsc{PWG+PY8} Var. Up} & \multicolumn{3}{c|}{\textsc{PWG+PY8} Var. Down} & \multicolumn{3}{c|}{a\textsc{MC@NLO+PY8}} & \multicolumn{3}{c|}{\textsc{Sherpa}} & \multicolumn{3}{c|}{\textsc{PWG+H7}} \\
& \multicolumn{2}{c}{$\chi^{2}$/NDF} &  ~$p$-value  & \multicolumn{2}{c}{$\chi^{2}$/NDF} &  ~$p$-value  & \multicolumn{2}{c}{$\chi^{2}$/NDF} &  ~$p$-value  & \multicolumn{2}{c}{$\chi^{2}$/NDF} &  ~$p$-value  & \multicolumn{2}{c}{$\chi^{2}$/NDF} &  ~$p$-value  & \multicolumn{2}{c}{$\chi^{2}$/NDF} &  ~$p$-value  \\
\hline
$ |y^{t,1}| \textrm{ vs } m^{\ttbar}$ & {\ } 10.9 & 11 & 0.46  &  {\ } 11.8 & 11 & 0.38  &  {\ } 10.7 & 11 & 0.47  &  {\ } 13.6 & 11 & 0.26  &  {\ } 11.5 & 11 & 0.41  &  {\ } 9.4 & 11 & 0.58 \\
$ |y^{t,2}| \textrm{ vs } m^{\ttbar}$ & {\ } 22.1 & 11 & 0.02  &  {\ } 20.4 & 11 & 0.04  &  {\ } 24.6 & 11 & 0.01  &  {\ } 23.8 & 11 & 0.01  &  {\ } 23.0 & 11 & 0.02  &  {\ } 20.9 & 11 & 0.03 \\
$ |y^{t,2}| \textrm{ vs } |y^{t,1}| $ & {\ } 6.7 & 16 & 0.98  &  {\ } 6.7 & 16 & 0.98  &  {\ } 7.1 & 16 & 0.97  &  {\ } 8.0 & 16 & 0.95  &  {\ } 5.5 & 16 & 0.99  &  {\ } 6.6 & 16 & 0.98 \\
$ |y^{\ttbar}| \textrm{ vs } m^{\ttbar}$ & {\ } 16.9 & 11 & 0.11  &  {\ } 18.4 & 11 & 0.07  &  {\ } 15.5 & 11 & 0.16  &  {\ } 17.7 & 11 & 0.09  &  {\ } 16.3 & 11 & 0.13  &  {\ } 16.4 & 11 & 0.13 \\
$ p_\mathrm{T}^{t,1} \textrm{ vs } m^{\ttbar}$ & {\ } 16.5 & 10 & 0.09  &  {\ } 15.4 & 10 & 0.12  &  {\ } 26.2 & 10 & $<$0.01  &  {\ } 37.1 & 10 & $<$0.01  &  {\ } 21.5 & 10 & 0.02  &  {\ } 14.3 & 10 & 0.16 \\
$ p_\mathrm{T}^{t,1} \textrm{ vs } p_\mathrm{T}^{t,2} $ & {\ } 18.6 & 12 & 0.10  &  {\ } 30.8 & 12 & $<$0.01  &  {\ } 21.6 & 12 & 0.04  &  {\ } 26.9 & 12 & $<$0.01  &  {\ } 37.1 & 12 & $<$0.01  &  {\ } 14.6 & 12 & 0.26 \\
$ p_\mathrm{T}^{t,2} \textrm{ vs } m^{\ttbar} $ & {\ } 16.5 & 13 & 0.23  &  {\ } 15.4 & 13 & 0.29  &  {\ } 24.0 & 13 & 0.03  &  {\ } 17.7 & 13 & 0.17  &  {\ } 16.7 & 13 & 0.21  &  {\ } 14.8 & 13 & 0.32 \\
$ p_\mathrm{T}^{\ttbar} \textrm{ vs } m^{\ttbar }$ & {\ } 17.5 & 12 & 0.13  &  {\ } 34.1 & 12 & $<$0.01  &  {\ } 16.4 & 12 & 0.17  &  {\ } 41.8 & 12 & $<$0.01  &  {\ } 23.8 & 12 & 0.02  &  {\ } 15.4 & 12 & 0.22 \\
\hline
\end{tabular}}}
\end{table}
 
\begin{table}[ht]
\caption{ Comparison of the measured parton-level normalised double-differential cross-sections with the predictions from several MC generators. For each prediction, a $\chi^2$ and a $p$-value are calculated using the covariance matrix of the measured spectrum. The number of degrees of freedom (NDF) is equal to the number of bins in the distribution minus one.}
\label{tab:chisquare:relative:parton_2D}
\footnotesize
\centering
\renewcommand\arraystretch{\arrstretch}
\resizebox*{0.7\textwidth}{!}{
\noindent\makebox[\textwidth]{
\begin{tabular}{|c|  r @{/} l c  | r @{/} l c  | r @{/} l c  | r @{/} l c  | r @{/} l c  | r @{/} l c |}
\hline
Observable  & \multicolumn{3}{c|}{\textsc{PWG+PY8}} & \multicolumn{3}{c|}{\textsc{PWG+PY8} Var. Up} & \multicolumn{3}{c|}{\textsc{PWG+PY8} Var. Down} & \multicolumn{3}{c|}{a\textsc{MC@NLO+PY8}} & \multicolumn{3}{c|}{\textsc{Sherpa}} & \multicolumn{3}{c|}{\textsc{PWG+H7}} \\
& \multicolumn{2}{c}{$\chi^{2}$/NDF} &  ~$p$-value  & \multicolumn{2}{c}{$\chi^{2}$/NDF} &  ~$p$-value  & \multicolumn{2}{c}{$\chi^{2}$/NDF} &  ~$p$-value  & \multicolumn{2}{c}{$\chi^{2}$/NDF} &  ~$p$-value  & \multicolumn{2}{c}{$\chi^{2}$/NDF} &  ~$p$-value  & \multicolumn{2}{c}{$\chi^{2}$/NDF} &  ~$p$-value  \\
\hline
$ |y^{t,1}| \textrm{ vs } m^{\ttbar}$ & {\ } 12.1 & 10 & 0.28  &  {\ } 13.2 & 10 & 0.21  &  {\ } 11.8 & 10 & 0.30  &  {\ } 15.4 & 10 & 0.12  &  {\ } 12.9 & 10 & 0.23  &  {\ } 10.6 & 10 & 0.39 \\
$ |y^{t,2}| \textrm{ vs } m^{\ttbar}$ & {\ } 24.4 & 10 & $<$0.01  &  {\ } 22.8 & 10 & 0.01  &  {\ } 26.6 & 10 & $<$0.01  &  {\ } 26.7 & 10 & $<$0.01  &  {\ } 25.5 & 10 & $<$0.01  &  {\ } 23.3 & 10 & $<$0.01 \\
$ |y^{t,2}| \textrm{ vs } |y^{t,1}| $ & {\ } 7.0 & 15 & 0.96  &  {\ } 6.9 & 15 & 0.96  &  {\ } 7.7 & 15 & 0.94  &  {\ } 8.2 & 15 & 0.91  &  {\ } 5.8 & 15 & 0.98  &  {\ } 6.9 & 15 & 0.96 \\
$ |y^{\ttbar}| \textrm{ vs } m^{\ttbar}$ & {\ } 18.0 & 10 & 0.06  &  {\ } 19.7 & 10 & 0.03  &  {\ } 16.4 & 10 & 0.09  &  {\ } 18.6 & 10 & 0.05  &  {\ } 17.0 & 10 & 0.07  &  {\ } 17.5 & 10 & 0.06 \\
$ p_\mathrm{T}^{t,1} \textrm{ vs } m^{\ttbar}$ & {\ } 22.0 & 9 & $<$0.01  &  {\ } 19.8 & 9 & 0.02  &  {\ } 33.9 & 9 & $<$0.01  &  {\ } 45.2 & 9 & $<$0.01  &  {\ } 28.4 & 9 & $<$0.01  &  {\ } 18.5 & 9 & 0.03 \\
$ p_\mathrm{T}^{t,1} \textrm{ vs } p_\mathrm{T}^{t,2} $ & {\ } 22.3 & 11 & 0.02  &  {\ } 33.6 & 11 & $<$0.01  &  {\ } 28.5 & 11 & $<$0.01  &  {\ } 40.5 & 11 & $<$0.01  &  {\ } 46.0 & 11 & $<$0.01  &  {\ } 19.1 & 11 & 0.06 \\
$ p_\mathrm{T}^{t,2} \textrm{ vs } m^{\ttbar} $ & {\ } 17.9 & 12 & 0.12  &  {\ } 16.8 & 12 & 0.16  &  {\ } 25.7 & 12 & 0.01  &  {\ } 19.1 & 12 & 0.09  &  {\ } 18.1 & 12 & 0.11  &  {\ } 16.2 & 12 & 0.18 \\
$ p_\mathrm{T}^{\ttbar} \textrm{ vs } m^{\ttbar }$ & {\ } 19.7 & 11 & 0.05  &  {\ } 33.3 & 11 & $<$0.01  &  {\ } 18.9 & 11 & 0.06  &  {\ } 48.5 & 11 & $<$0.01  &  {\ } 25.2 & 11 & $<$0.01  &  {\ } 18.3 & 11 & 0.07 \\
\hline
\end{tabular}}}
\end{table}
 
\clearpage
 
\subsection{Discussion of individual observables}
\label{sec:results:obs}
 
In the following sections, trends in specific observables are discussed.
Particle-level results are shown for selected normalised single-differential cross-sections in Section~\ref{sec:results:particle}.
Similarly, selected parton-level results are discussed in Section~\ref{sec:results:parton}.
 
While a large variety of observables have been considered in this analysis, the present section focuses on selected variables either to illustrate features when data are compared to predictions or to achieve a comparison with other measurements performed by the ATLAS experiment in the \ljets decay channel of the top-quark pair~\cite{TOPQ-2018-15}.
 
\subsubsection{Results at particle level}
\label{sec:results:particle}
 
\begin{figure}[htbp]
\centering
\subfloat[]{\includegraphics[width=0.5\textwidth]{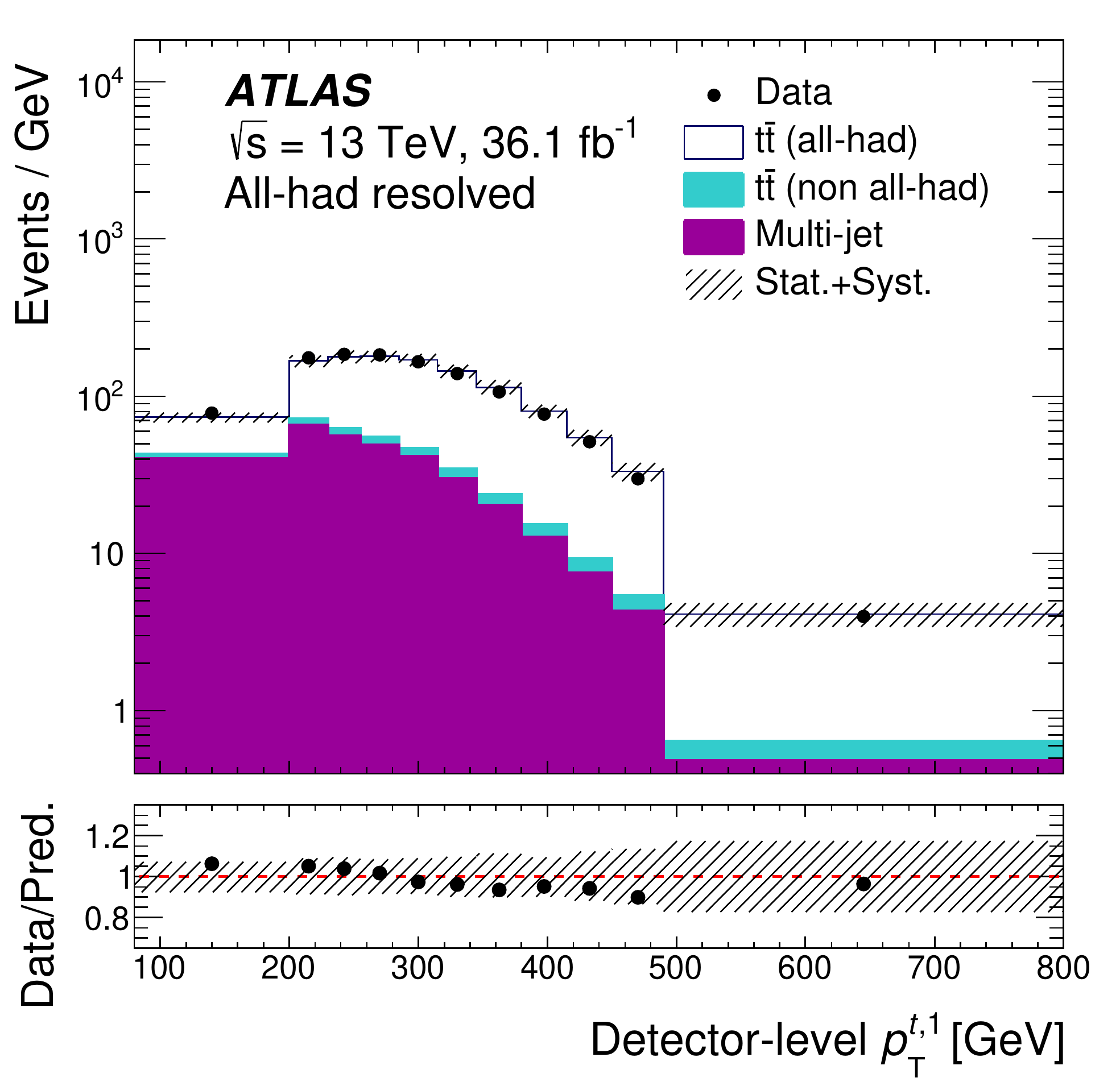}\label{fig:datamc:t1_pt}}
\subfloat[]{\includegraphics[width=0.5\textwidth]{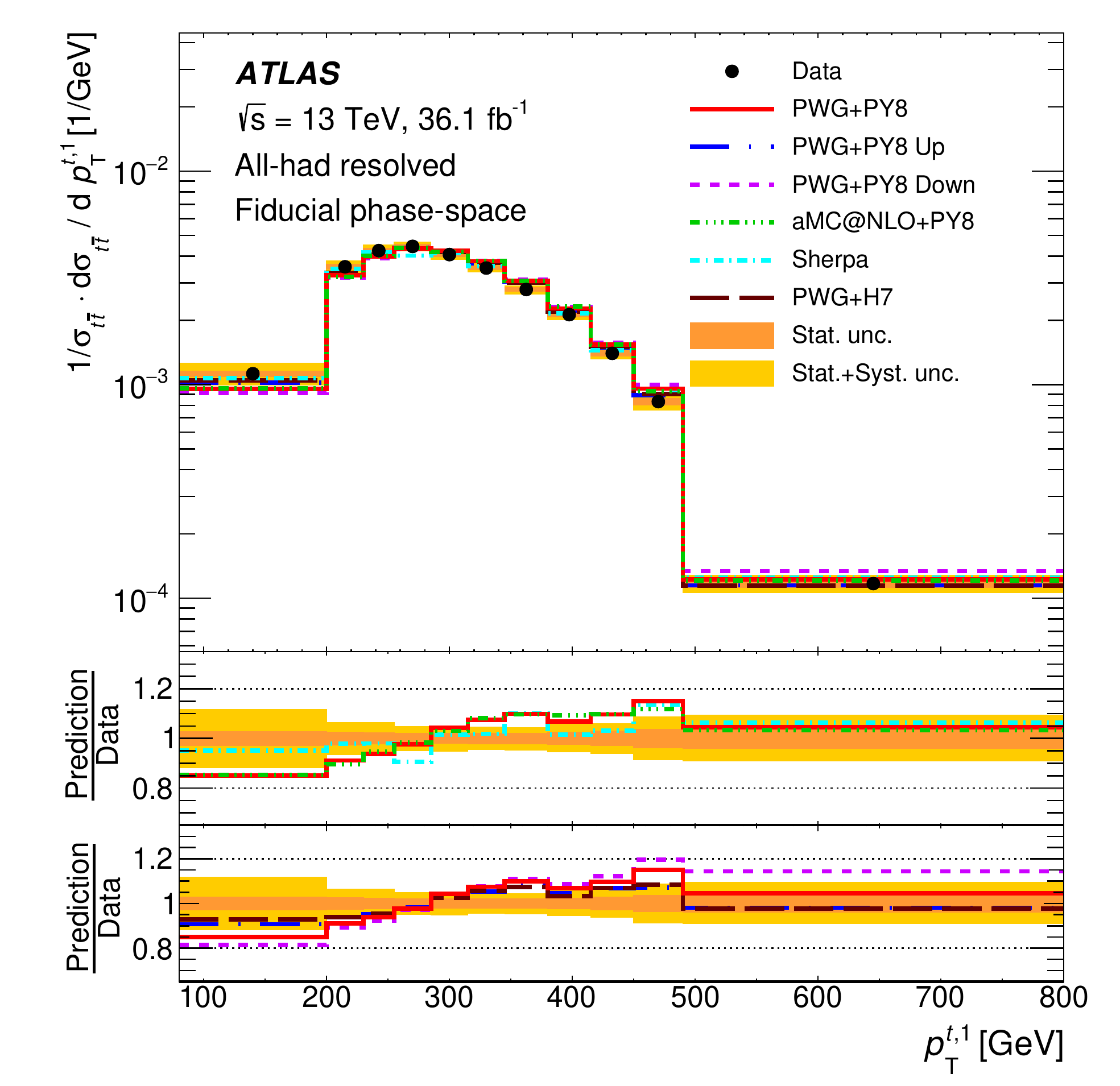}\label{fig:unf_rel:t1_pt}}
 
\subfloat[]{\includegraphics[width=0.5\textwidth]{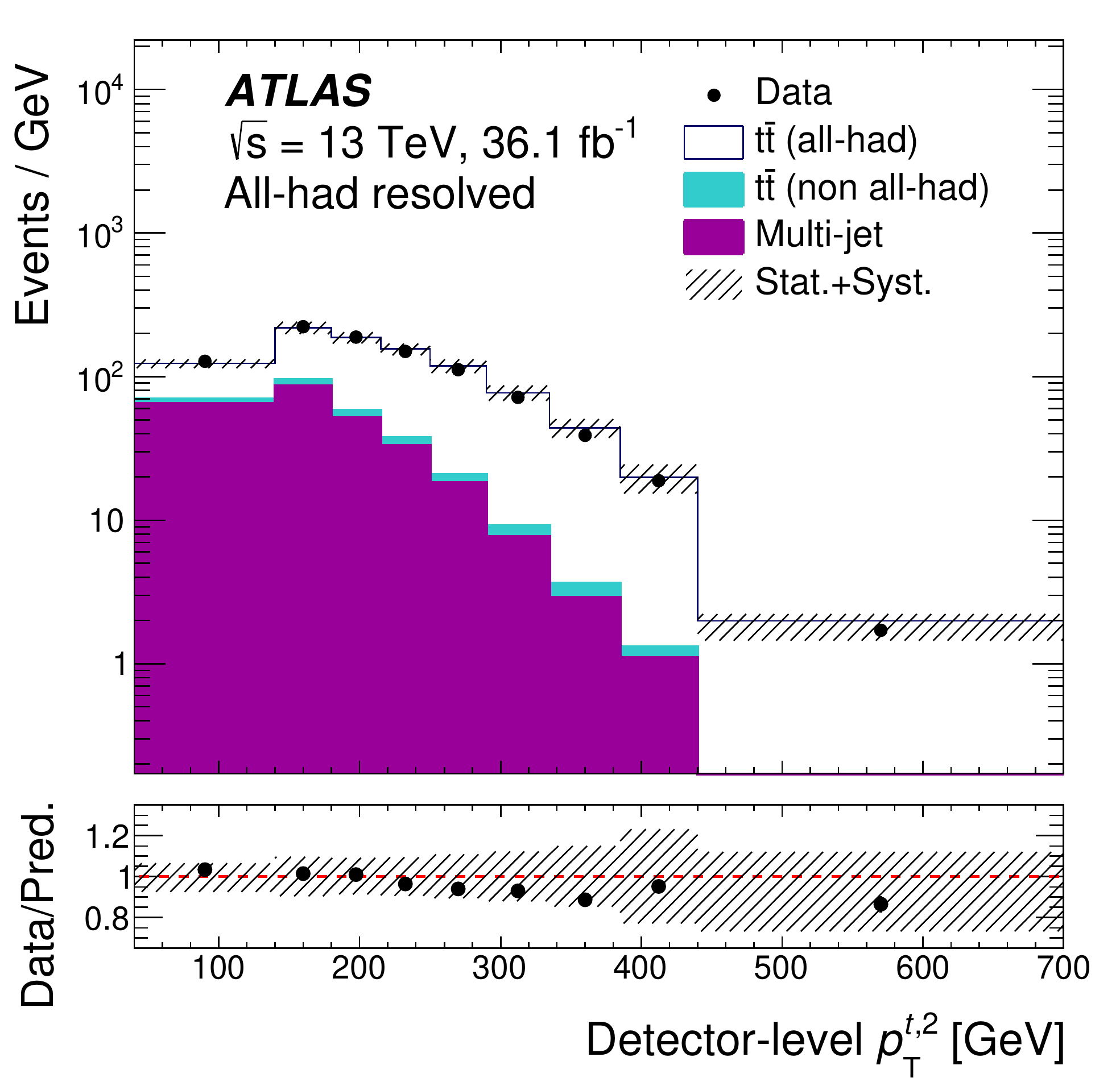}\label{fig:datamc:t2_pt}}
\subfloat[]{\includegraphics[width=0.5\textwidth]{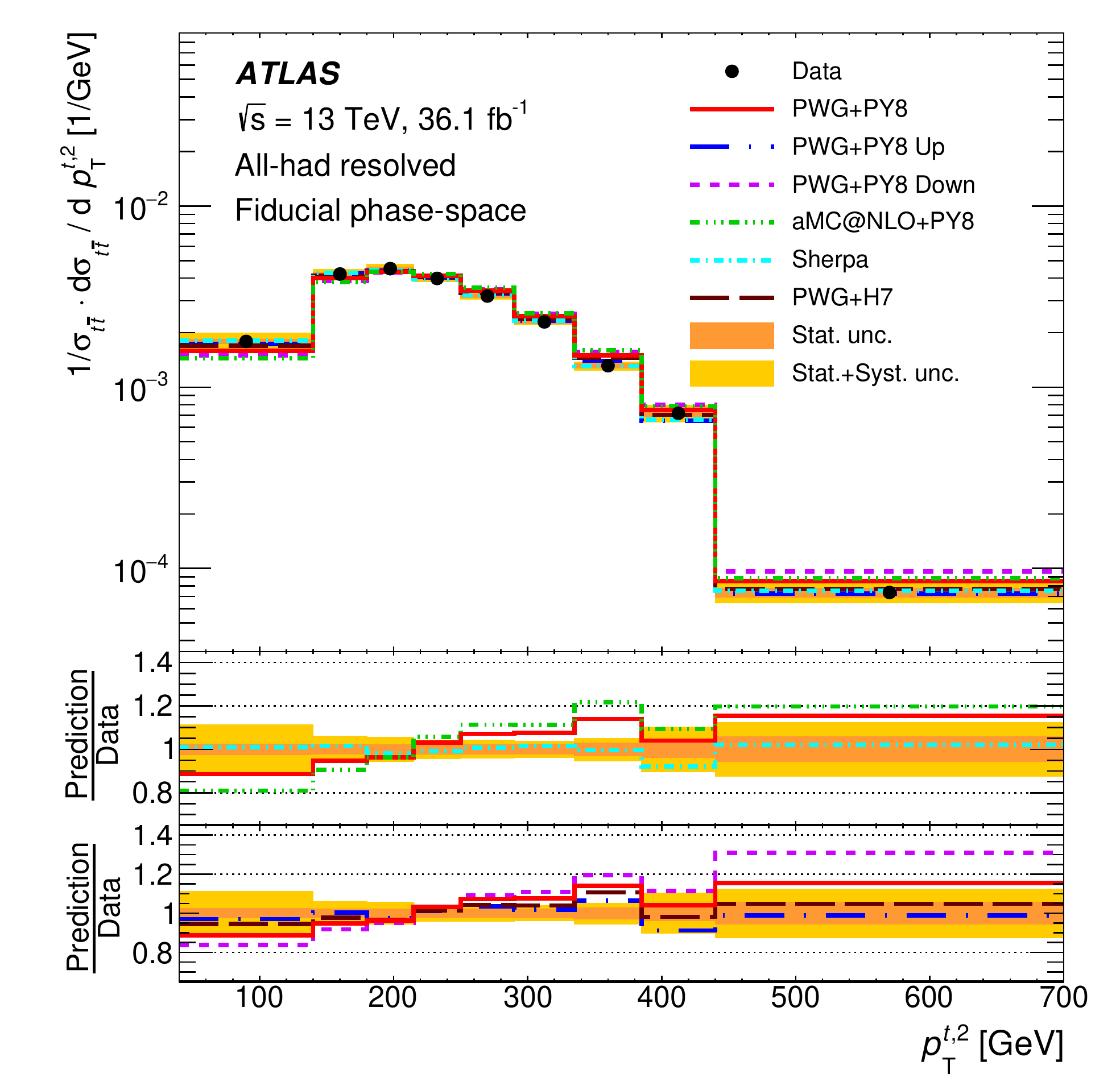}\label{fig:unf_rel:t2_pt}}
\caption{Comparison of the ATLAS data with the fully simulated nominal SM predictions for the~(a) leading and~(c) sub-leading top-quark transverse momenta. \dataMCplotcaption\,
Single-differential normalised cross-section measurements, unfolded at particle level, as a function of the~(b) leading and~(d) sub-leading top-quark transverse momenta. Overflow events are included in the last bin of every distribution shown. The unfolded data are compared with theoretical predictions. \unfoldedplotcaption\,
}
\label{fig:results:particle:top12}
\end{figure}

\begin{figure}[htbp]
\centering
\subfloat[]{\includegraphics[width=0.5\textwidth]{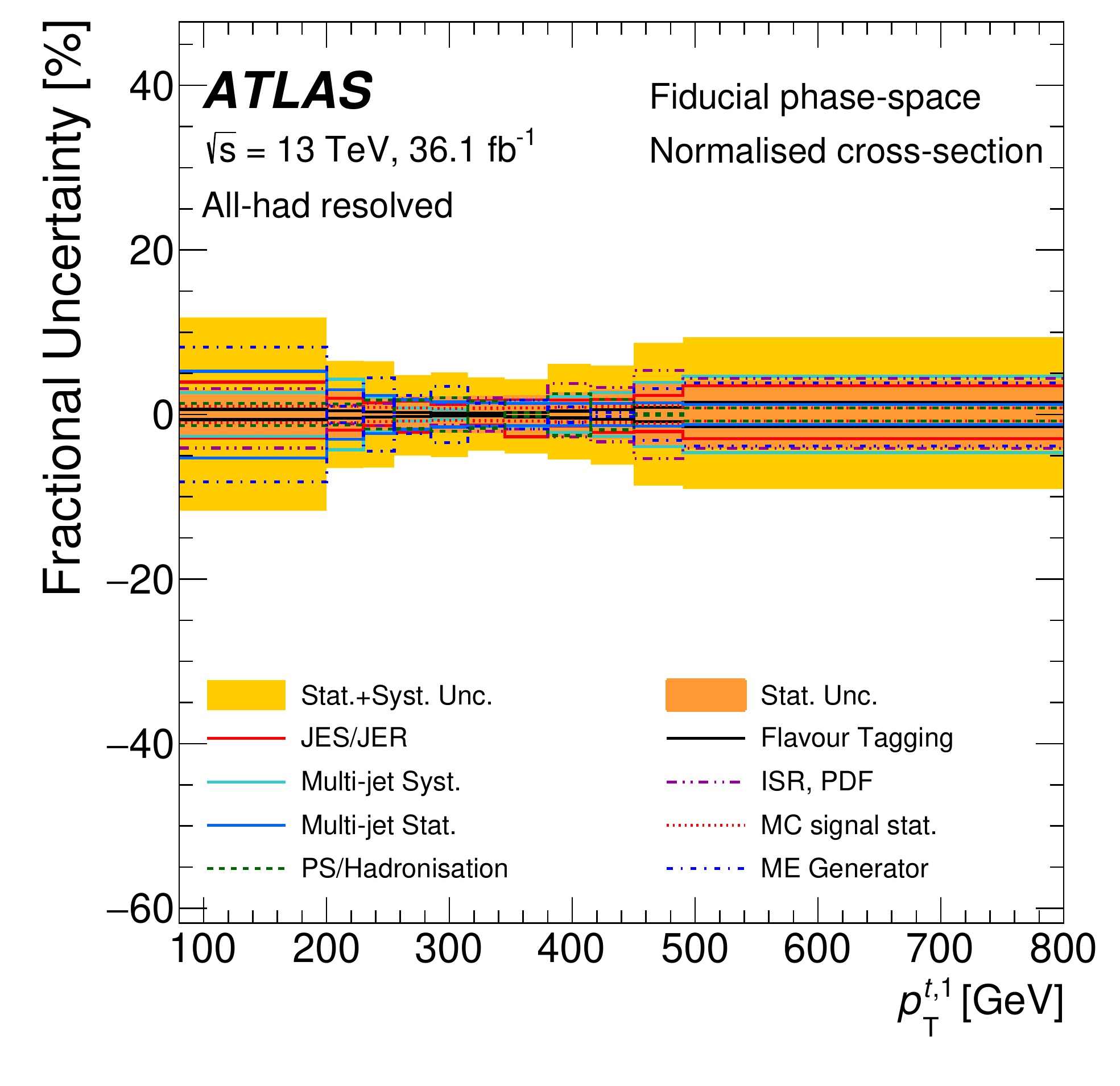}\label{fig:unc_rel:t1_pt}}
\subfloat[]{\includegraphics[width=0.5\textwidth]{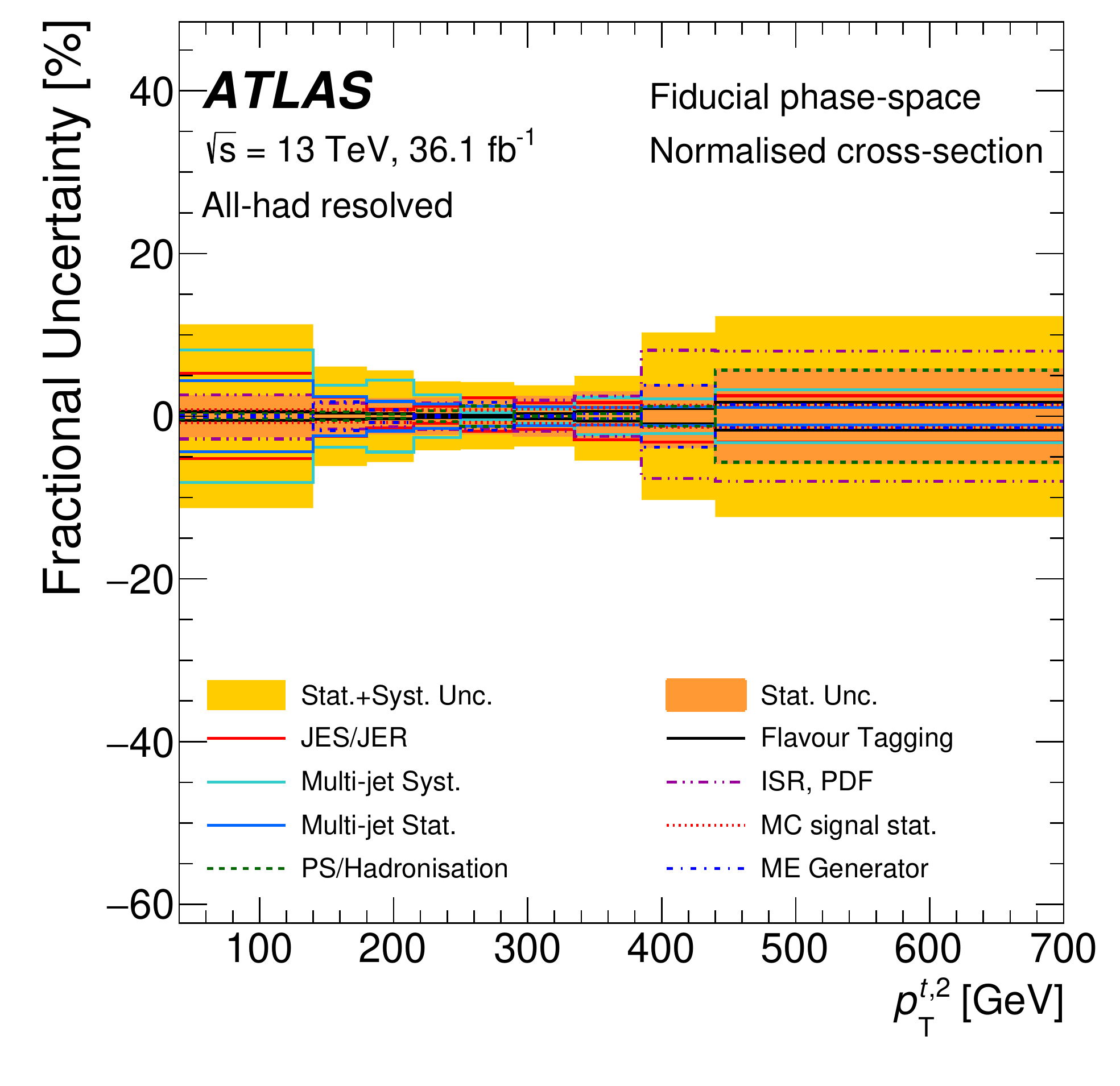}\label{fig:unc_rel:t2_pt}}
\caption{Fractional uncertainties for the normalised single-differential distributions unfolded at particle level as a function of the~(a) leading and~(b) sub-leading top-quark transverse momenta. \reluncertplotcaption\,}
\label{fig:unc:particle:top12}
\end{figure}

\begin{figure}[htbp]
\centering
\subfloat[]{\includegraphics[width=0.5\textwidth]{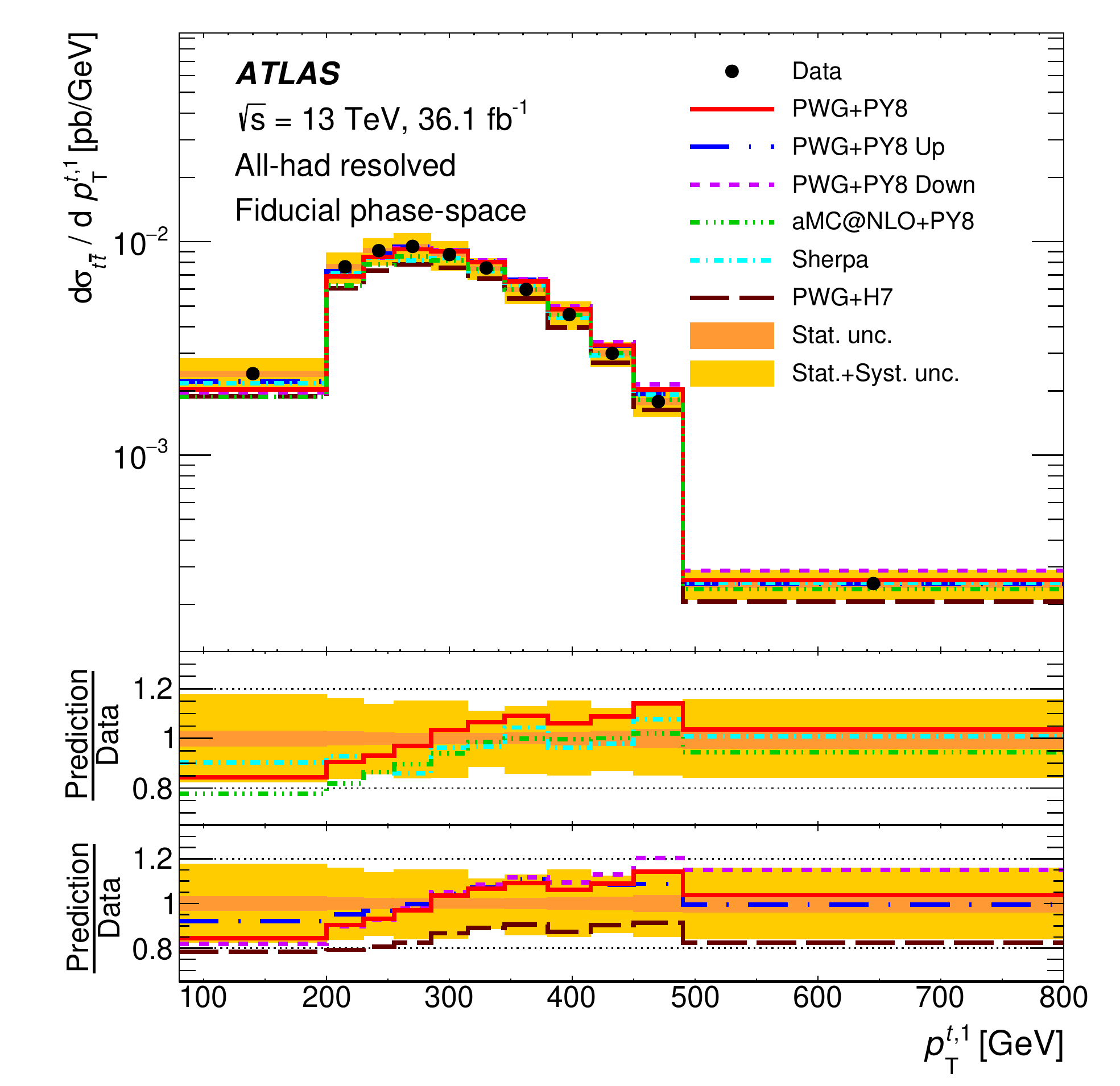}\label{fig:unf_abs:t1_pt}}
\subfloat[]{\includegraphics[width=0.5\textwidth]{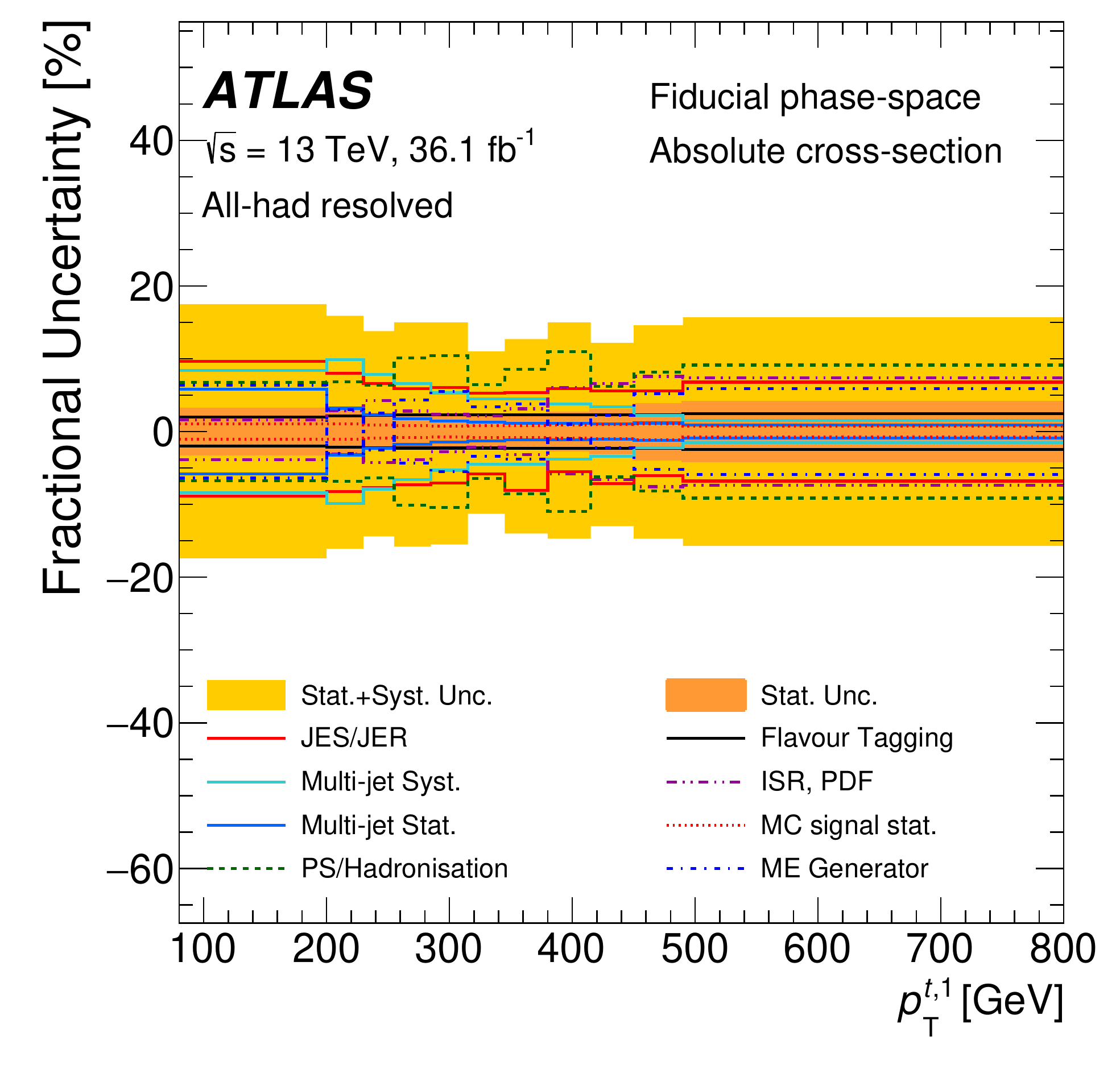}\label{fig:unc_abs:t1_pt}}
\caption{Particle-level single-differential absolute cross-section measurement~(a), as a function of the leading top-quark transverse momentum. The unfolded data are compared with theoretical predictions. \unfoldedplotcaption\,
Fractional uncertainties~(b) for the absolute single-differential cross-sections as a function of the leading top-quark transverse momentum. \reluncertplotcaption\,}
\label{fig:results:particle:absoluteTop1}
\end{figure}

\begin{figure}[htbp]
\centering
\subfloat[]{\includegraphics[width=0.5\textwidth]{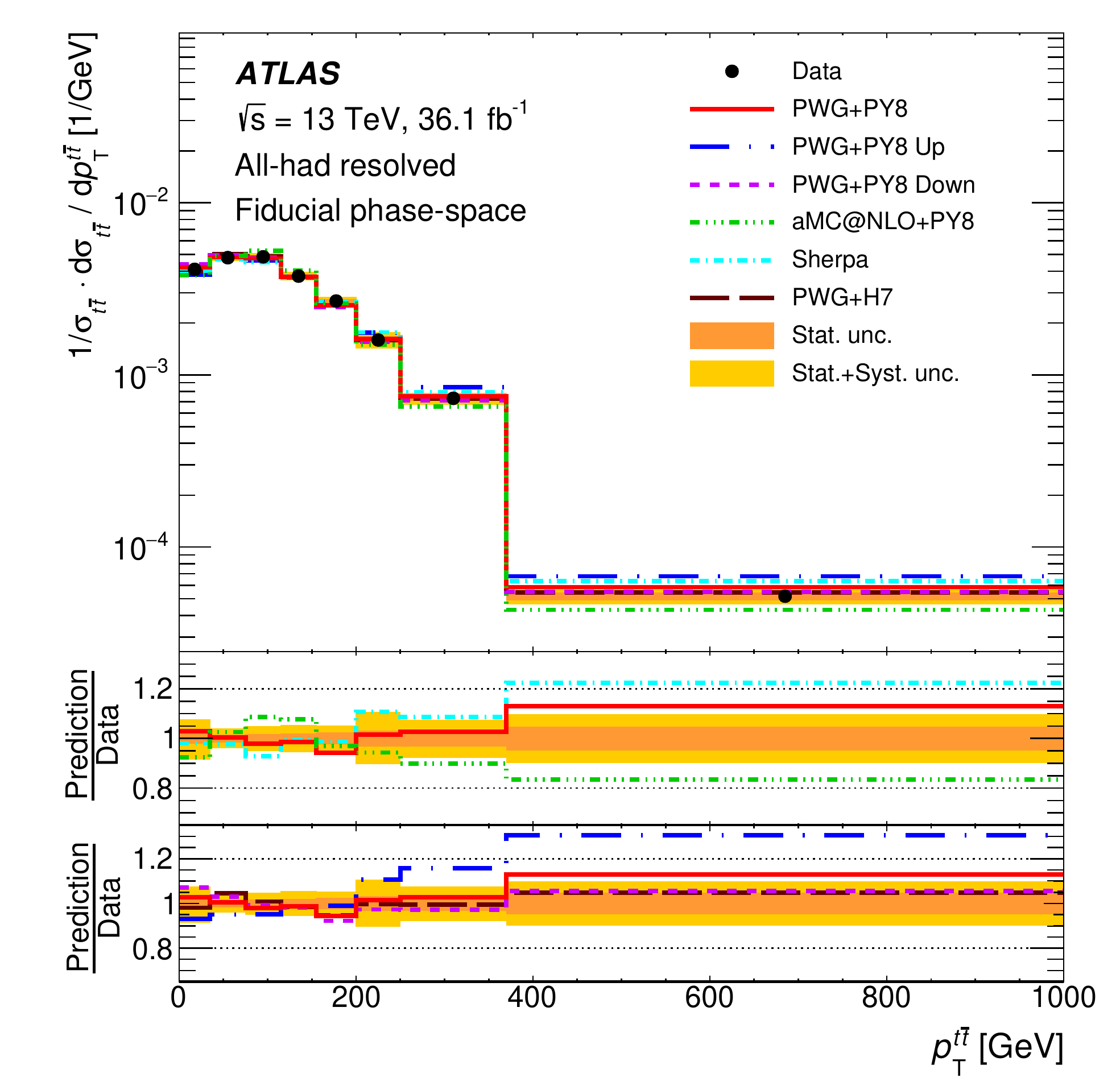}\label{fig:unf_rel:tt_pt}}
\subfloat[]{\includegraphics[width=0.5\textwidth]{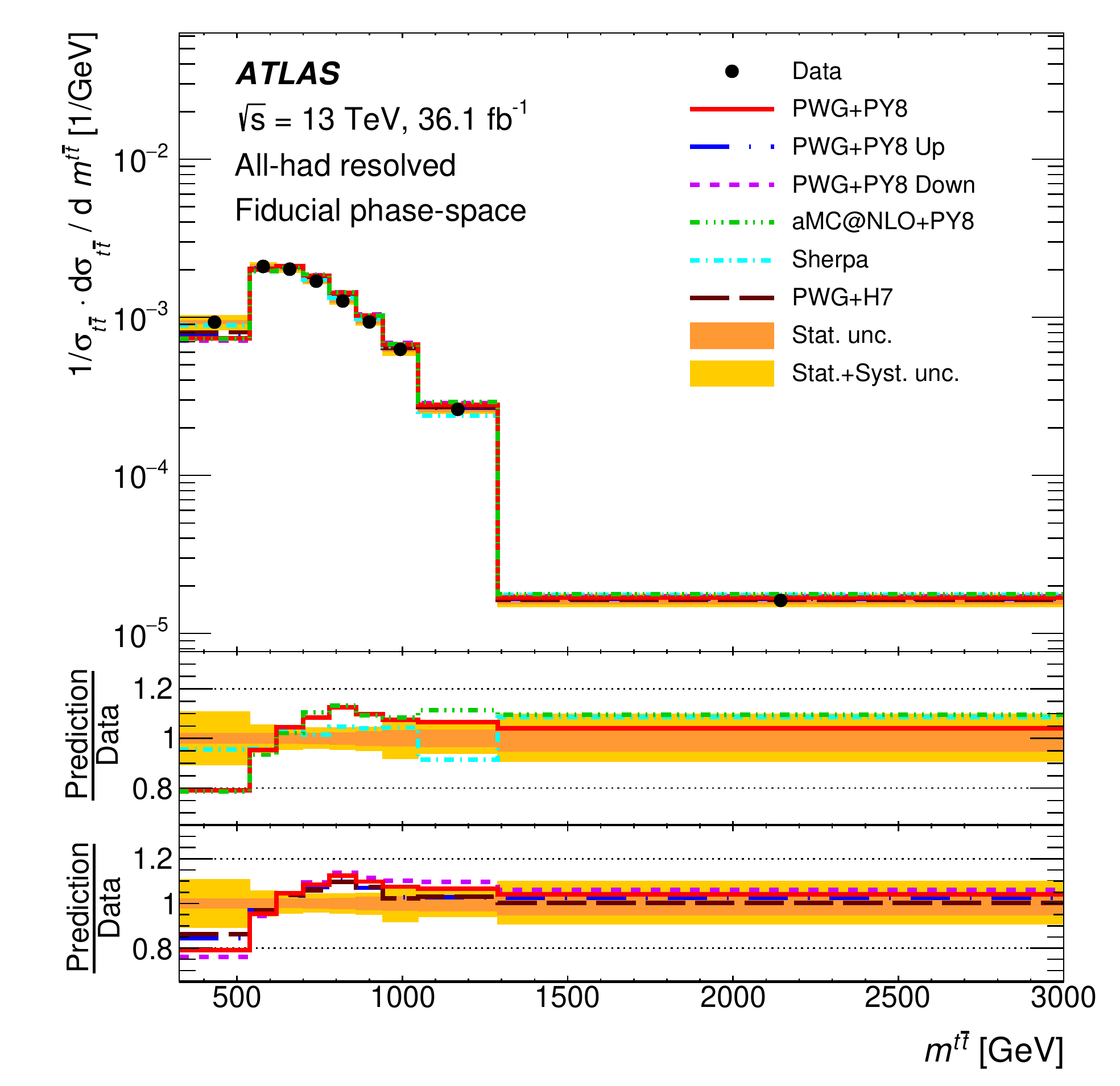}\label{fig:unf_rel:tt_m}}
\caption{Particle-level normalised single-differential cross-sections as a function of the~(a) transverse momentum of the $\ttbar$ system and of the~(b) $\ttbar$ system mass, compared with different MC predictions. \unfoldedplotcaption\,}
\label{fig:results:particle:topPair}
\end{figure}

\begin{figure}[htbp]
\centering
\includegraphics[width=0.5\textwidth]{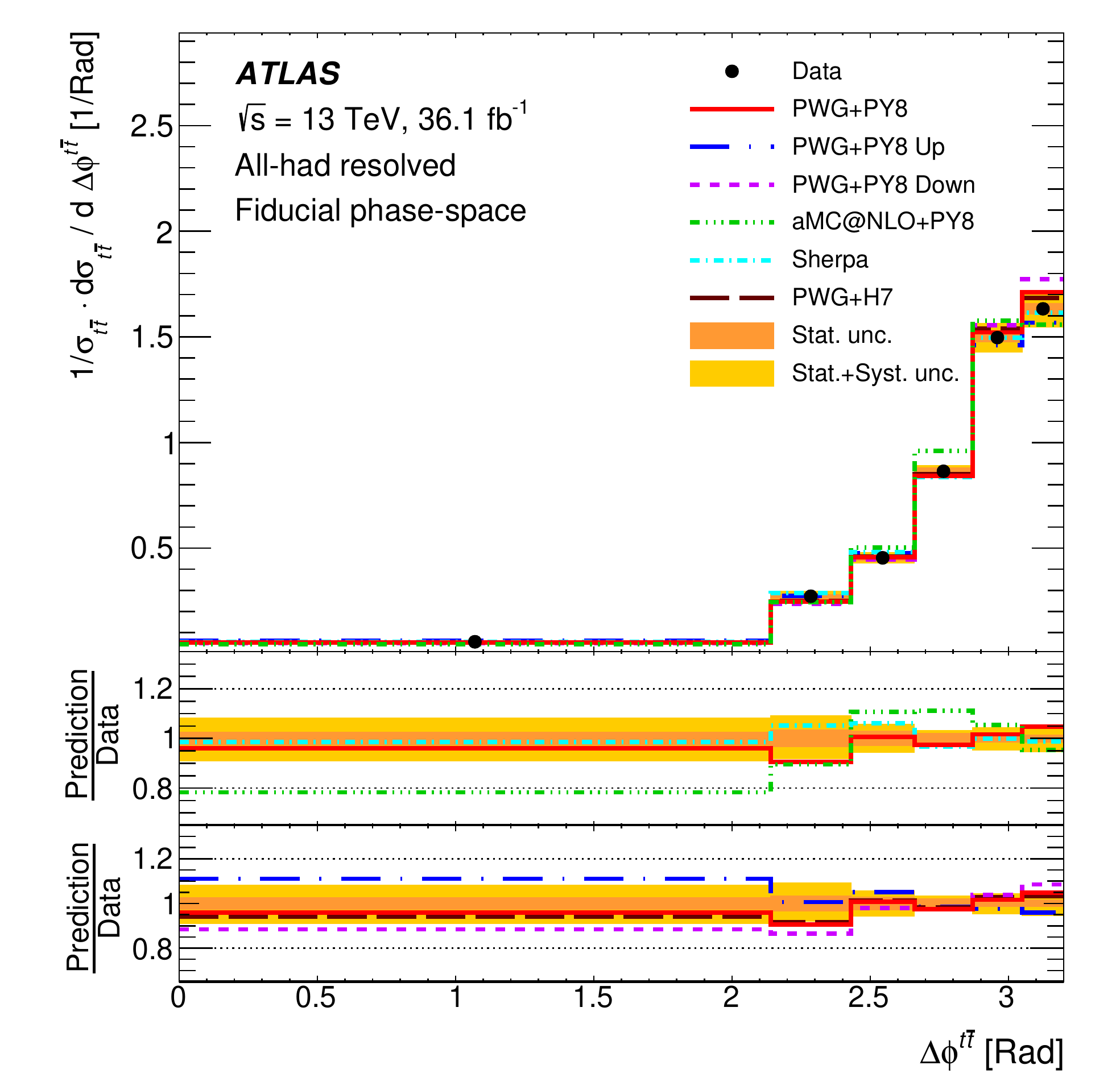}
\caption{Particle-level normalised single-differential cross-sections as a function of the azimuthal separation $\deltaPhittbar$ between the two top-quark candidates. \unfoldedplotcaption\,}
\label{fig:results:particle:DPhi}
\end{figure}

\begin{figure}[htbp]
\centering
\includegraphics[width=0.5\textwidth]{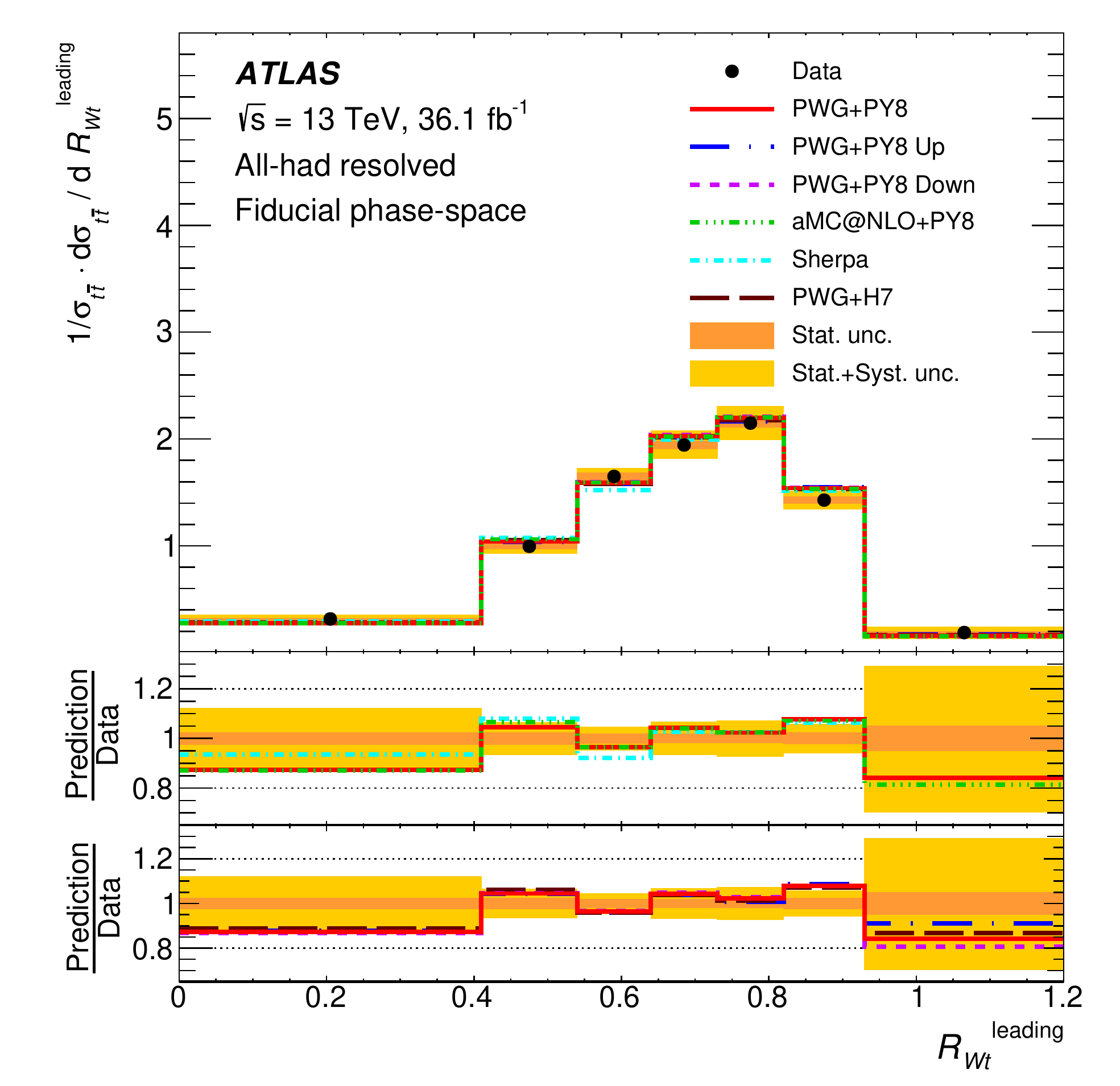}
\caption{Particle-level normalised single-differential cross-section as a function of the ratio of the $W$-boson \pt to its parent top-quark \pt for the leading top quark, \RWtl. \unfoldedplotcaption\,}
\label{fig:results:particle:RWtl}
\end{figure}

\begin{figure}[htbp]
\centering
\subfloat[]{\includegraphics[width=0.5\textwidth]{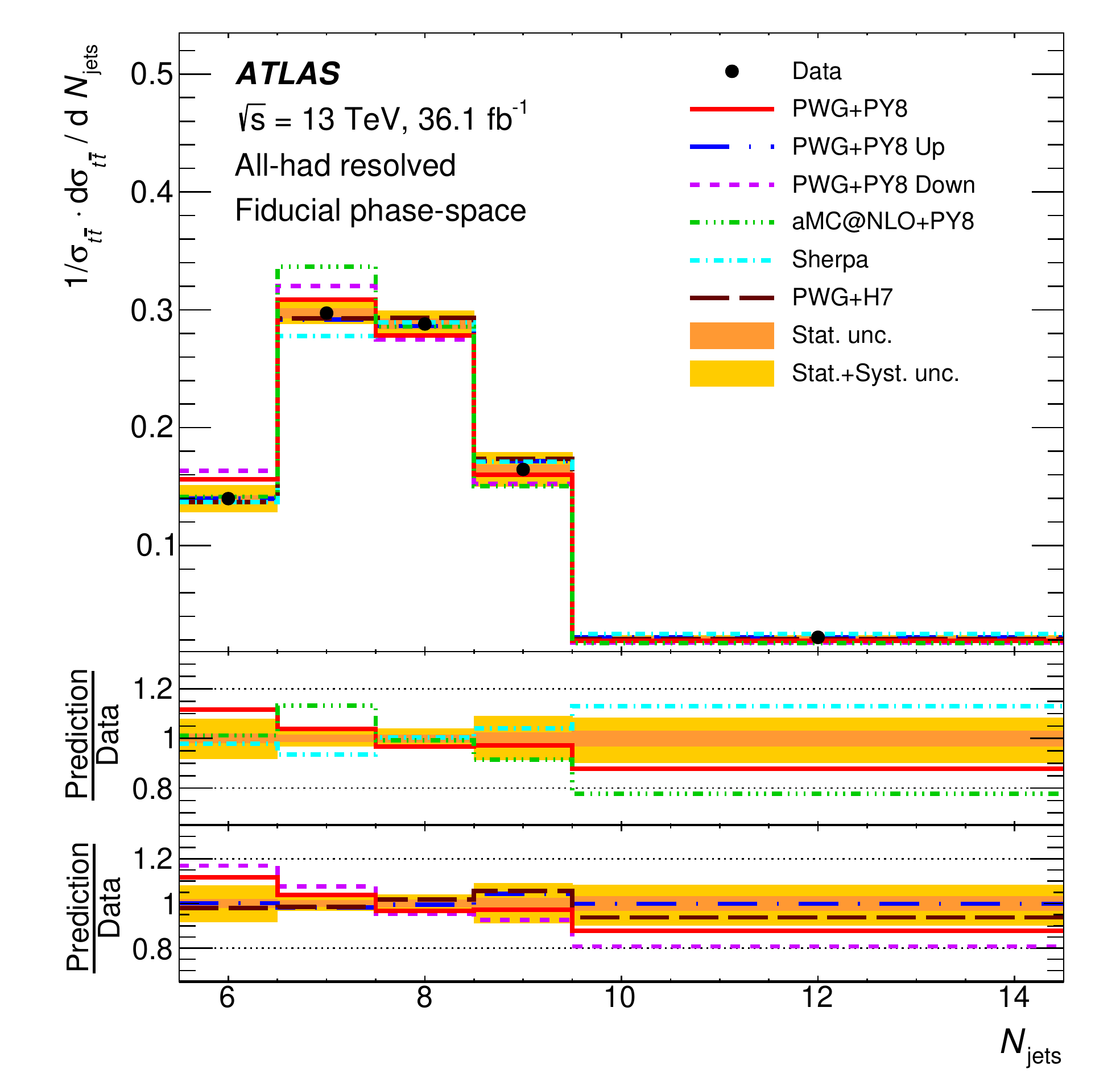}\label{fig:unf_rel:jets_n}}
\subfloat[]{\includegraphics[width=0.5\textwidth]{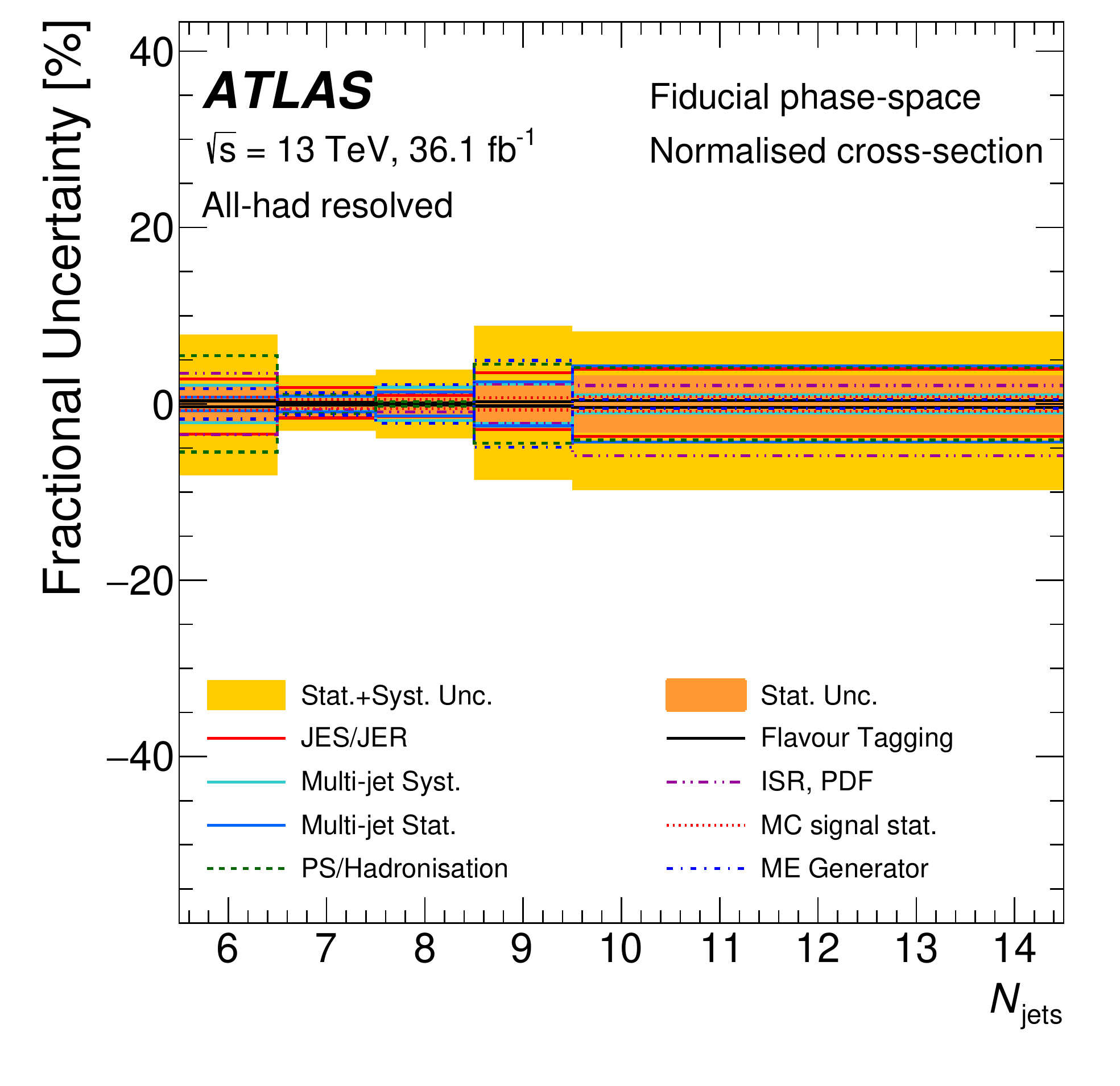}\label{fig:unc_rel:jets_n}}
\caption{
Single-differential normalised cross-section measurements~(a), unfolded at particle level, as a function of the jet multiplicity. \unfoldedplotcaption\,
Fractional uncertainties~(b) for the normalised single-differential distributions unfolded at particle level as a function of the jet multiplicity. \reluncertplotcaption\,}
\label{fig:results:particle:Njets}
\end{figure}

\begin{figure}[htbp]
\centering
\subfloat[]{\includegraphics[width=0.5\textwidth]{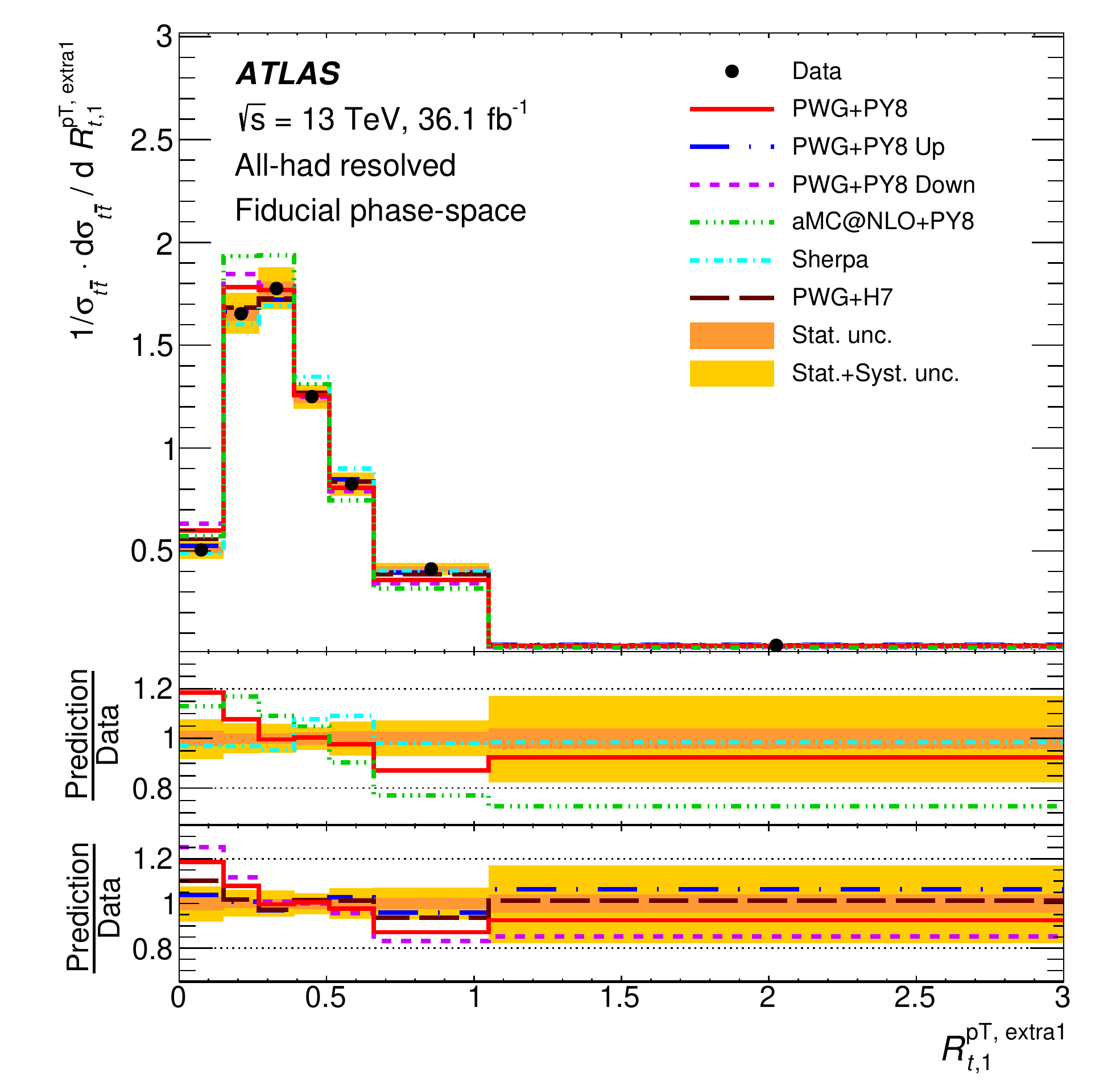}\label{fig:unf_rel:Rpt_e1t1}}
\subfloat[]{\includegraphics[width=0.5\textwidth]{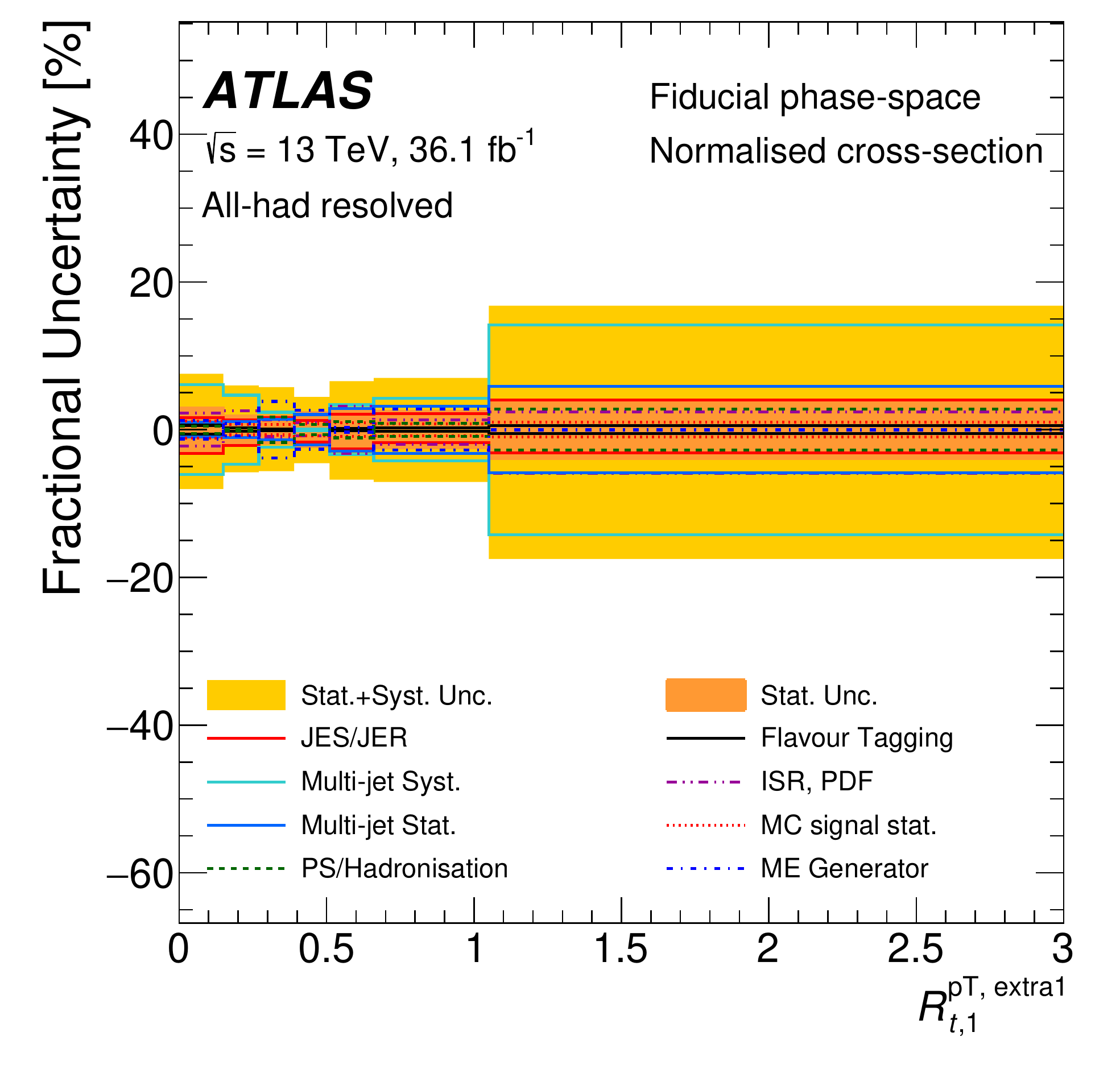}\label{fig:unc_rel:Rpt_e1t1}}
\\
\subfloat[]{\includegraphics[width=0.5\textwidth]{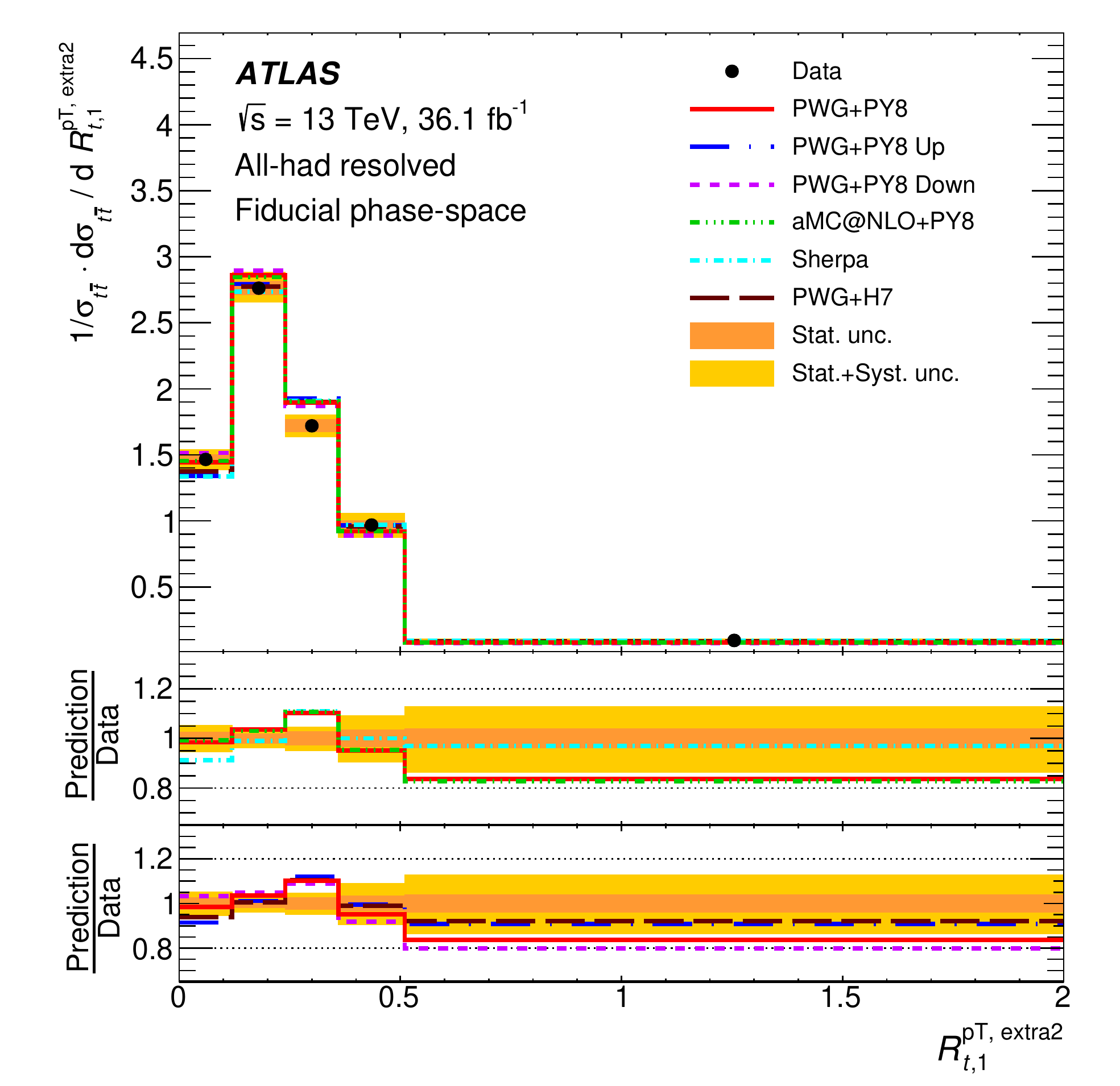}\label{fig:unf_rel:Rpt_e1t2}}
\subfloat[]{\includegraphics[width=0.5\textwidth]{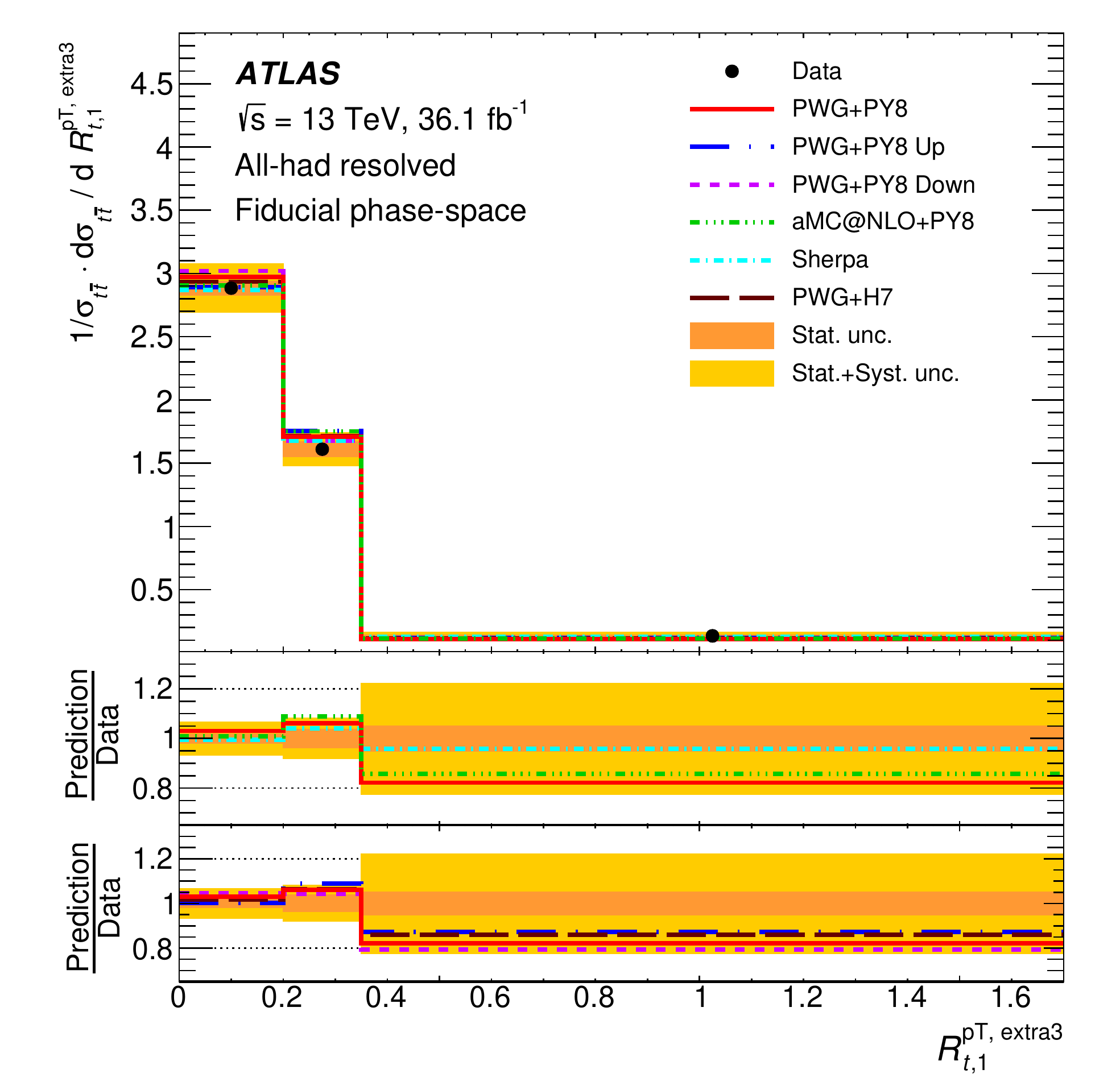}\label{fig:unf_rel:Rpt_e1t3}}
\caption{
Particle-level single-differential normalised cross-sections (a,c,d) as a function of the ratio of the leading (a), sub-leading (c) and sub-subleading (d) extra jet \pt to the leading top-quark \pt. \unfoldedplotcaption\,
Fractional uncertainties~(b) for the normalised single-differential cross-sections as a function of the ratio of leading extra jet \pt to the leading top-quark \pt. \reluncertplotcaption\,
}
\label{fig:results:particle:RExtraJetTop}
\end{figure}

\begin{figure}[htbp]
\centering
\subfloat[]{\includegraphics[width=0.5\textwidth]{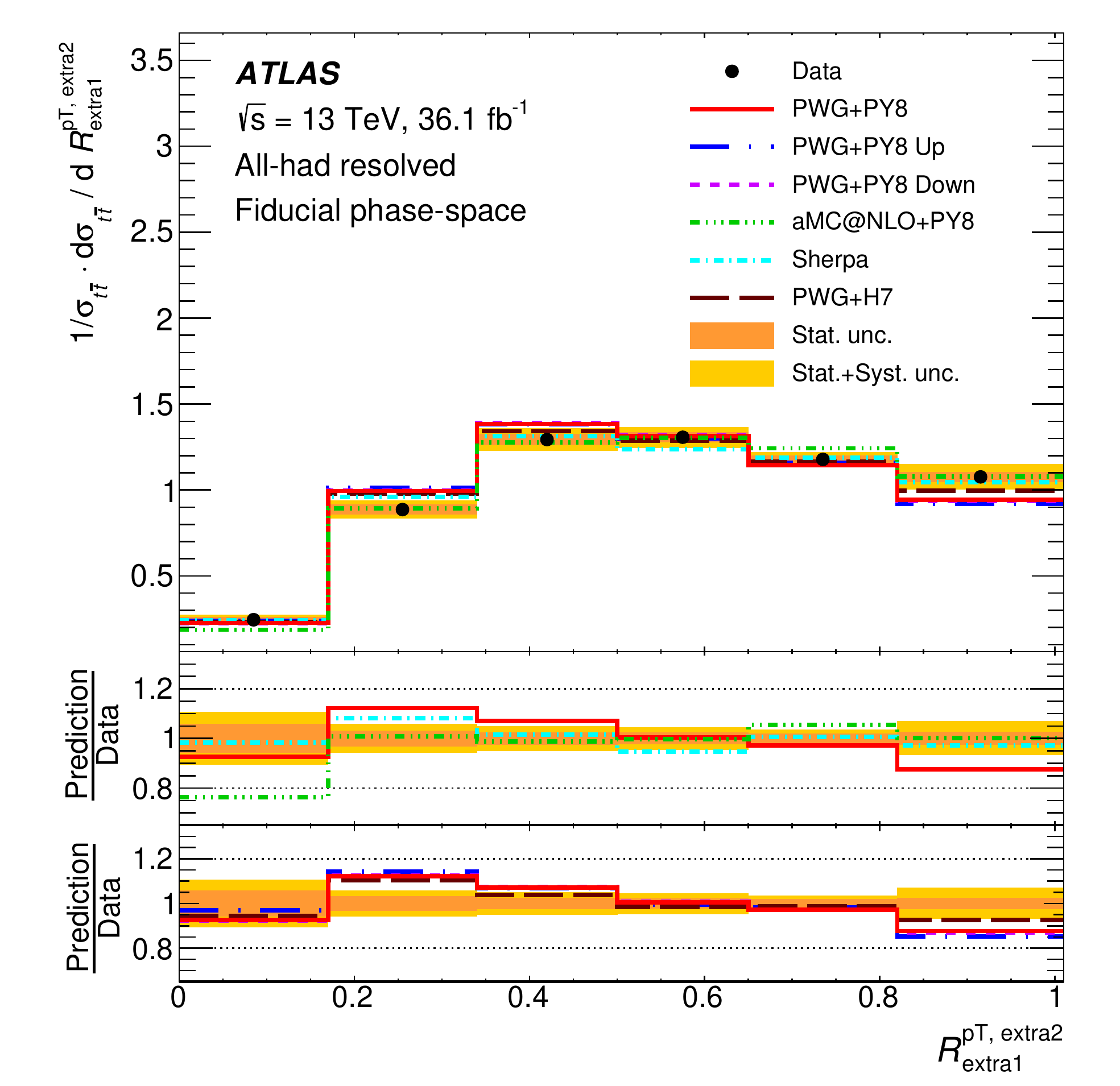}\label{fig:unf_rel:Rpt_e2e1}}
\subfloat[]{\includegraphics[width=0.5\textwidth]{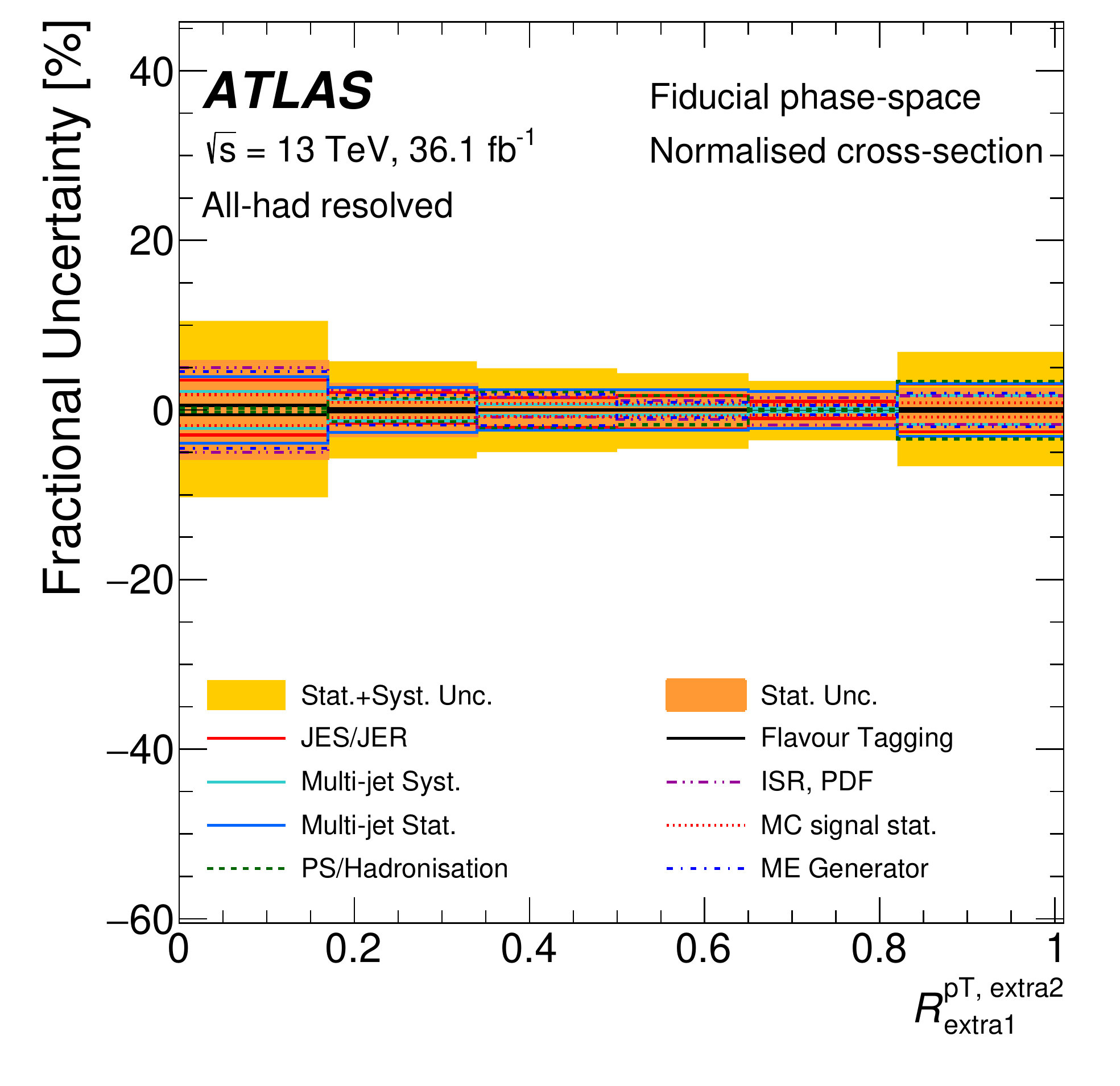}\label{fig:unc_rel:Rpt_e2e1}}
\caption{
Particle-level single-differential normalised cross-sections~(a) as a function of the ratio of the sub-leading extra jet \pt to the leading extra jet \pt. \unfoldedplotcaption\,
Fractional uncertainties~(b) for the normalised single-differential cross-sections as a function of the ratio of sub-leading extra jet \pt to the leading extra jet \pt. \reluncertplotcaption\,
}
\label{fig:results:particle:Rx2x1}
\end{figure}

\begin{figure}[htbp]
\centering
\includegraphics[width=0.5\textwidth]{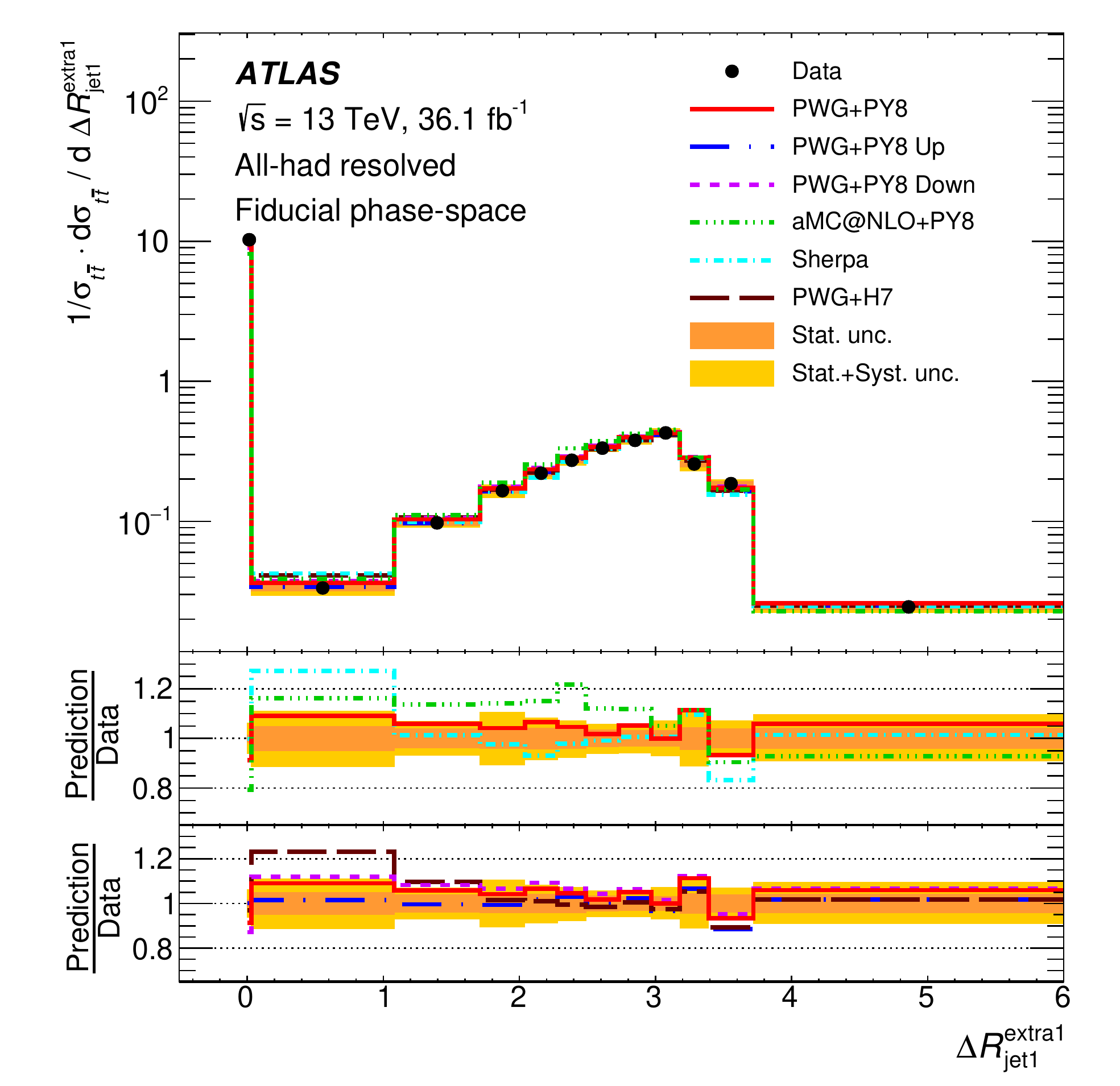}
\caption{
Particle-level normalised single-differential cross-section as a function of the angular separation $\Delta R$ between the leading extra jet and leading jet. \unfoldedplotcaption\,}
\label{fig:results:particle:ExtraJetDR}
\end{figure}

Single-differential cross-sections are presented in Figures~\ref{fig:results:particle:top12}--\ref{fig:results:particle:ExtraJetDR} for selected observables. These observables fall into two categories:
\begin{itemize}
\item Kinematic variables that are characteristic of the top-quark candidates or the top-quark pair system, specifically the transverse momentum of the leading and sub-leading top-quark candidates, the top-quark pair \pT and mass, the azimuthal angle between the two top-quark candidates, and the ratio of the \pT of the daughter $W$ boson of the leading top-quark candidate to the \pT of its parent.
\item Variables that compare the identified extra jets with the $\ttbar$ system kinematic properties, such as the ratio of the leading extra jet \pT to the top-quark \pT and of the sub-leading extra jet \pT to the leading extra jet \pT, as well as the $\Delta R$ between the leading extra jet and the leading jet. These observables explicitly differentiate between jets from the top-quark pair system and additional radiation.
\end{itemize}

\Fig{\ref{fig:results:particle:top12}} shows the measured normalised differential cross-sections as a function of the leading and sub-leading top-quark transverse momenta.
For illustration, these are shown alongside the detector-level distributions.
The detector-level distributions indicate good signal purity, with e.g. a background contamination of 30\% or less for $\pttl>200~\GeV$. The breakdown of the systematic uncertainties is shown on Figure~\ref{fig:unc:particle:top12}. The less collimated top-quark decays at low top-quark transverse momenta lead to a smaller signal--background separation, which in the case of the sub-leading top-quark \pt distribution causes the background uncertainty to be dominant, whereas the radiation and PDF uncertainties are most important at large \pt.
In the case of the leading top-quark \pt, the dominant uncertainties are from theoretical sources at low \pt, mainly from the matrix element calculation, while at high \pt no individual uncertainty source dominates.
 
Similar trends are seen for the two observables in Figure~\ref{fig:results:particle:top12}, showing that the event generators predict a harder \pt spectrum than observed in data.
The slopes in the lower panels are significant compared to the uncertainty bands.
The data are mostly consistent with the predictions from \SHERPA, \PHHSEVEN and \PHPYEIGHT with increased radiation.
 
\clearpage
The particle-level absolute differential cross-section measurement for the leading top-quark \pt is shown in \Fig{\ref{fig:results:particle:absoluteTop1}} for comparison with the normalised measurement shown in Figure~\ref{fig:unf_rel:t1_pt}.
It can clearly be seen that the normalisation substantially reduces the total uncertainties, particularly those originating from the parton shower and hadronisation, improving sensitivity to mismodelling of the distributions.
 
Two features of the \ttbar system are shown in \Fig{\ref{fig:results:particle:topPair}}, namely the top-quark pair transverse momentum and the top-quark pair mass.
The \ptttbar\ distribution mostly agrees well with the nominal \PHPYEIGHT prediction, but there are substantial deviations from the \MCNPYEIGHT and \SHERPA predictions, which predict spectra that are respectively too soft and too hard at high $\ptttbar$.
The \PHPYEIGHT Var3cUp variation is also harder than the data, which is at odds with what is seen for the top-quark transverse momenta in \Fig{\ref{fig:results:particle:top12}}, where this generator configuration reproduces the data better than the nominal \PHPYEIGHT configuration.
On the other hand, the \mttbar\ distribution shows the same features as already observed in the individual top-quark \pt distributions, since for central top-quark production the mass is dominated by the top-quark transverse momenta.
 
\clearpage
Given that in the all-hadronic channel the four-momenta of both top quarks are fully reconstructed, angular distributions are important observables to study in this channel.
One such observable is the azimuthal separation \dPhittbar\ between the top quarks, as shown in \Fig{\ref{fig:results:particle:DPhi}}, which may be sensitive to BSM couplings~\cite{quark_contact}, and is influenced by the \ptttbar\ distribution.
Deviations from the data are observed for both \PHPYEIGHT alternative samples and for \MCNPYEIGHT, but for the nominal \PHPYEIGHT configuration good agreement is observed.
 
\clearpage
Kinematic correlations in the top-quark decay process are probed, for example, by the ratio of the transverse momenta of either $W$ boson to its associated top quark.
For the leading top quark, this distribution is shown in \Fig{\ref{fig:results:particle:RWtl}}.
All MC predictions show poor agreement with data for this observable, with $p$-values at or below the 10\% level.
The data indicate a slightly higher proportion of events in the first and last bins where \RWtl\ has a value close to 0 or 1.
 
One of the chief goals of this paper is to characterise the modelling of jet radiation accompanying \ttbar production.
The unfolded jet multiplicity distribution is shown in \Fig{\ref{fig:results:particle:Njets}}.
From \Fig{\ref{fig:bkgest:Njets}}, the signal purity is seen to be relatively good for $\njets<10$, and the background uncertainties in the normalised cross-section (\Fig{\ref{fig:unc_rel:jets_n}}) are small compared with theoretical uncertainties.
Thus, conclusions can be safely drawn about the properties of up to three emissions, and these are discussed below.
 
Matrix element, ISR and parton shower/hadronisation uncertainties are dominant in most jet multiplicity bins, while jet energy scale and resolution uncertainties are large both for the seven-jet bin and for events with at least 10 jets.
The data indicate that more radiation is produced than predicted by the nominal \PHPYEIGHT configuration, while the Var3cUp variation and \PHHSEVEN are more consistent with the data.
While \SHERPA and \MCNPYEIGHT also reproduce the data fairly well, there is disagreement in the 7-jet bin which corresponds to events with a single hard emission.
 
\Fig{\ref{fig:results:particle:RExtraJetTop}} shows the differential cross-section as a function of two \pt ratios computed with the jets originating outside the \ttbar decay.
The \pt of the first ISR emission and the leading top-quark \pt are both important scales for the \ttbar production process.
Their ratio \Rptxonetl\ compares these two scales and shows a significant departure from the data for a number of generators.
Systematic uncertainties in the background estimation are dominant for the \Rptxonetl\ distribution but are substantially smaller than the observed deviation.
The background uncertainty is comparable to the modelling uncertainties for \Rptxtwotl\ but is also dominant for \Rptxthreetl.
It is observed that the leading extra jet's \pt spectrum has a most probable value at around a quarter of the leading top-quark \pt and exceeds the leading top-quark \pt at a low rate.
Consistent with other observations, the nominal \PHPYEIGHT configuration produces a first emission that is too soft with respect to the data, as does \MCNPYEIGHT.
The second emission \pt peaks slightly lower than the first.
While reproduced better than the leading emission, all simulations produce too many events with \Rptxtwotl\ close to 0.3 and too few elsewhere.
The third emission \pt does not show significant deviations from the data.
 
In events with substantial ISR, the leading extra jet may provide the relevant scale for the process. The distribution of \RptxtwoX\ (\Fig{\ref{fig:results:particle:Rx2x1}}) tests the second emission modelling relative to the leading extra jet \pt and shows a minor trend.
Cancellation of systematic uncertainties, notably those in the background prediction, across the spectrum results in small uncertainties in the \RptxtwoX\ distribution in all bins.
The sub-leading extra jet \pt is broadly peaked just below half the leading extra jet \pt with a skew towards higher values.
 
\clearpage
\Fig{\ref{fig:results:particle:ExtraJetDR}} shows the separation in $\Delta R$ between the first emission and the leading jet in the event, which may or may not originate from the decay of one of the top quarks.
A significant peak is observed at $\Delta R = 0$, demonstrating that in events with at least one extra jet the leading jet is most often from ISR rather than a top-quark decay product.
The distribution of $\Delta R$ for events in which the leading jet is associated with one of the top quarks has a tendency towards large values, close to $\pi$.
In other words, in events where the leading jet is a decay product of one of the top quarks, the first hard emission tends to be emitted in a direction that balances the leading top-quark jet.
 
Significant mismodelling of this distribution is observed in \SHERPA, \MCNPYEIGHT and \PHHSEVEN, all of which overestimate how frequently the leading jet is a decay product of one of the top quarks.
For such events, the extra jet is also typically emitted too close to the leading top-quark jet.
The same trend is seen to a lesser degree for the nominal \PHPYEIGHT configuration, whereas the Var3cUp variation reproduces the data well.
 
\clearpage
 
\begin{figure}[htbp]
\centering
\subfloat[]{\includegraphics[width=0.4\textwidth]{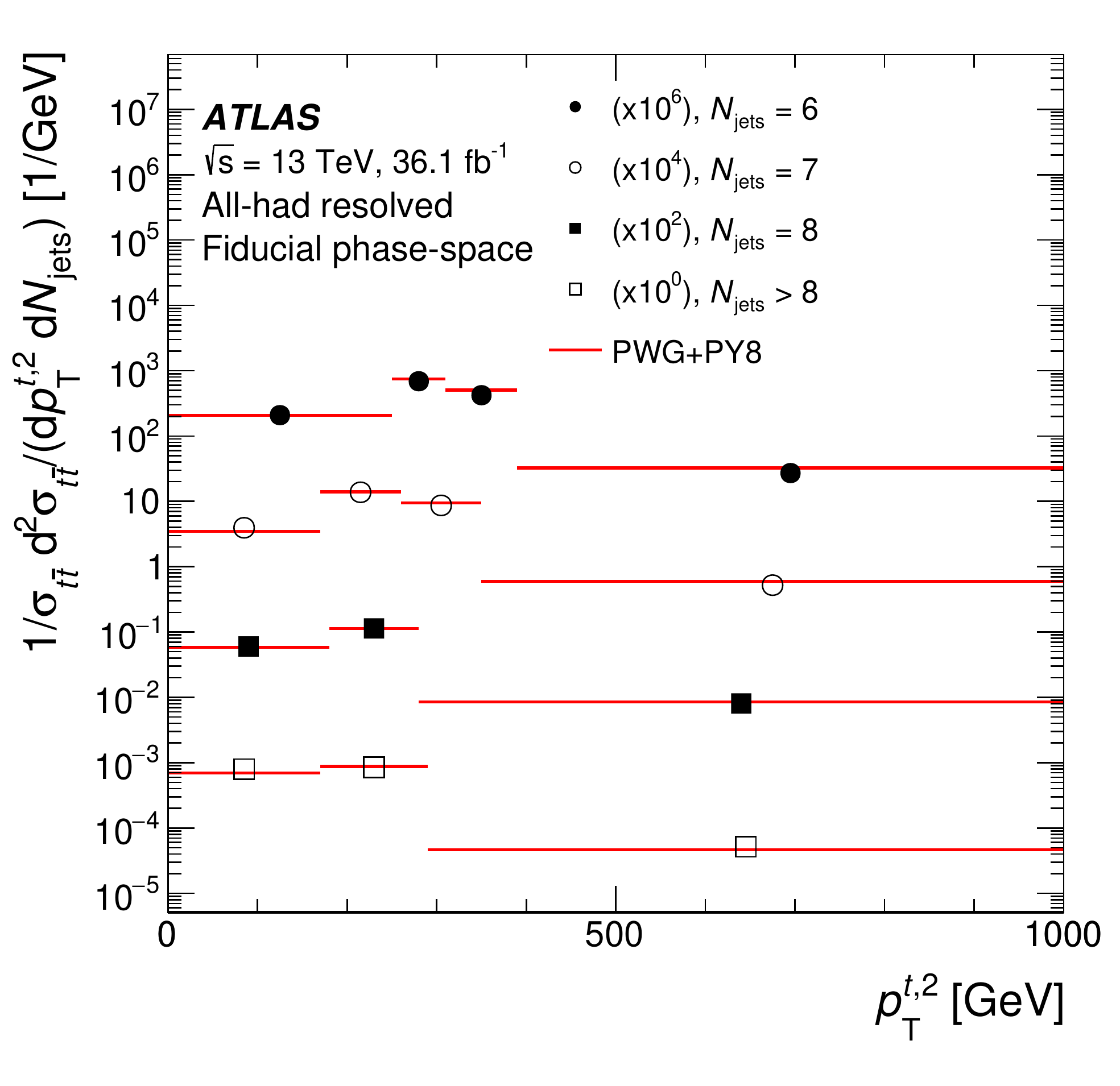}\label{fig:unf_rel:t2_pt_jet_n}}
\\
\subfloat[]{\includegraphics[width=0.7\textwidth]{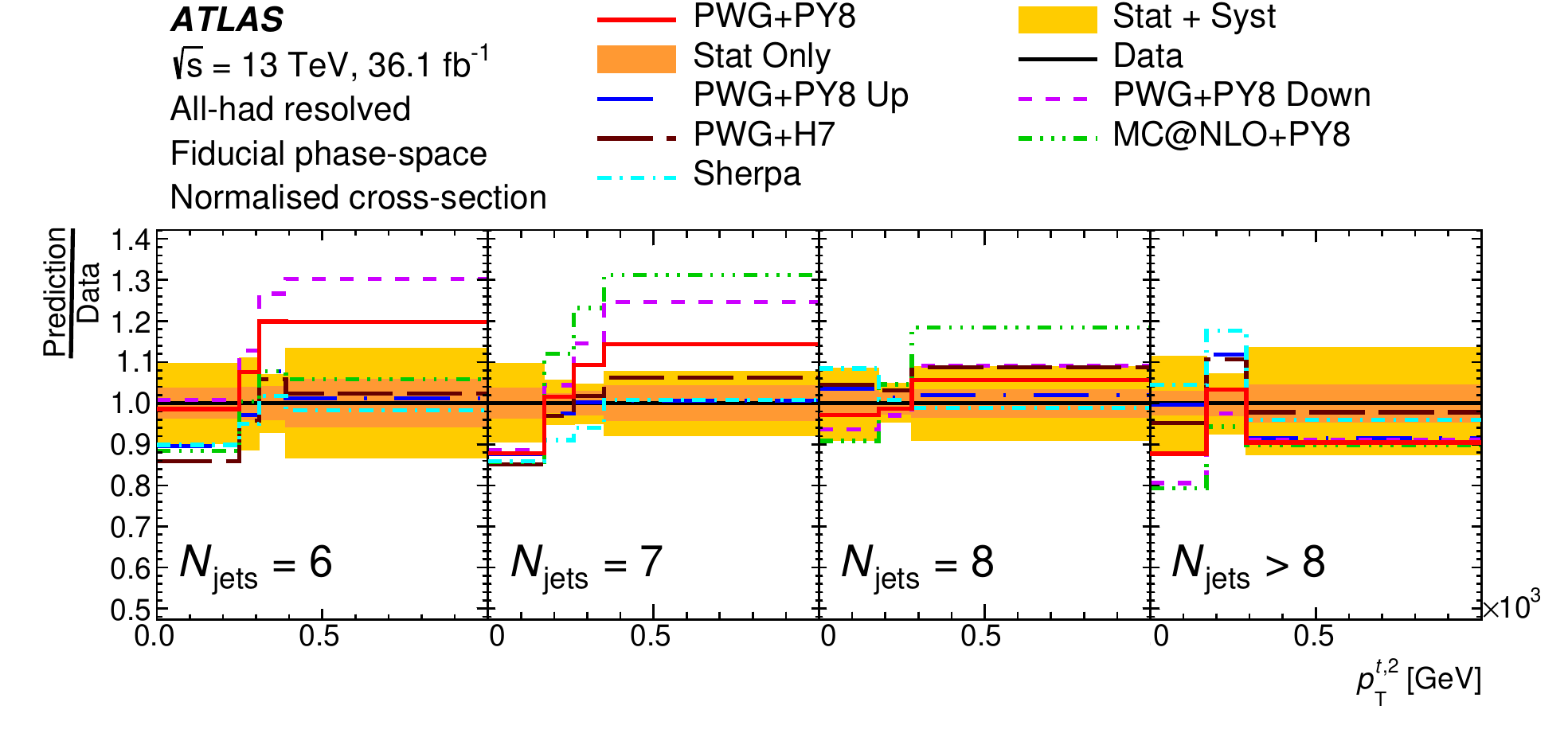}\label{fig:comp_rel:t2_pt_jet_n}}
 
\caption{Particle-level double-differential normalised cross-section~(a) as a function of the sub-leading top-quark transverse momentum \pttsl\ in bins of the jet multiplicity \njets, compared with the nominal \PHPYEIGHT prediction without uncertainties. Data points are placed at the centre of each bin. Different markers are used to distinguish the four bins in \njets, while \pttsl\ is shown on the horizontal axis.
The ratio~(b) of the measured cross-section to different MC predictions. \twodratioplotcaption\,}
\label{fig:results:particle:DoubleTop2}
\end{figure}

\begin{figure}[htbp]
\centering
\subfloat[]{\includegraphics[width=0.4\textwidth]{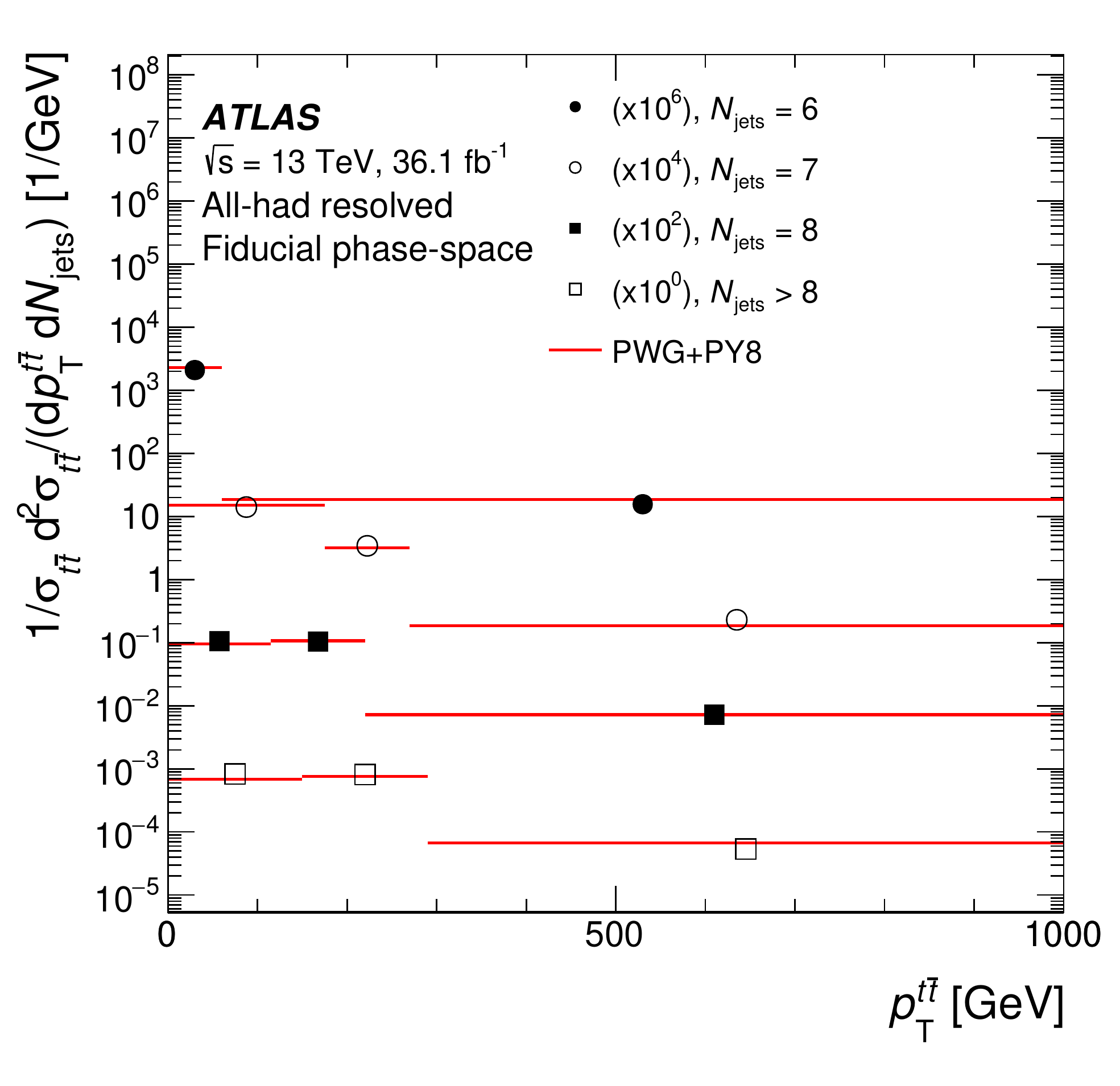}\label{fig:unf_rel:tt_pt_jet_n}}
\\
\subfloat[]{\includegraphics[width=0.7\textwidth]{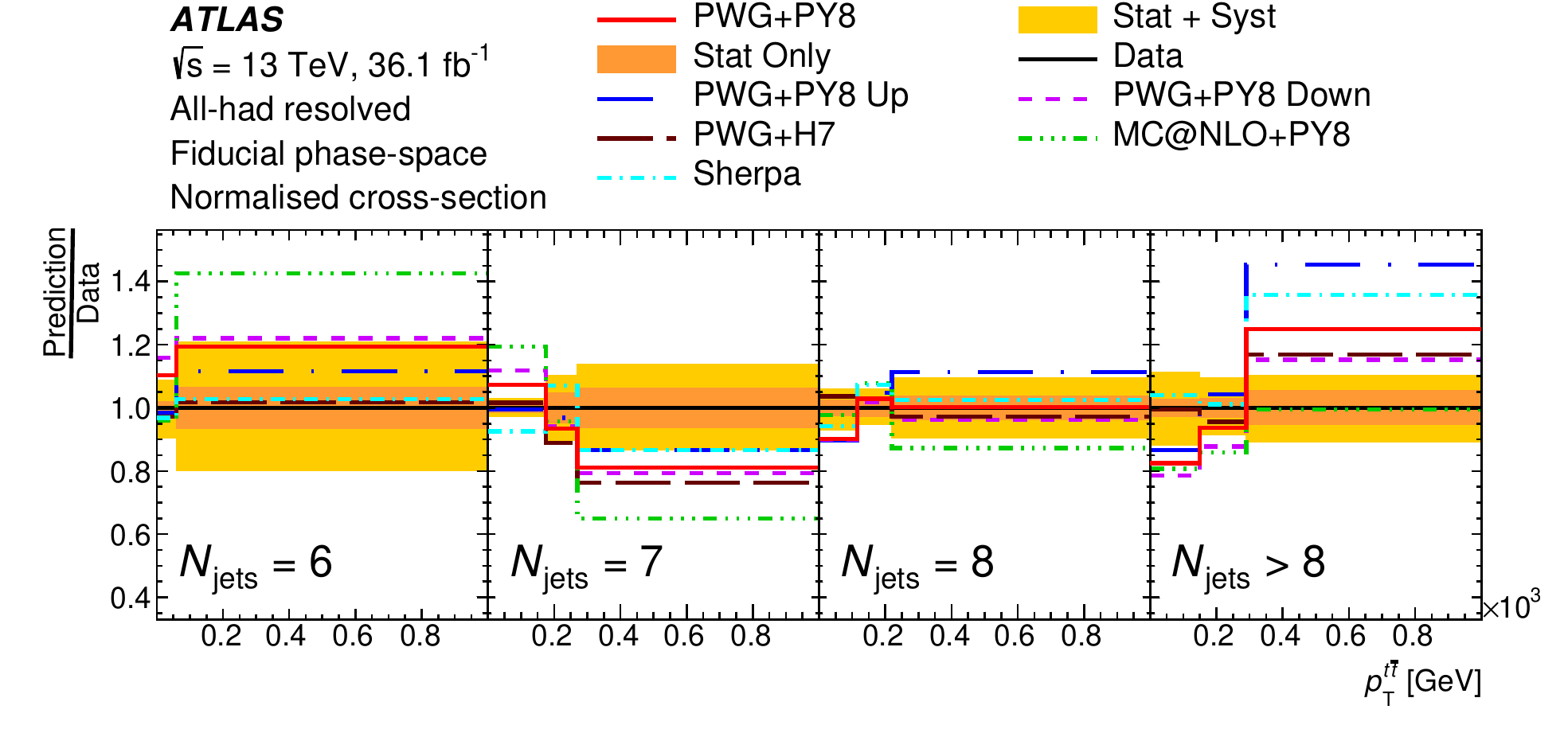}\label{fig:comp_rel:tt_pt_jet_n}}
 
\caption{Particle-level double-differential normalised cross-section~(a) as a function of the $\ttbar$ system transverse momentum $\ptttbar$ in bins of the jet multiplicity \njets, compared with the nominal \PHPYEIGHT prediction without uncertainties. Data points are placed at the centre of each bin. Different markers are used to distinguish the four bins in \njets, while \ptttbar\ is shown on the horizontal axis.
The ratio~(b) of the measured cross-section to different MC predictions. \twodratioplotcaption\,}
\label{fig:results:particle:DoubleTopPair}
\end{figure}

Normalised double-differential cross-sections as a function of the sub-leading top-quark \pt and the $\pt$ of the $\ttbar$ system in bins of jet multiplicity are presented in Figures~\ref{fig:results:particle:DoubleTop2} and \ref{fig:results:particle:DoubleTopPair}, respectively.
For low jet multiplicities, which are relatively pure in signal, the dominant uncertainties are from jet energy scales, PDFs and the modelling of the \ttbar radiation.
As the parton shower modelling is particularly important at larger jet multiplicities, the corresponding uncertainty grows to be the most significant for both observables.
 
In both observables, the six- and seven-jet bins show the clearest signs of mismodelling.
The MC predictions tend to be too hard for the sub-leading top-quark \pt, as was observed in the single-differential measurement.
For the \ttbar transverse momentum, on the other hand, different trends are seen, where the predictions are typically too soft in the seven-jet bin, where a single hard emission is produced, but too hard in the other bins.
 
\begin{figure}[htbp]
\centering
\subfloat[]{\includegraphics[width=0.4\textwidth]{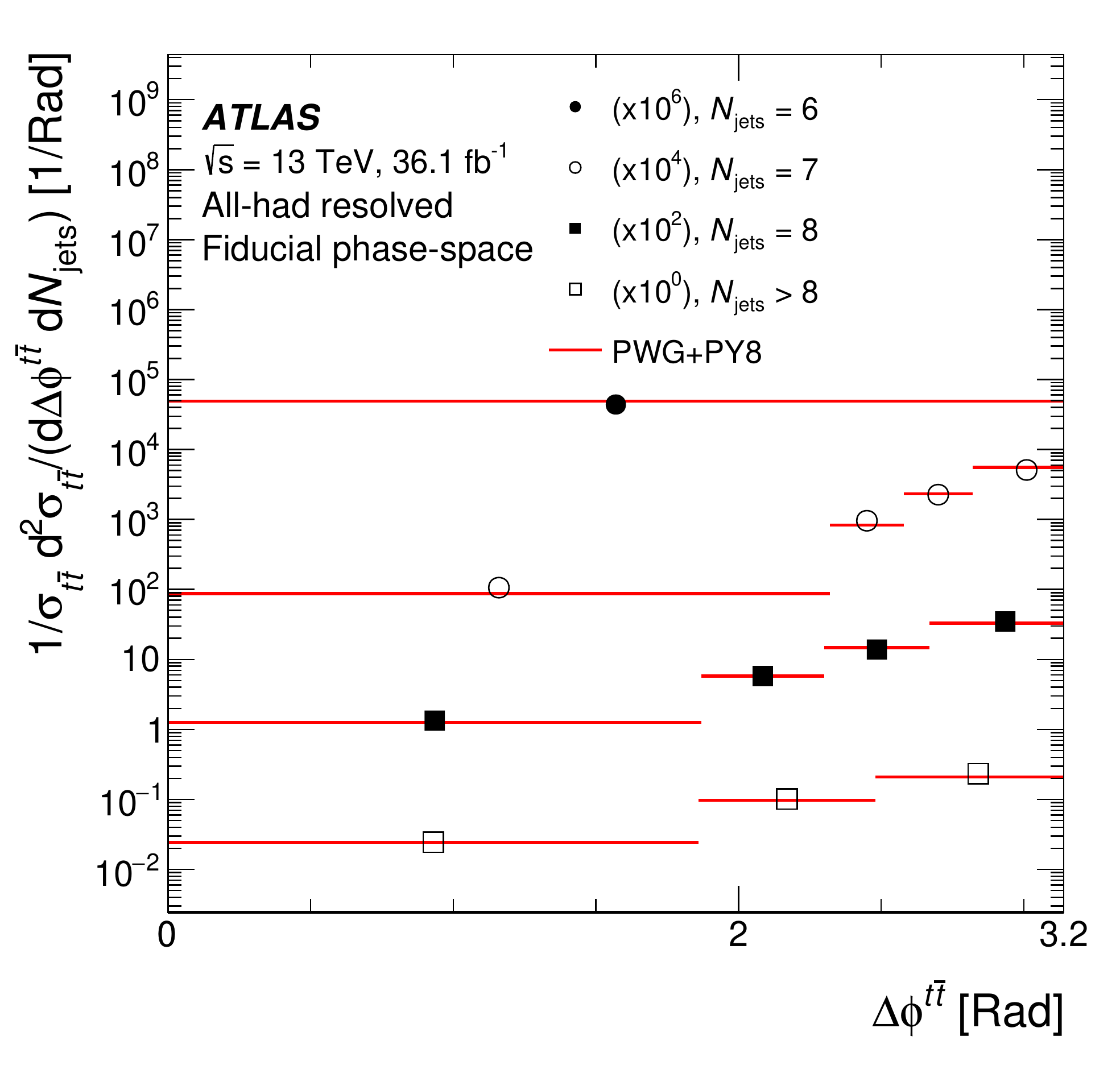}\label{fig:unf_rel:DeltaPhi_jet_n}}
\\
\subfloat[]{\includegraphics[width=0.7\textwidth]{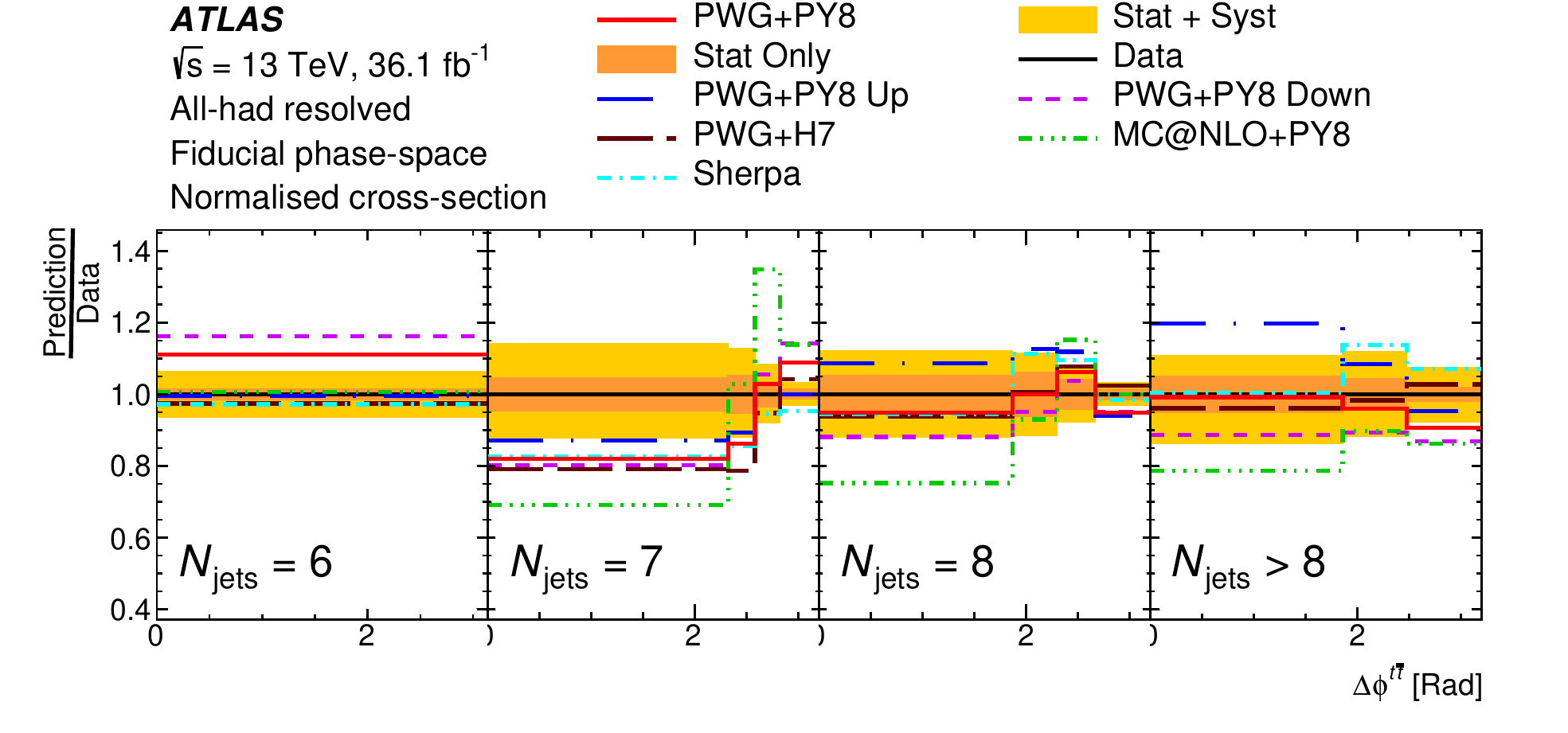}\label{fig:comp_rel:DeltaPhi_jet_n}}
\caption{Particle-level double-differential normalised cross-section~(a) as a function of the azimuthal separation \deltaPhittbar\  between the top quark and the antitop quark in bins of the jet multiplicity \njets, compared with the nominal \PHPYEIGHT prediction without uncertainties. Data points are placed at the centre of each bin. Different markers are used to distinguish the four bins in \njets, while \deltaPhittbar\ is shown on the horizontal axis.
The ratio~(b) of the measured cross-section to different MC predictions. \twodratioplotcaption}
\label{fig:results:particle:DoubleDeltaPhi}
\end{figure}

\Fig{\ref{fig:results:particle:DoubleDeltaPhi}} shows the normalised double-differential cross-section as a function of the \ttbar azimuthal separation \deltaPhittbar~ and the jet multiplicity.
Whereas the single-differential distribution (\Fig{\ref{fig:results:particle:DPhi}}) shows mostly good agreement between data and simulation, the decomposition of the distribution into different jet multiplicity bins is poorly modelled by all event generators.
The common trend is that in the seven-jet bin, the data indicates that the top-quark pair should be more back-to-back, while for events with at least two additional emissions the top-quark pair should be less separated.
This observation correlates clearly with the results seen in the double-differential measurement of \ptttbar\ versus \njets~(\Fig{\ref{fig:results:particle:DoubleTopPair}}); events where the two top quarks are more back-to-back will result in a system with a smaller \ptttbar\, as observed in the distribution in the seven-jet bin, while in the case of smaller angular distance the \ptttbar\ would be higher as observed in the distribution of the last \njets{} bin of \Fig{\ref{fig:results:particle:DoubleTopPair}}.
 
\clearpage
\subsubsection{Results at parton level in the full phase space}
\label{sec:results:parton}
 
\begin{figure}[htbp]
\centering
\subfloat[]{\includegraphics[width=0.5\textwidth]{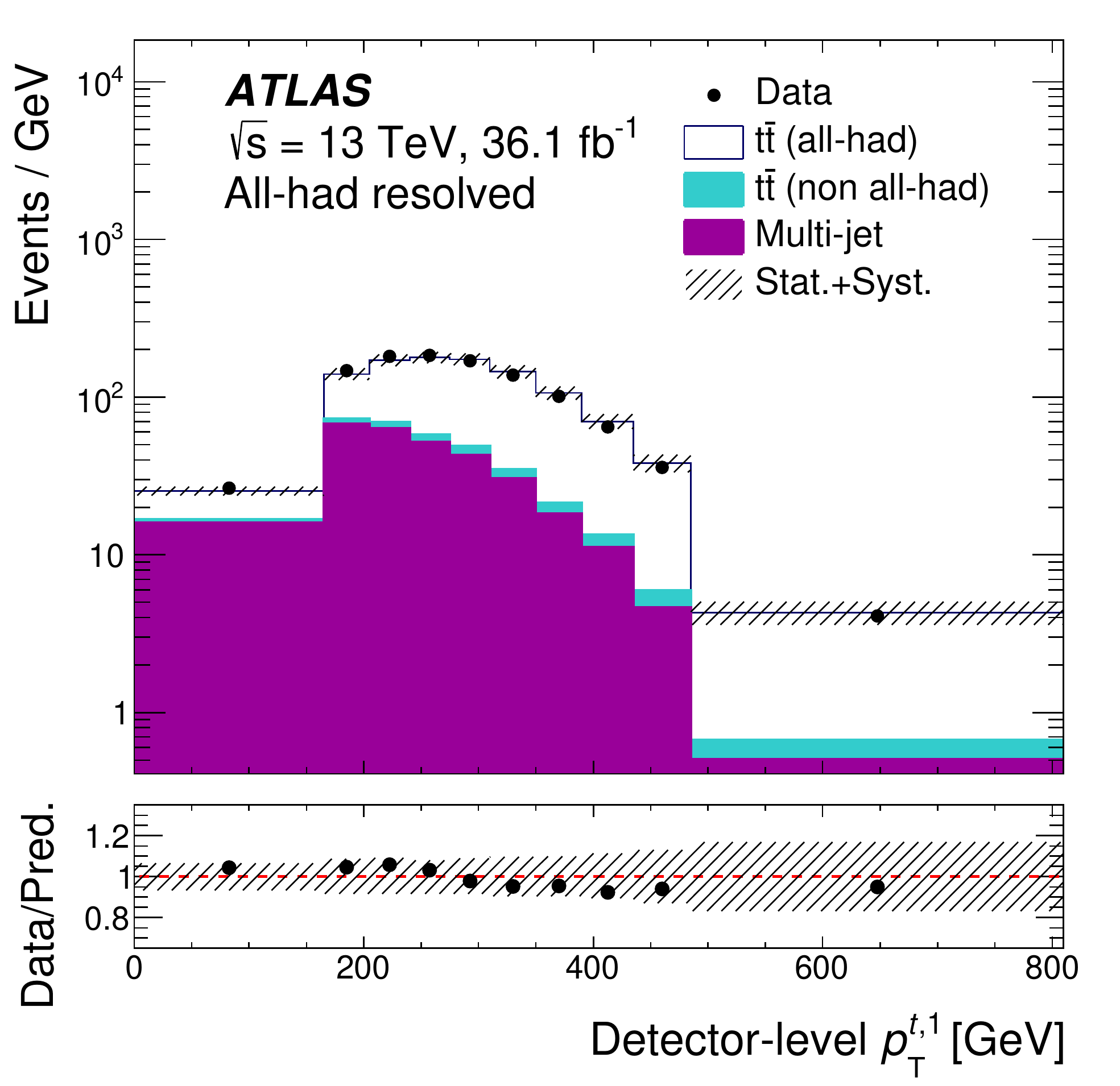}\label{fig:datamc_truth:t1_pt}}
\subfloat[]{\includegraphics[width=0.5\textwidth]{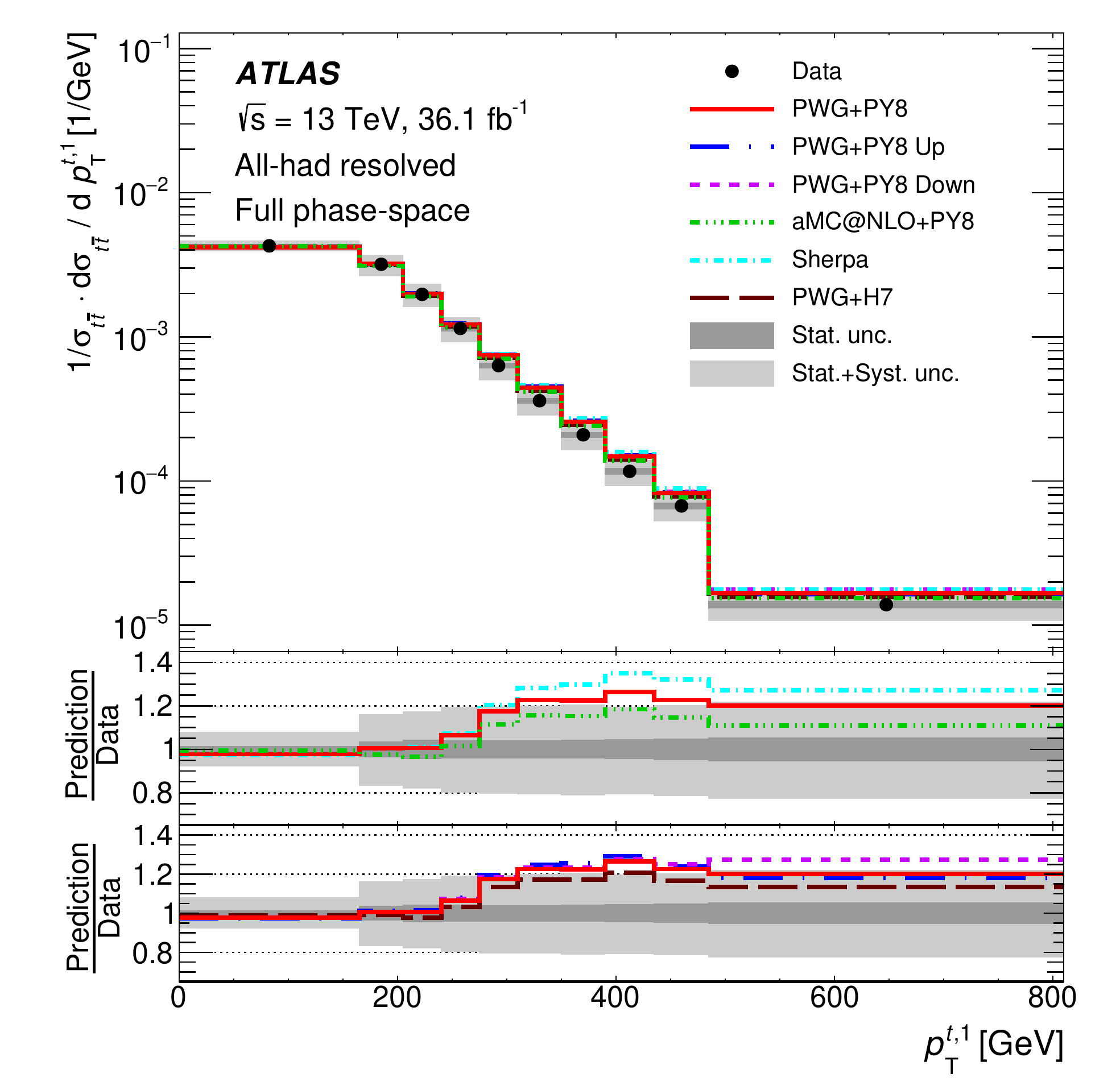}\label{fig:unf_rel_truth:t1_pt}}
 
\subfloat[]{\includegraphics[width=0.5\textwidth]{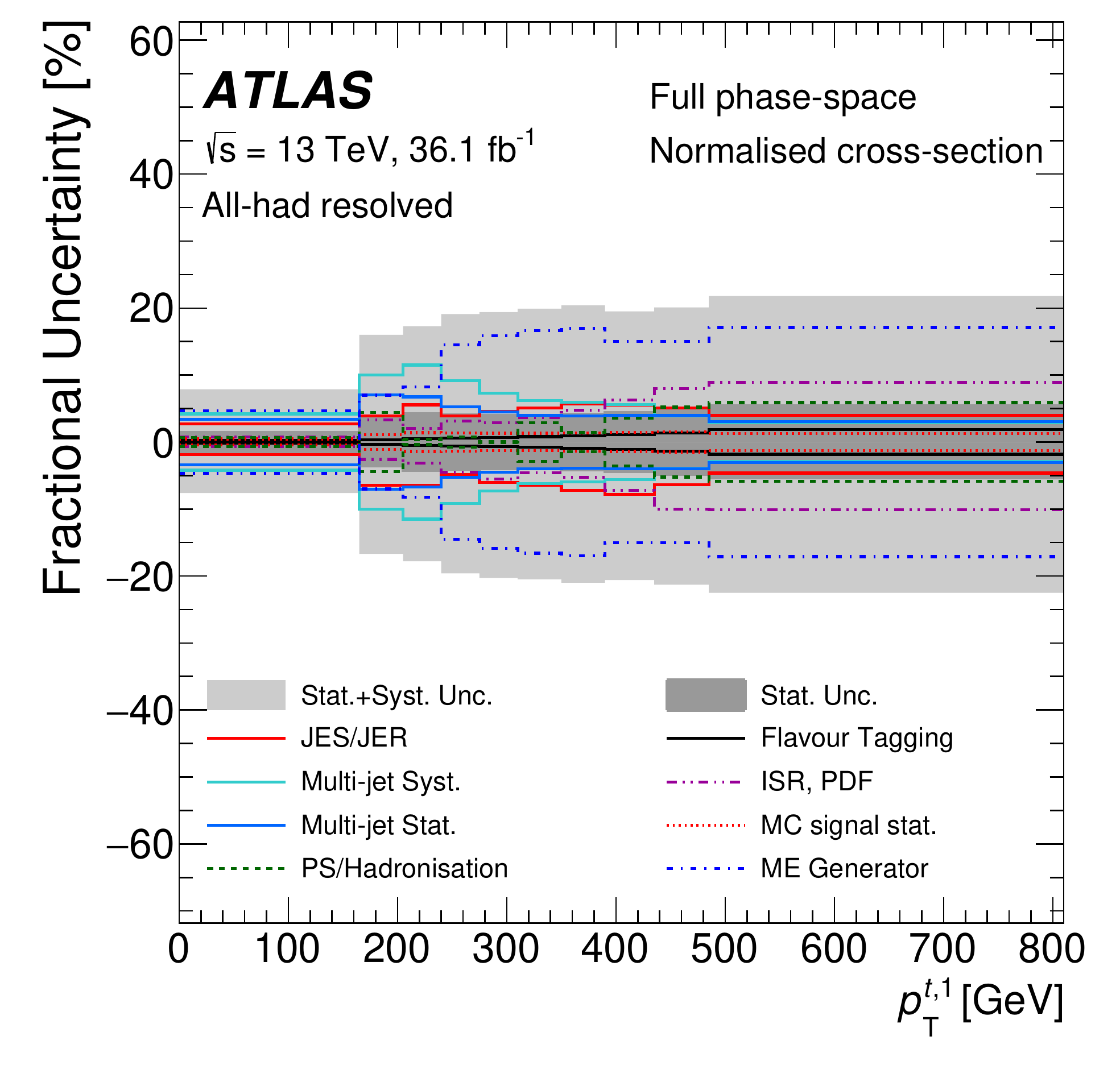}\label{fig:unc_rel_truth:t1_pt}}
\caption{Comparison of the ATLAS data with the fully simulated nominal SM predictions~(a) for the leading top-quark transverse momentum. \dataMCplotcaption\ Overflow events are included in the last bin of every distribution shown.
Single-differential normalised cross-section measurements, unfolded at parton level~(b), as a function of the leading top-quark transverse momentum. The unfolded data are compared with theoretical predictions. \unfoldedplotcaption\,
Fractional uncertainties for the normalised single-differential distributions unfolded at parton level~(c) as a function of the leading top-quark transverse momentum. \reluncertplotcaption\,
}
\label{fig:results:parton:top12}
\end{figure}

\begin{figure}[htbp]
\centering
\subfloat[]{\includegraphics[width=0.5\textwidth]{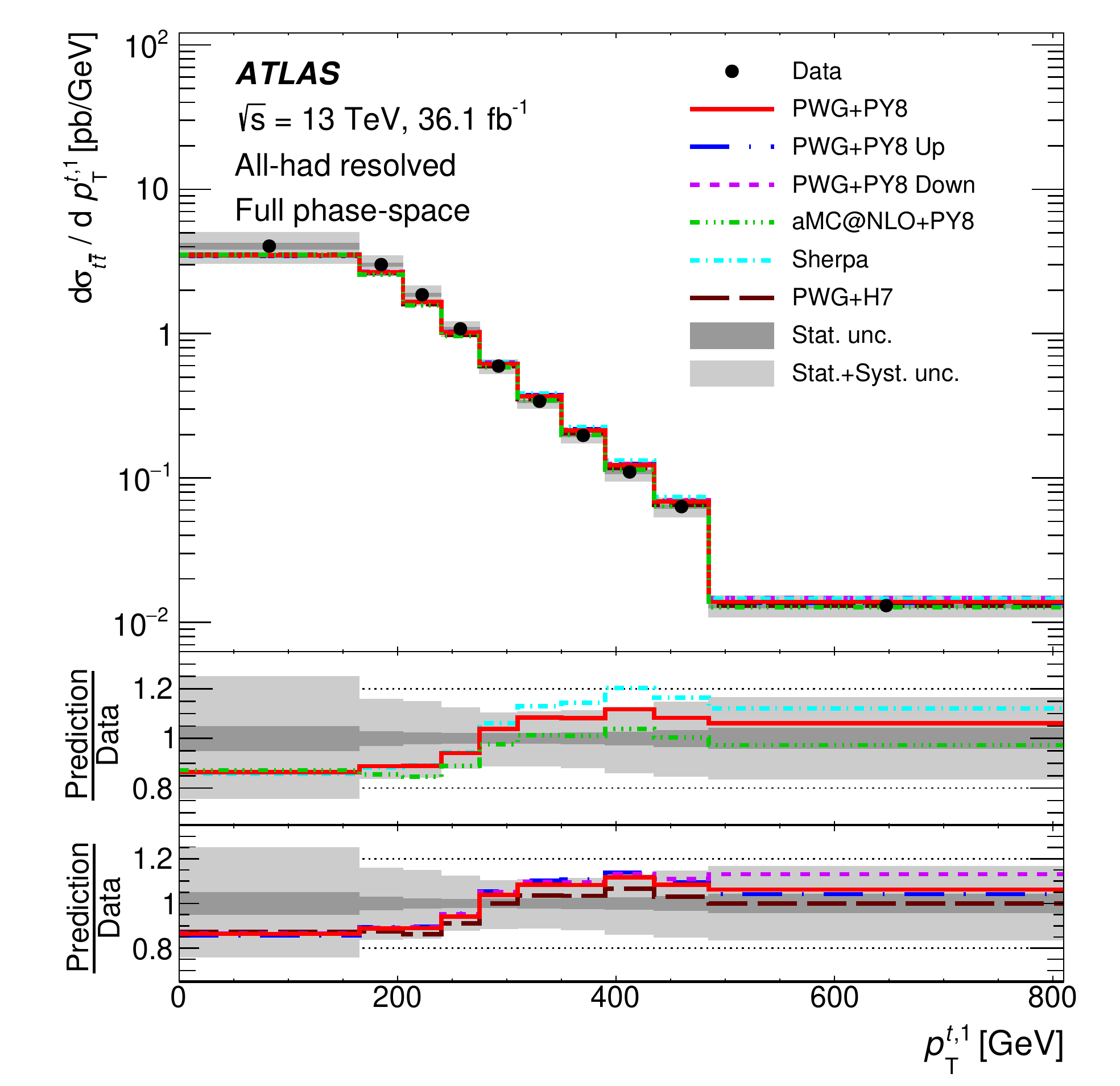}\label{fig:unf_abs_truth:t1_pt}}
\subfloat[]{\includegraphics[width=0.5\textwidth]{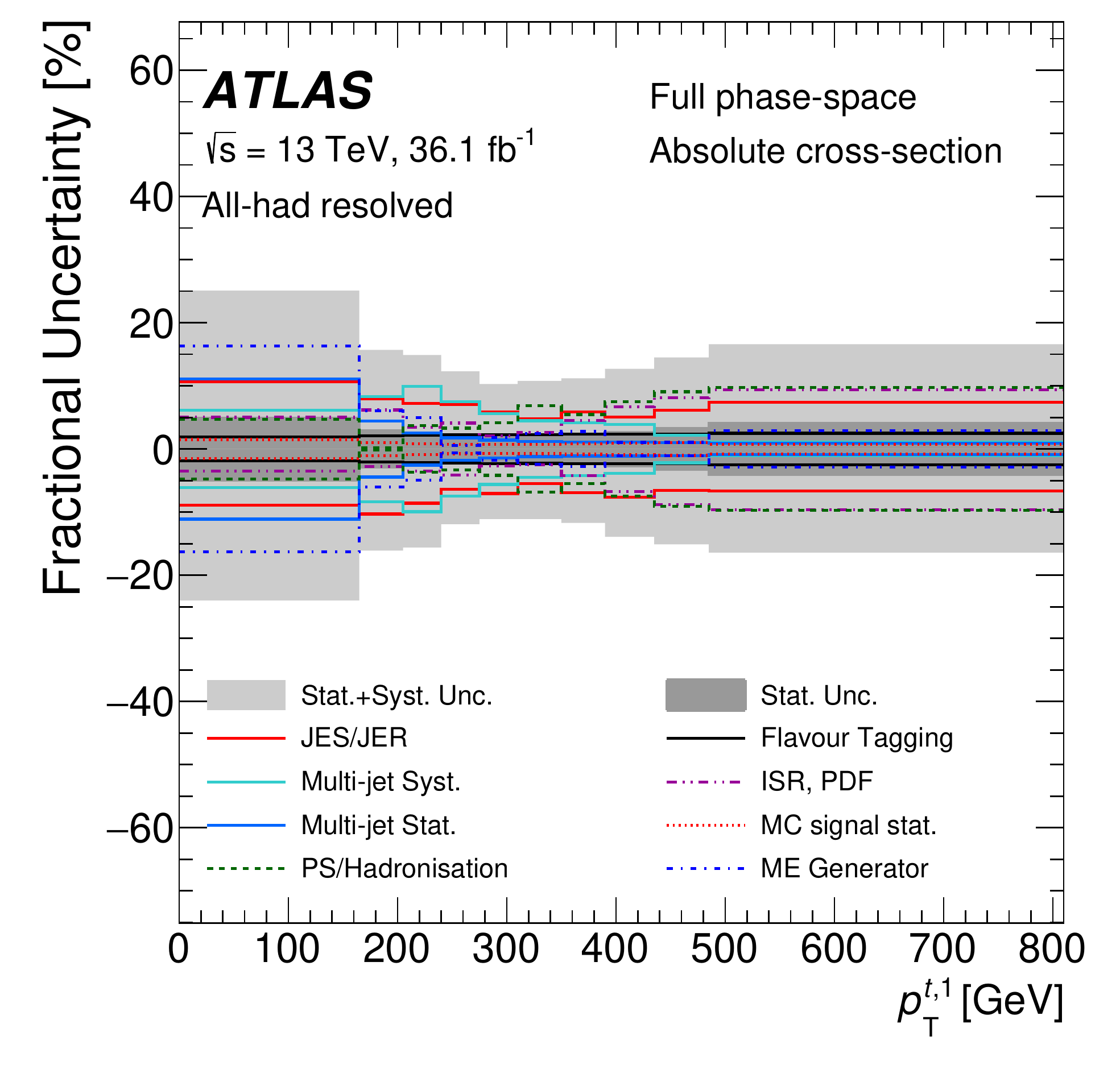}\label{fig:unc_abs_truth:t1_pt}}
\caption{
Single-differential absolute cross-section measurements, unfolded at parton level~(a), as a function of the leading top-quark transverse momentum.
The unfolded data are compared with theoretical predictions. \unfoldedplotcaption\,
Fractional uncertainties for the absolute single-differential distributions unfolded at parton level~(b) as a function of the leading top-quark transverse momentum. \reluncertplotcaption\,
}
\label{fig:results:parton:absoluteTop1}
\end{figure}

At parton level, the normalised single-differential cross-section unfolded to the full phase space as a function of the transverse momentum of the leading top quark is presented in \Fig{\ref{fig:results:parton:top12}}.
The corresponding absolute differential cross-section is shown in \Fig{\ref{fig:results:parton:absoluteTop1}} for comparison.
The normalised measurement is once more characterised by significant cancellations in the uncertainties (primarily the $b$-tagging and parton shower ones).
However, the normalisation procedure inflates the hard-scatter uncertainty at large \pt, due to the normalisation being influenced mostly by bins at low transverse momentum for which the absolute differential cross-section is affected by a large hard-scatter uncertainty.
Even so, the trends are similar to those observed in the particle-level measurements, with the data being best described by \PHHSEVEN.

\begin{figure}[htbp]
\centering
\subfloat[]{\includegraphics[width=0.4\textwidth]{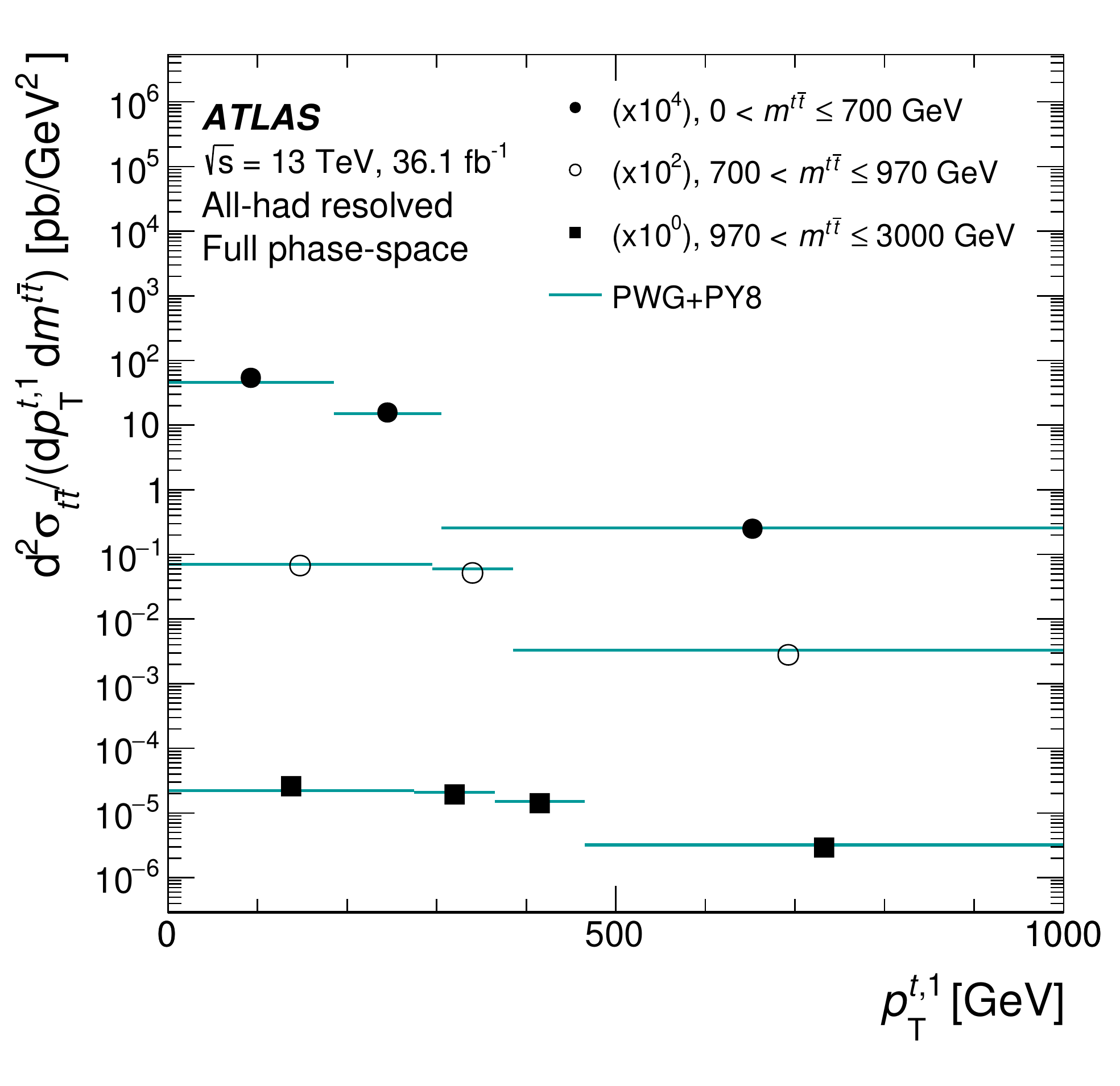}\label{fig:unf_abs_truth:t1_pt_m}}
\\
\subfloat[]{\includegraphics[width=0.7\textwidth]{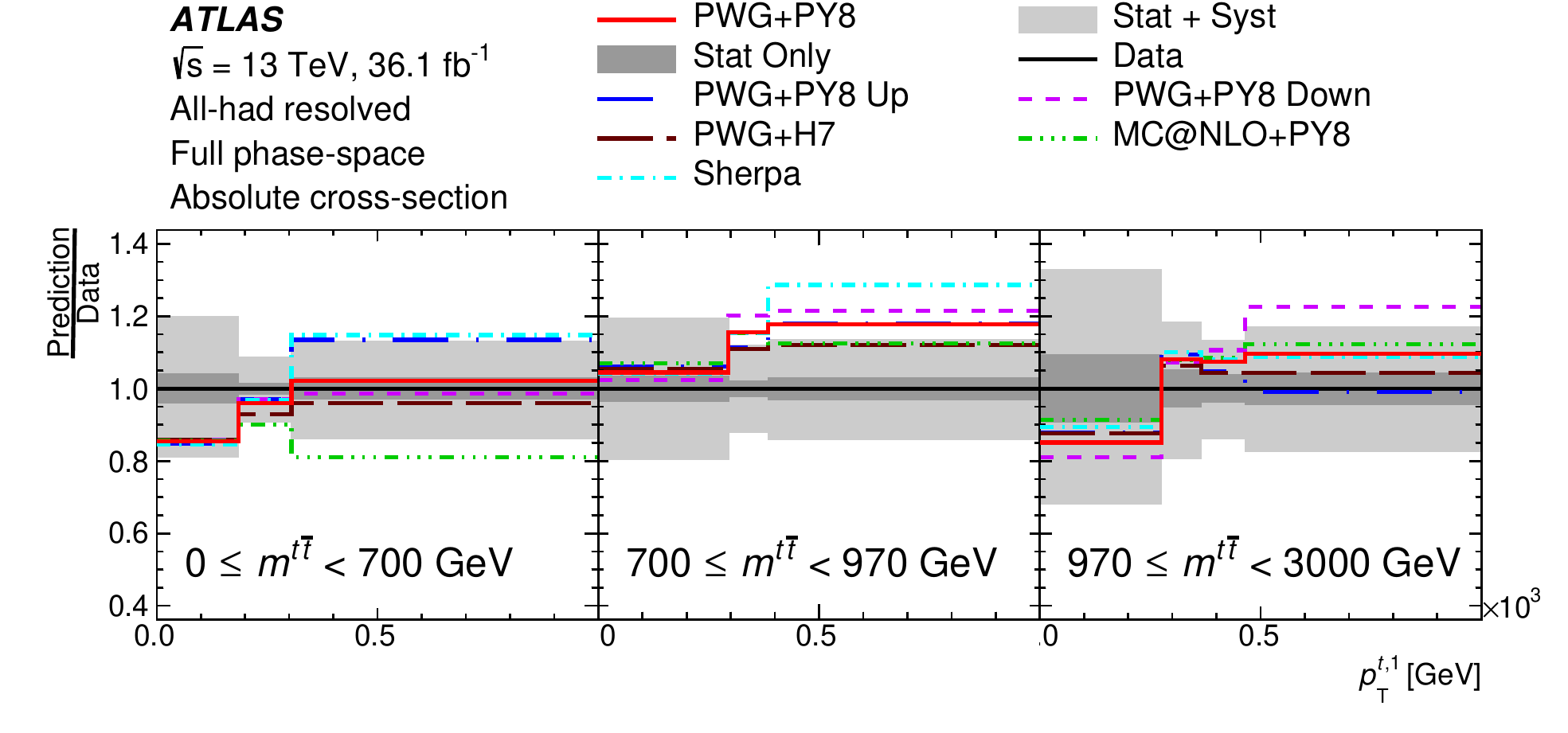}\label{fig:comp_abs_truth:t1_pt_m}}
\caption{Parton-level double-differential absolute cross-section~(a) as a function of the leading top-quark transverse momentum in bins of the $\ttbar$ system mass, compared with the nominal \PHPYEIGHT prediction without uncertainties. Data points are placed at the centre of each bin. Different markers are used to distinguish the three bins in \mttbar, while \pttl\ is shown on the horizontal axis.
The ratio~(b) of the measured cross-section to different MC predictions. \twodratioplotcaption\,}
\label{fig:results:parton:DoubleTop1PT}
\end{figure}

\begin{figure}[htbp]
\centering
\subfloat[]{\includegraphics[width=0.4\textwidth]{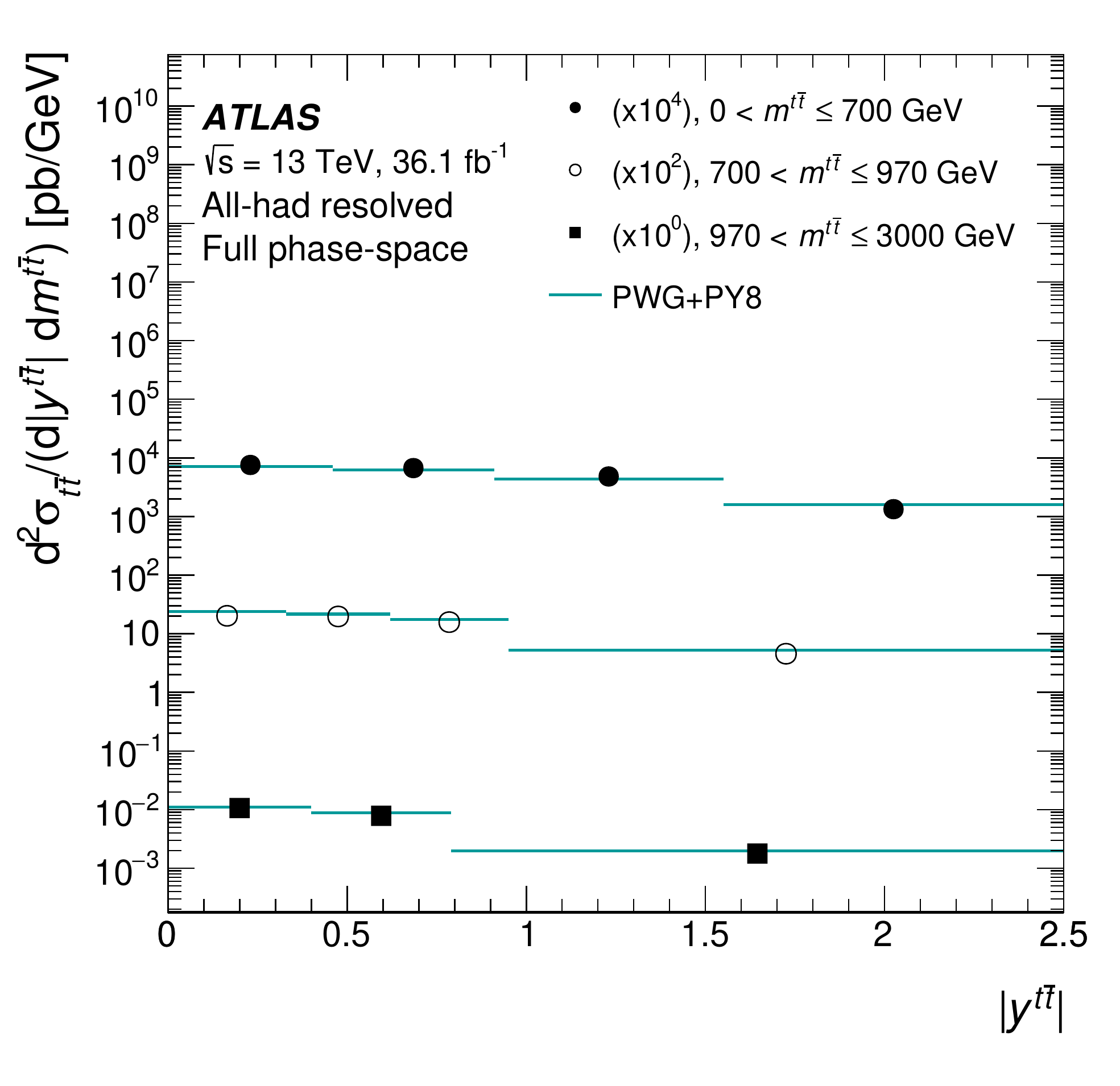}\label{fig:unf_abs_truth:tt_y_m}}
\\
\subfloat[]{\includegraphics[width=0.7\textwidth]{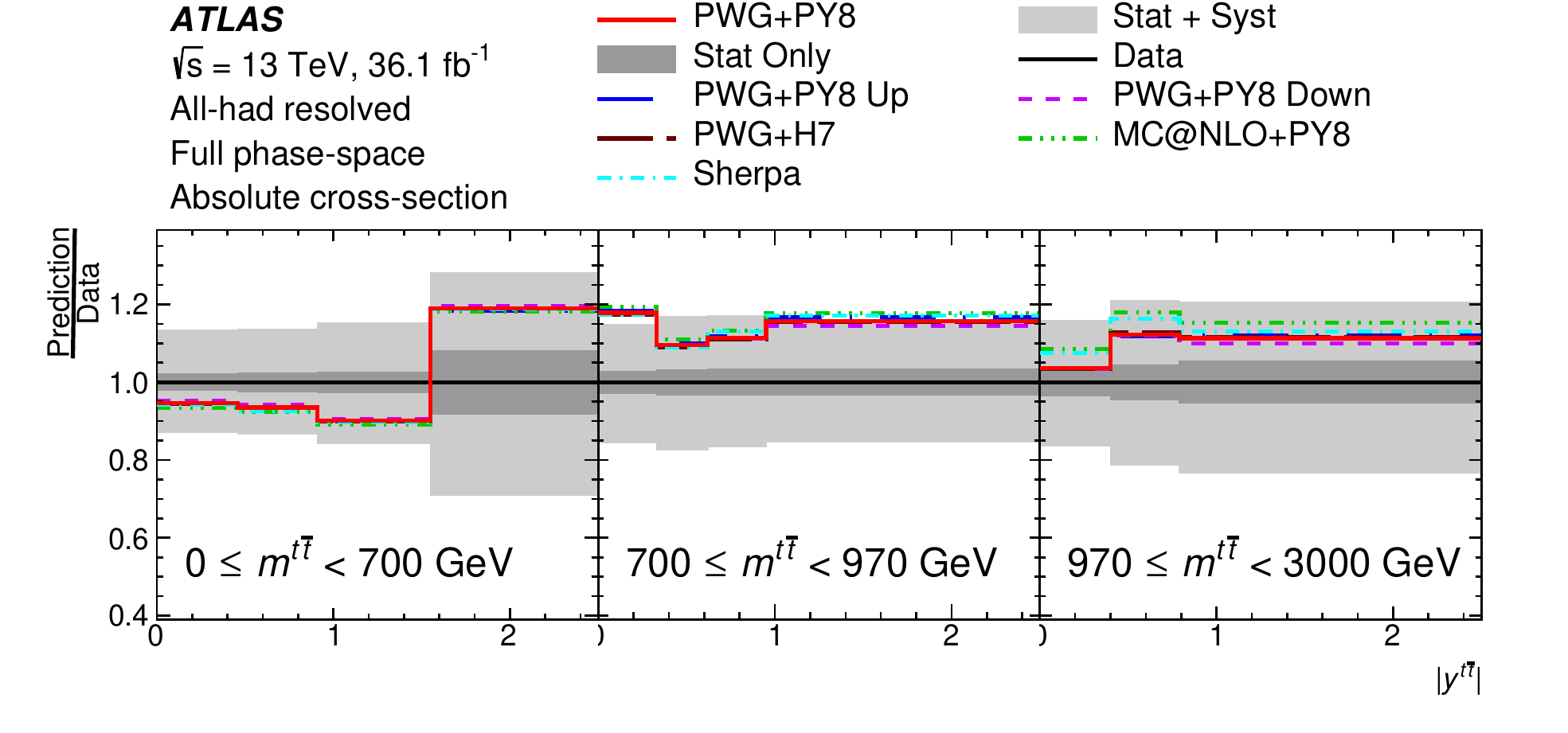}\label{fig:comp_abs_truth:tt_y_m}}
\caption{Parton-level double-differential absolute cross-section~(a) as a function of the $\ttbar$ system rapidity in bins of the $\ttbar$ system mass, compared with the nominal \PHPYEIGHT prediction without uncertainties. Data points are placed at the centre of each bin. Different markers are used to distinguish the three bins in \mttbar, while \yttbar\ is shown on the horizontal axis.
The ratio~(b) of the measured cross-section to different MC predictions. \twodratioplotcaption}
\label{fig:results:parton:DoubleTop1Rap}
\end{figure}

\clearpage
Two absolute double-differential cross-section measurements are shown at parton level: the dependence of the leading top-quark transverse momentum on the top-quark pair mass in \Fig{\ref{fig:results:parton:DoubleTop1PT}} and the dependence of the leading top-quark rapidity on the top-quark pair mass in \Fig{\ref{fig:results:parton:DoubleTop1Rap}}.
The main trend that is observed in the former distributions is, once more, that the event generators predict a harder leading top-quark \pt than is seen in the data.
This feature appears in all \mttbar\ bins. By contrast, the rapidity is fairly well modelled in all bins.
 
\FloatBarrier
 
\subsection{Total cross-section}
\label{sec:TotalXsParticle}
\begin{figure*}[t]
\centering
 
\includegraphics[width=0.5\textwidth]{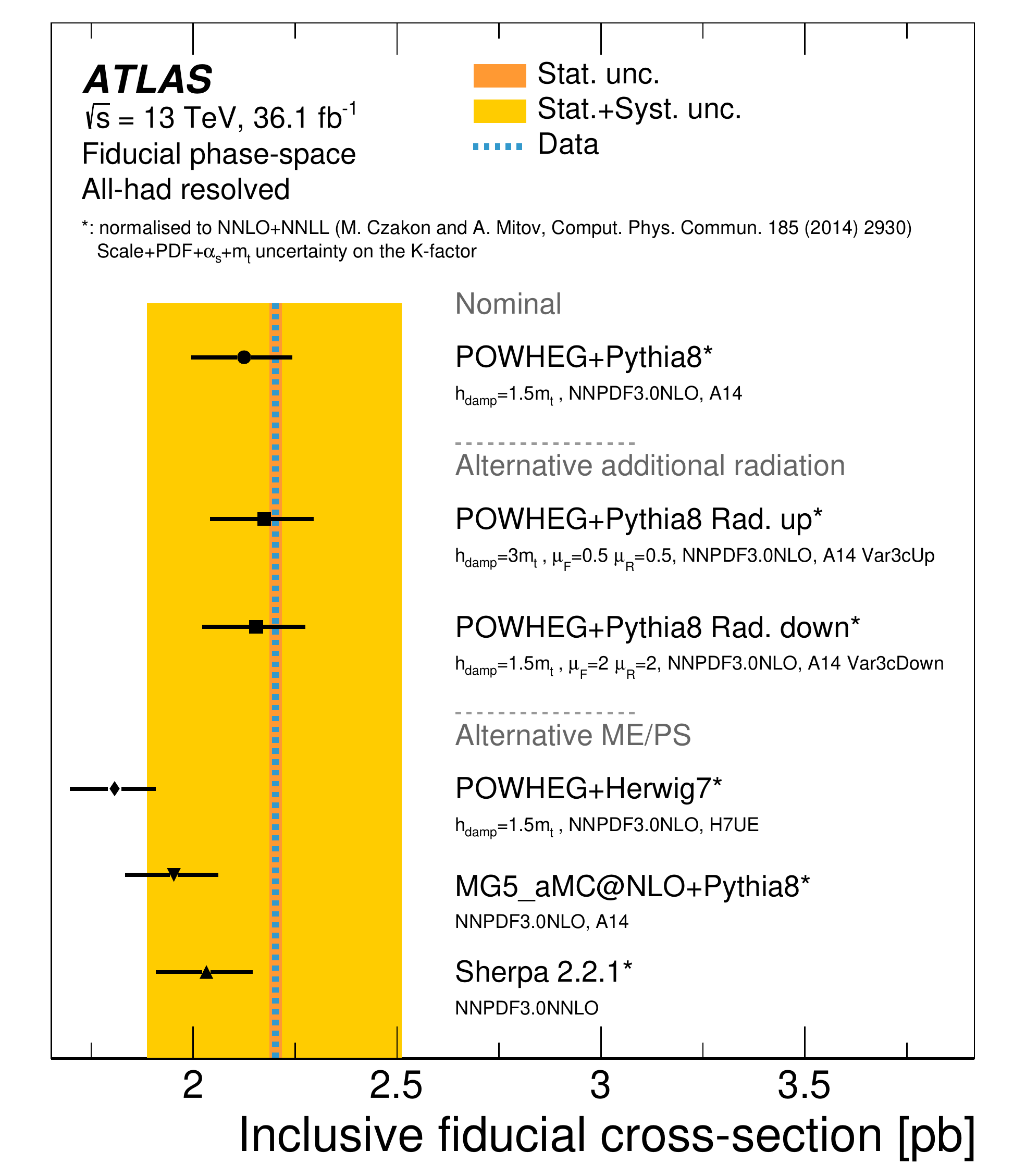}
\caption{Comparison of the measured inclusive fiducial cross-section with the predictions from several MC generators. The yellow band represents the total uncertainty in the data. The orange band represents the statisical uncertainty only. The uncertainty in the cross-section predicted by each NLO MC generator only includes the uncertainty (due to scales, PDFs, $\mt$ and $\alphas$)  affecting the $K$-factor used in the normalisation.}
\label{fig:results_particle:totalXs}
 
\end{figure*}

The total cross-section in the fiducial phase space is found to be $\sigma_{t\bar{t}} = 2.20~\pm~0.31~\mathrm{(stat. + syst.)}$~pb where the total uncertainty is 14\% and the statistical uncertainty is 0.5\%. This value is compatible with the MC predictions described previously which are shown in Figure~\ref{fig:results_particle:totalXs} and in Table~\ref{tab:total_xsec}. The total cross-section as predicted by each NLO MC generator is normalised to the NNLO+NNLL prediction as quoted in Ref.~\cite{Toppp}. Only the uncertainty affecting the $K$-factor used in the normalisation is considered for the predictions.
 
\begin{table}[t]
\caption{Comparison of the measured inclusive fiducial cross-section with the predictions from several MC generators. The uncertainty in the cross-section predicted by each NLO MC generator only includes the uncertainty (due to PDFs, $\mt$ and $\alphas$) affecting the $K$-factor used in the normalisation.}
\label{tab:total_xsec}
\centering
\begingroup
\aboverulesep=0ex
\belowrulesep=0ex
\setlength{\tabcolsep}{10pt} 
\renewcommand{\arraystretch}{1.3} 
\begin{tabular}{|l|l|l|}
\hline
Sample & Fiducial cross-section [pb] \\
\hline
\textsc{Pwg+Py8} & 2.13$^{+0.13}_{-0.12}$ \\
\hline
\textsc{Pwg+Py8} Rad. Up & 2.17$^{+0.13}_{-0.12}$ \\
\textsc{Pwg+Py8} Rad. Down & 2.15$^{+0.13}_{-0.12}$ \\
\hline
\textsc{Pwg+H7} & 1.81$^{+0.11}_{-0.10}$\\
\textsc{MadGraph5}\_aMC@NLO & 1.95$^{+0.12}_{-0.11}$ \\
\textsc{Sherpa} 2.2.1 & 2.03$^{+0.12}_{-0.11}$ \\
\hline
Data & $2.20 \pm 0.31 (\mathrm{stat.+syst.})$ \\
\hline
\end{tabular}
\endgroup
\end{table}

 
The total cross-section in the full phase space, accounting for all decay modes, is measured to be $\sigma_{t\bar{t}} = 864~\pm~127~\mathrm{(stat.+syst.)}$~pb where the total uncertainty is 15\% and the statistical uncertainty is 0.5\%. This cross-section is compatible with a value of $\sigma_{t\bar{t}} = 832^{+20}_{-29}(\mathrm{scale})~\pm 35~(\mathrm{PDF}, \alphas)$~pb as calculated with the Top++2.0 program at NNLO in perturbative QCD, including soft-gluon resummation to NNLL~\cite{Toppp,POWHEG:topq,Toppp:EW,Baernreuther:2012ws,Czakon:2012zr,Czakon:2012pz,Czakon:2013goa} and assuming $\mt$ = 172.5~\GeV{}.
\FloatBarrier
\subsection{Compatibilty with other differential cross-section measurements}
\label{sec:results:other}
\subsubsection{Comparison of results with the \ljets channel}
 
Many of the observables measured in this paper were also measured in the \ljets final state of top-quark pair production~\cite{TOPQ-2018-15}. Before comparing the results, it is important to note that there are several differences between the two approaches. The object selection, which is driven by the triggers, is significantly different since the results presented in this paper are based on a selection of at least six jets with a \pt greater than $55~\GeV$ while the \ljets analysis requires a lepton with $\pt>20~\GeV$ and all jets to have $\pt>25~\GeV$. The extrapolation to the full phase space used for the parton-level results is therefore much bigger for the all-hadronic channel and the size of the available data sample is smaller. However, the all-hadronic channel allows full event reconstruction from well-measured objects, leading to better resolution for the observables, and in particular angular distributions and measurements of extra jets relative to the top-quark pair system.  These effects combine in non-trivial ways, and it is therefore difficult to say a priori which analysis can provide the highest discrimination between models.
 
The particle-level results of both analyses are generally compatible in terms of the level of agreement observed between data and predictions, with some differences identified where variables are better described in either the \ljets channel or the all-hadronic channel. A summary of values with result from both analyses is shown in Table~\ref{tab:chisquare:absolute:particle:ljets_comparison} for an easier comparison. For example, consistent mismodelling is observed in both analyses for the \ptttbar\ distribution. The \HTttbar\ distribution is strongly correlated with the top-quark \pt distributions; it is poorly modelled by all the MC predictions in the all-hadronic channel, while in the $\ell$+jets channel, good agreement is observed between data and all the MC predictions. Mismodelling between data and some of the MC predictions is observed for the \mttbar\ observable in the all-hadronic channel, while good modelling is observed for this variable in the $\ell$+jets channel. The $\ell$+jets channel analysis showed mismodelling in \deltaPhittbar\ for some MC predictions, whereas in the all-hadronic channel this mismodelling is not apparent in the single-differential distribution but appears when measured in bins of jet multiplicity.
 
\begin{table}[ht]
\caption{ Selection of measured particle-level absolute single-differential cross-sections for all-hadronic and single lepton analyses for predictions from several MC generators. For each prediction, a $\chi^2$ and a $p$-value are calculated using the covariance matrix of the measured spectrum. The number of degrees of freedom (NDF) is equal to the number of bins in the distribution.}
\label{tab:chisquare:absolute:particle:ljets_comparison}
\footnotesize
\centering
\renewcommand\arraystretch{\arrstretch}
\resizebox*{0.7\textwidth}{!}{
\noindent\makebox[\textwidth]{
\begin{tabular}{|c|c|  r @{/} l c  | r @{/} l c  | r @{/} l c  | r @{/} l c  | r @{/} l c  | r @{/} l c |}
\hline
Observable & Analysis  & \multicolumn{3}{c|}{\textsc{PWG+PY8}} & \multicolumn{3}{c|}{\textsc{PWG+PY8} Var. Up} & \multicolumn{3}{c|}{\textsc{PWG+PY8} Var. Down} & \multicolumn{3}{c|}{\textsc{Sherpa}} & \multicolumn{3}{c|}{\textsc{PWG+H7}} \\
& & \multicolumn{2}{c}{$\chi^{2}$/NDF} &  ~$p$-value  & \multicolumn{2}{c}{$\chi^{2}$/NDF} &  ~$p$-value  & \multicolumn{2}{c}{$\chi^{2}$/NDF}  &  ~$p$-value  & \multicolumn{2}{c}{$\chi^{2}$/NDF} &  ~$p$-value  & \multicolumn{2}{c}{$\chi^{2}$/NDF} &  ~$p$-value  \\
\hline
$ p_\mathrm{T}^{\ttbar} $ & all-had & {\ } 4.8 & 8 & 0.78   & {\ } 39.7 & 8 & $<$0.01   & {\ } 7.2 & 8 & 0.51      &  {\ } 15.7 & 8 & 0.05     & {\ } 5.0 & 8 & 0.75 \\
&  l-jets & {\ } 23.1 & 11 & 0.02 & {\ } 196.0 & 11 & $<$0.01 & {\ } 16.9 & 11 & 0.11    &  {\ } 88.0 & 11 & $<$0.01 & {\ } 33.4 & 11 & $<$0.01 \\
\hline
$  H_\mathrm{T}^{\ttbar}$ & all-had & {\ } 23.4 & 11 & 0.02 & {\ } 22.8 & 11 & 0.02     & {\ } 32.8 & 11 & $<$0.01 &  {\ } 13.7 & 11 & 0.25    & {\ } 8.2 & 11 & 0.69 \\
&  l-jets & {\ } 11.1 & 18 & 0.89 & {\ } 17.7 & 18 & 0.48     & {\ } 10.5 & 18 & 0.91    &  {\ } 11.9 & 18 & 0.85    & {\ } 11.4 & 18 & 0.88 \\
\hline
$            m^{\ttbar} $ & all-had & {\ } 17.0 & 9 & 0.05  & {\ } 12.5 & 9 & 0.19      & {\ } 22.4 & 9 & $<$0.01  &  {\ } 7.6 & 9 & 0.57      & {\ } 10.6 & 9 & 0.30 \\
&  l-jets & {\ } 17.8 & 16 & 0.34 & {\ } 16.4 & 16 & 0.43     & {\ } 20.2 & 16 & 0.21    &  {\ } 17.1 & 16 & 0.38    & {\ } 15.5 & 16 & 0.49 \\
\hline
$    \Delta\phi^{\ttbar}$ & all-had & {\ } 4.4 & 6 & 0.62   & {\ } 4.3 & 6 & 0.63       & {\ } 11.5 & 6 & 0.08     &  {\ } 3.9 & 6 & 0.69      & {\ } 3.9 & 6 & 0.69 \\
&  l-jets &{ \ } 3.0 & 7 & 0.89   & {\ } 57.7 & 7 & $<$0.01   & {\ } 12.3 & 7 & 0.09     &  {\ } 22.1 & 7 & $<$0.01  & {\ } 4.7 & 7 & 0.70\\
\hline
\end{tabular}}}
\end{table}
 
At parton level, the $\ptttbar$ and $\mttbar$ distributions are poorly described by most of the MC predictions in both the $\ell$+jets and all-hadronic channels. Unlike at particle-level, good agreement between data and all the MC predictions is observed in both channels for the $\HTttbar$ variable. This is due to larger uncertainties, a consequence of the extrapolation from the small fiducial region to the full phase space. A summary of the results from the two analyses is presented in Table~\ref{tab:chisquare:absolute:parton:ljets_comparison}.
 
\begin{table}[ht]
\caption{ Selection of measured parton-level absolute single-differential cross-sections for all-hadronic and single lepton analyses for predictions from several MC generators. For each prediction, a $\chi^2$ and a $p$-value are calculated using the covariance matrix of the measured spectrum. The number of degrees of freedom (NDF) is equal to the number of bins in the distribution.}
\label{tab:chisquare:absolute:parton:ljets_comparison}
\footnotesize
\centering
\renewcommand\arraystretch{\arrstretch}
\resizebox*{0.7\textwidth}{!}{
\noindent\makebox[\textwidth]{
\begin{tabular}{|c|c|  r @{/} l c  | r @{/} l c  | r @{/} l c  | r @{/} l c  | r @{/} l c  | r @{/} l c |}
\hline
Observable  & Analysis & \multicolumn{3}{c|}{\textsc{PWG+PY8}} & \multicolumn{3}{c|}{\textsc{PWG+PY8} Var. Up} & \multicolumn{3}{c|}{\textsc{PWG+PY8} Var. Down} & \multicolumn{3}{c|}{\textsc{Sherpa}} & \multicolumn{3}{c|}{\textsc{PWG+H7}} \\
& & \multicolumn{2}{c}{$\chi^{2}$/NDF} &  ~$p$-value  & \multicolumn{2}{c}{$\chi^{2}$/NDF} & ~$p$-value  & \multicolumn{2}{c}{$\chi^{2}$/NDF} &  ~$p$-value  & \multicolumn{2}{c}{$\chi^{2}$/NDF} &  ~$p$-value  & \multicolumn{2}{c}{$\chi^{2}$/NDF} &  ~$p$-value  \\
\hline
$  H_\mathrm{T}^{\ttbar}$ & all-had & {\ } 13.6 & 11 & 0.26   & {\ } 11.0 & 11 & 0.44   & {\ } 20.0 & 11 & 0.05   & {\ } 16.7 & 11 & 0.12 & {\ } 11.0 & 11 & 0.45 \\
&  l-jets & {\ } 9.9 & 9 & 0.36     & {\ } 10.1 & 9 & 0.34    & {\ } 9.9 & 9 & 0.36     & {\ } 19.6 & 9 & 0.02  & {\ }  6.7 & 9 & 0.67\\
\hline
$            m^{\ttbar} $ & all-had & {\ } 9.3 & 9 & 0.41     & {\ } 10.1 & 9 & 0.34    & {\ } 8.8 & 9 & 0.46     & {\ } 9.1 & 9 & 0.43   & {\ } 8.8 & 9 & 0.46 \\
&  l-jets & {\ } 29.1 & 9 & $<$0.01 & {\ } 22.9 & 9 & $<$0.01 & {\ } 36.8 & 9 & $<$0.01 & {\ } 15.4 & 9 & 0.08  & {\ } 25.6 & 9 & $<$0.01 \\
\hline
$ p_\mathrm{T}^{\ttbar} $ & all-had & {\ } 1.5 & 5 & 0.91     & {\ } 12.6 & 5 & 0.03    & {\ } 2.2 & 5 & 0.83     & {\ } 9.6 & 5 & 0.09   & {\ } 3.0 & 5 & 0.70 \\
&  l-jets & {\ } 8.6 & 9 & 0.47     & {\ } 42.4 & 9 & $<$0.01 & {\ } 24.3 & 9 & $<$0.01 & {\ } 20.6 & 9 & 0.01  & {\ } 14.1 & 9 & 0.12\\
\hline
\end{tabular}}}

\end{table}
 
When considering the double-differential results at both particle and parton levels, both analyses show that none of the predictions can describe any of the measured distributions.
 
\begin{figure}[htbp]
\centering
\subfloat[]{\includegraphics[width=0.5\textwidth]{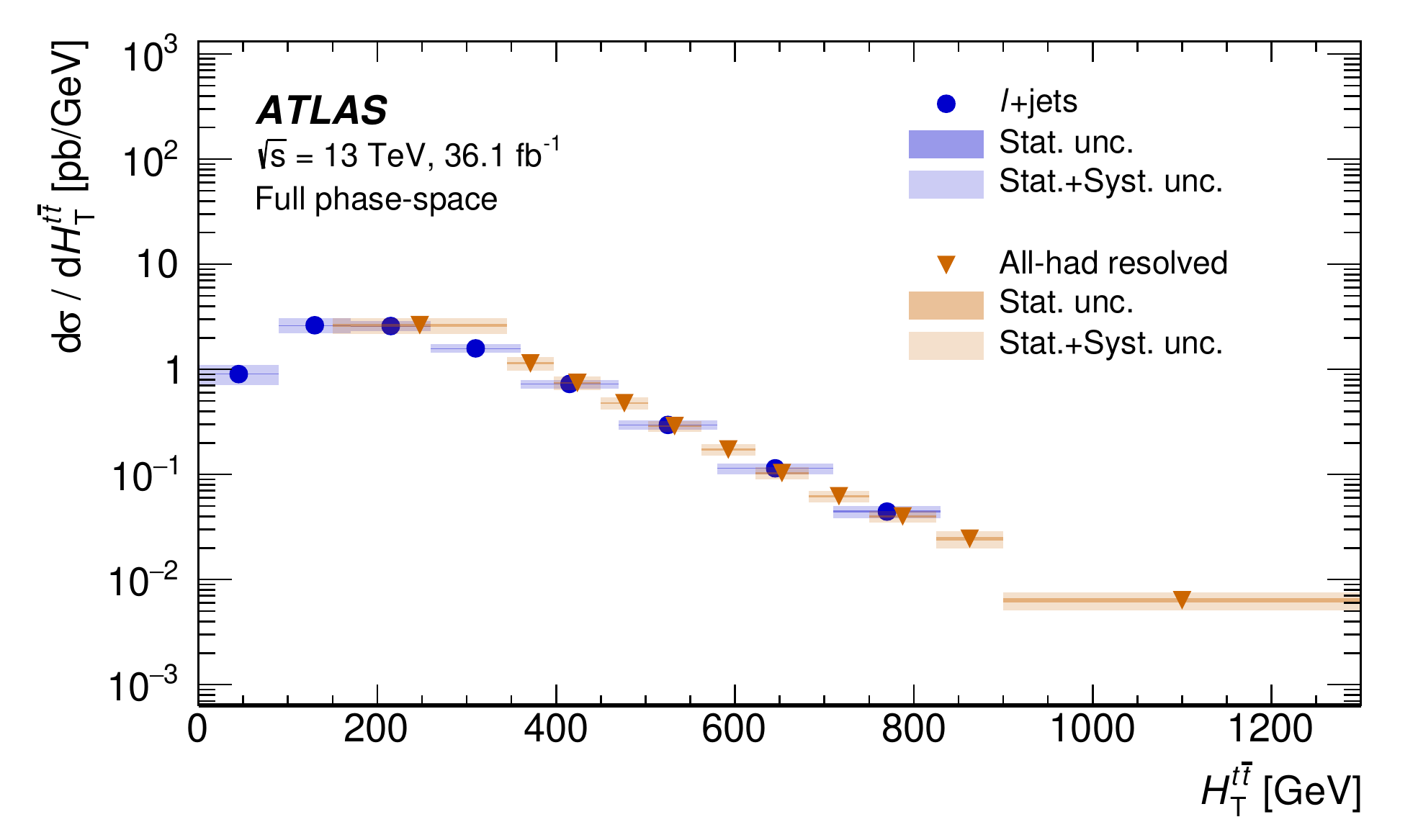}\label{fig:parton:comp_HT}}
\subfloat[]{\includegraphics[width=0.5\textwidth]{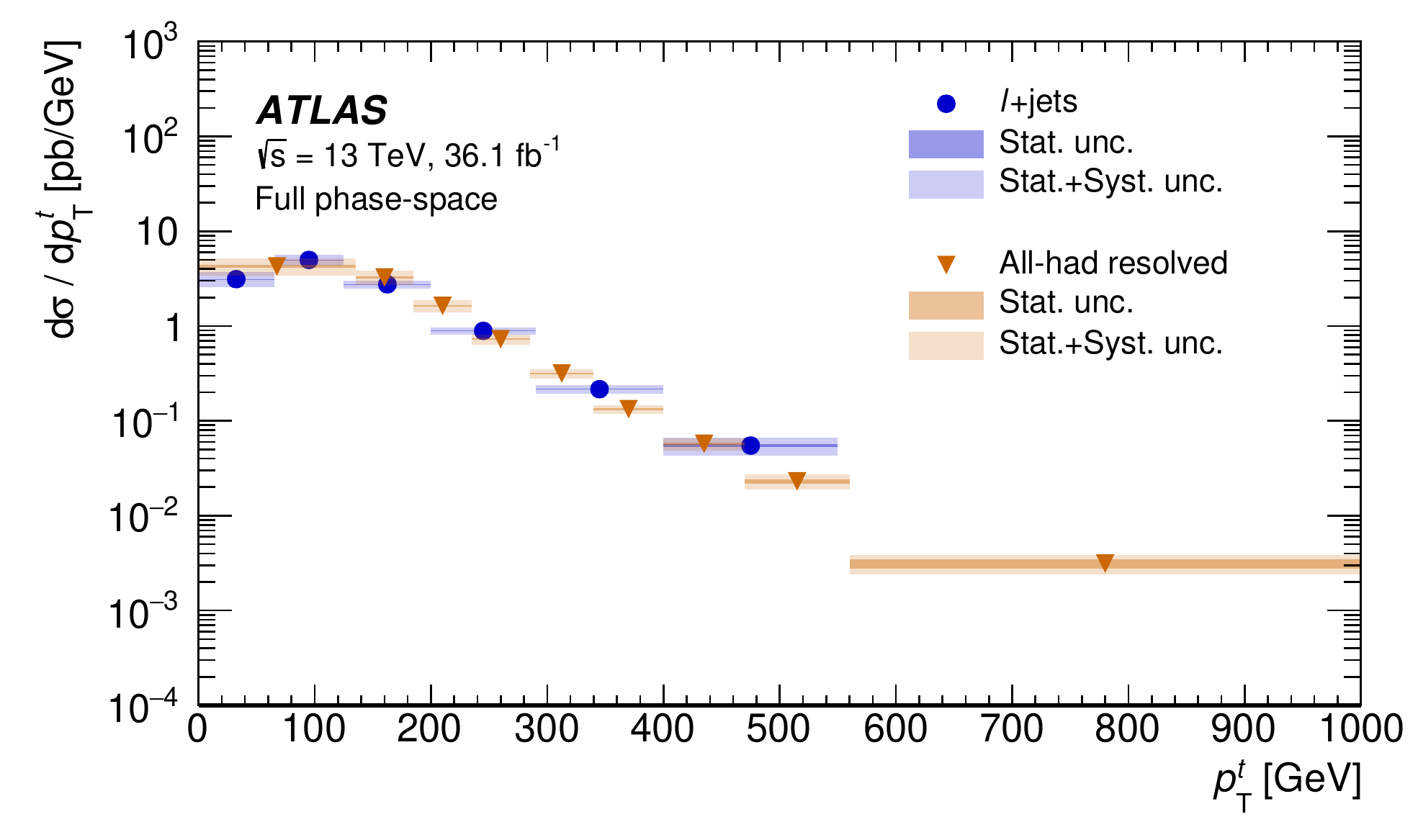}\label{fig:parton:comp_top}}
\caption{Comparison between the measured absolute single-differential cross-sections in the $\ell$+jets and all-hadronic channels as functions of (a) $\HTttbar$ and (b) the average top transverse momentum $\ptt$.}
\label{fig:results:parton:full_ljet_comparison}
\end{figure}

Figure~\ref{fig:results:parton:full_ljet_comparison} shows a comparison between the measured absolute differential cross-sections in the $\ell$+jets and all-hadronic channels, at parton level, for $\HTttbar$ and the average top transverse momentum $\ptt$.
The latter is determined by randomly picking one of the two top candidates in each event.
The two measurements are qualitatively consistent in the overlap region.
For $\HTttbar\gtrsim325\GeV$ and $\ptt\gtrsim130~\GeV$, where neither statistics nor signal purity are limiting, the all-hadronic measurement achieves a visibly better resolution than the $\ell$+jets channel.
 
\FloatBarrier
\subsubsection{Comparison of results with the all-hadronic channel in the boosted topology}
 
\begin{figure}[htbp]
\centering
\subfloat[]{\includegraphics[width=0.5\textwidth]{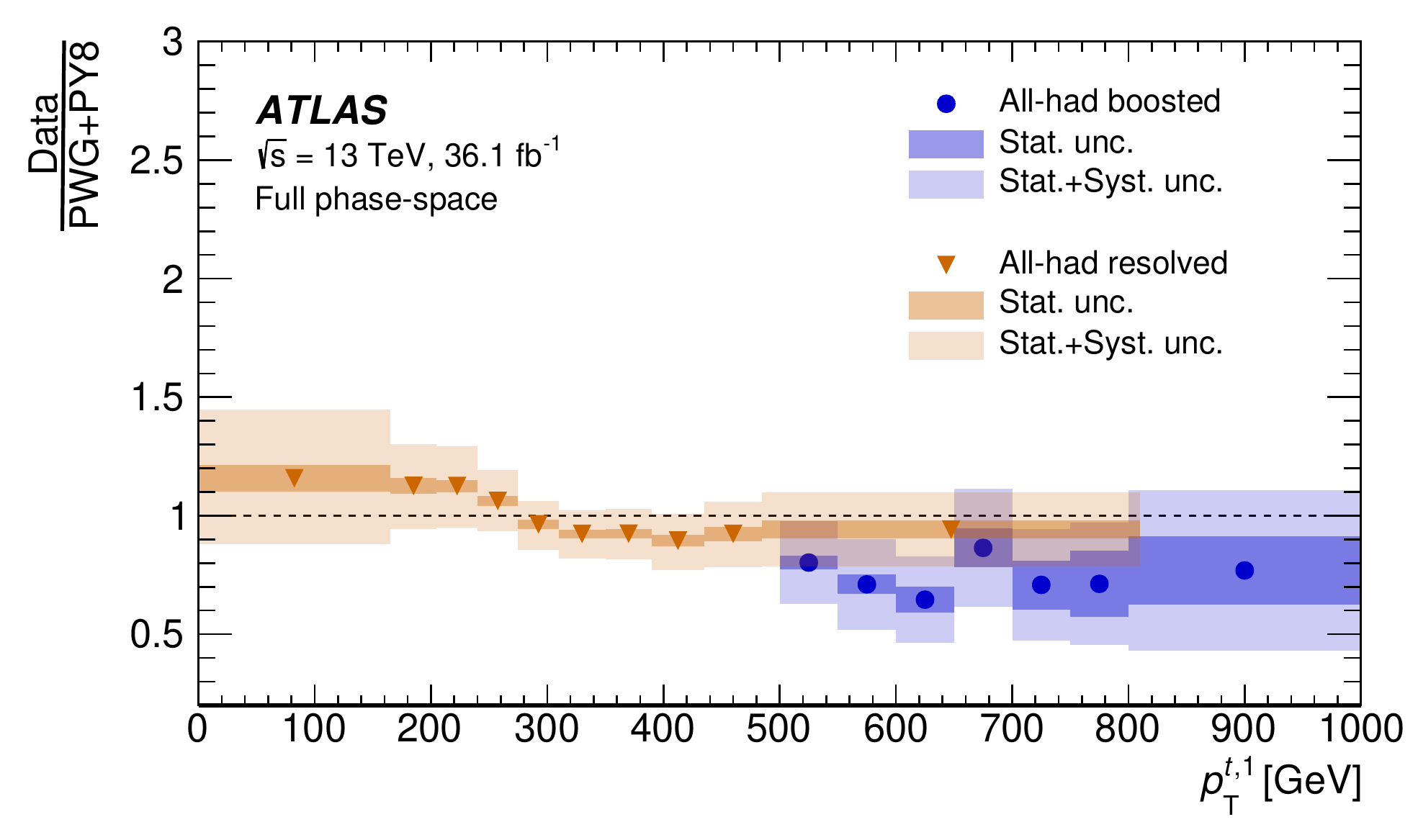}\label{fig:parton:comp_top1}}
\subfloat[]{\includegraphics[width=0.5\textwidth]{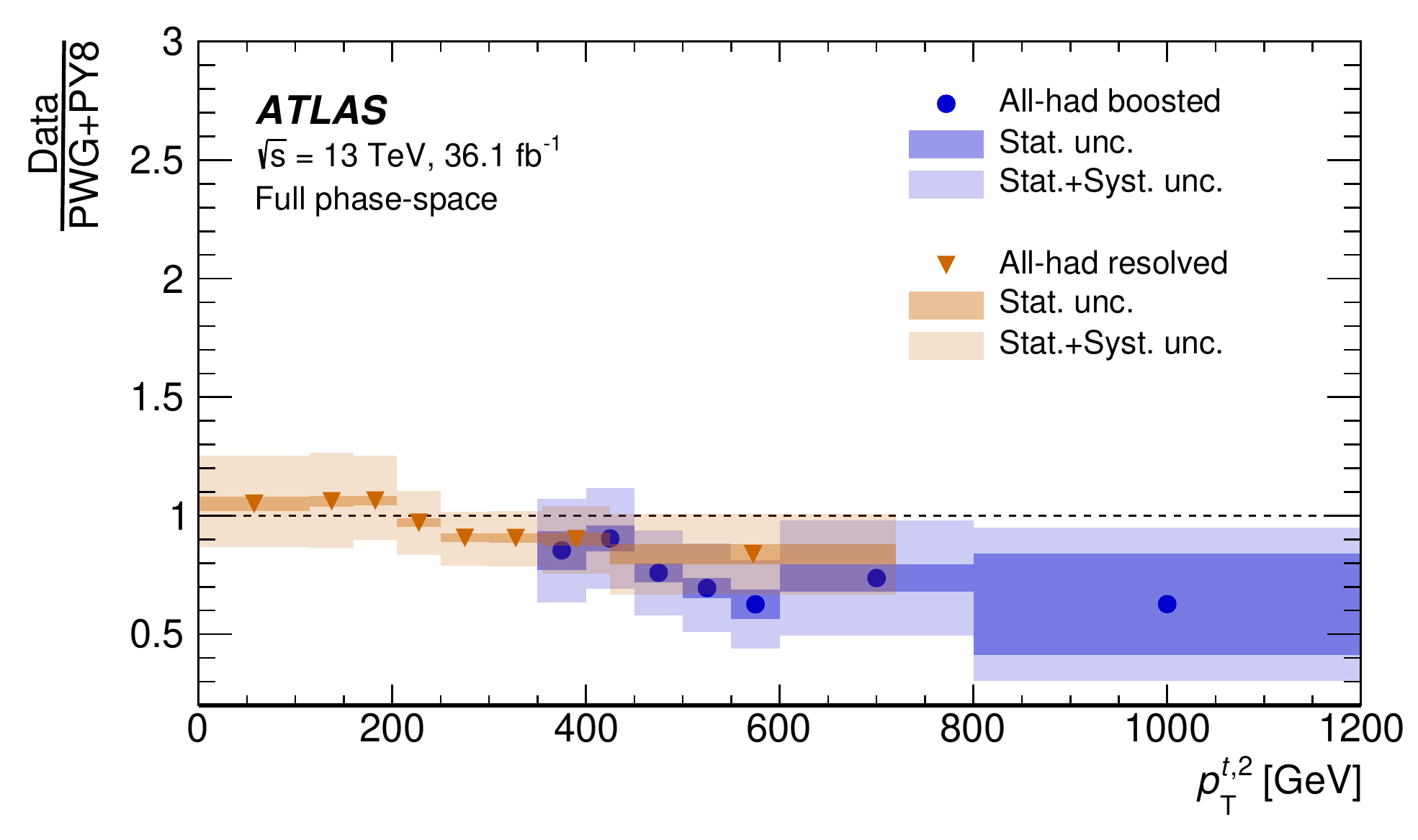}\label{fig:parton:comp_top2}}
\caption{The ratios of the measured absolute differential cross-sections at parton level to the predictions obtained with the \PHPYEIGHT\ MC generator in the all-hadronic resolved and boosted topologies as a function of the~(a) $\pttl$ and~(b) $\pttsl$ variables.}
\label{fig:results:parton:fullhad_comparison}
\end{figure}

Measurements of differential cross-sections in the all-hadronic channel have been performed in the boosted topology~\cite{TOPQ-2016-09}, motivating a comparison with the results of this analysis.
The all-hadronic resolved parton-level measurements are unfolded to the full phase space, while the measurements in the boosted topology are unfolded to a fiducial phase space, so a direct comparison of the differential measurements is not possible.
Instead, Figure~\ref{fig:results:parton:fullhad_comparison} shows the ratios of the measured absolute differential cross-sections at parton level to the predictions obtained with the \PHPYEIGHT  MC generator in the all-hadronic resolved and boosted topologies as a function of the $\pttl$ and $\pttsl$ variables.
It can be seen from the figures that the ratios between the data and the signal MC generator are qualitatively consistent between the two topologies in the overlap region.
 
\FloatBarrier
 
\section{Conclusions}
\label{sec:conclusion}
 
Comprehensive measurements of single- and double-differential cross-sections for the production of top-quark pairs are performed in the resolved topology of the all-hadronic channel using data from $pp$ collisions at 13~\TeV\ collected in 2015 and 2016 by the ATLAS detector at the CERN LHC and corresponding to an integrated luminosity of 36.1~fb$^{-1}$.
 
Absolute and normalised differential cross-sections are presented as functions of several kinematic variables, unfolded at the particle and parton levels. Several novel variables are introduced to better probe correlations between the kinematics of the top-quark pair and associated jet radiation. The results show sensitivity to different aspects of the tested MC predictions.
 
As several predictions in a number of variables have poor agreement with the data, these observations can be exploited to improve the top-quark MC modelling.
In particular, the double-differential cross-sections at the particle level will be extremely useful for improving the MC predictions in regions of the phase space with many additional jets, which are regions of interest for analyses of many rare processes.
 
The measurements at the parton level are compared with theory predictions obtained from NLO MC generators interfaced with parton shower and hadronisation models and can also be used in future measurements such as PDF and top-quark pole mass extraction.
The rapidities of the individual top quarks and of the top-quark pair are well modelled, while the leading top-quark transverse momentum and top-quark pair transverse momentum are found to be incompatible with several theoretical predictions. Furthermore, significant mismodelling is observed in the hardness of the additional jet emissions.
 
The comparison with other published results showed that the predictions are rather accurate and for the most part compatible in the different regions of phase space and for different channels. However, some tension remains and better models are indeed needed to reproduce the data for all observables in all channels. The Rivet~\cite{Buckley:2010ar} routine that will be made available will allow the use of the results presented in this paper to define better MC predictions for the next generation of ATLAS and LHC analyses.
 
 
\section*{Acknowledgements}
 

We thank CERN for the very successful operation of the LHC, as well as the
support staff from our institutions without whom ATLAS could not be
operated efficiently.
 
We acknowledge the support of ANPCyT, Argentina; YerPhI, Armenia; ARC, Australia; BMWFW and FWF, Austria; ANAS, Azerbaijan; SSTC, Belarus; CNPq and FAPESP, Brazil; NSERC, NRC and CFI, Canada; CERN; ANID, Chile; CAS, MOST and NSFC, China; COLCIENCIAS, Colombia; MSMT CR, MPO CR and VSC CR, Czech Republic; DNRF and DNSRC, Denmark; IN2P3-CNRS and CEA-DRF/IRFU, France; SRNSFG, Georgia; BMBF, HGF and MPG, Germany; GSRT, Greece; RGC and Hong Kong SAR, China; ISF and Benoziyo Center, Israel; INFN, Italy; MEXT and JSPS, Japan; CNRST, Morocco; NWO, Netherlands; RCN, Norway; MNiSW and NCN, Poland; FCT, Portugal; MNE/IFA, Romania; JINR; MES of Russia and NRC KI, Russian Federation; MESTD, Serbia; MSSR, Slovakia; ARRS and MIZ\v{S}, Slovenia; DST/NRF, South Africa; MICINN, Spain; SRC and Wallenberg Foundation, Sweden; SERI, SNSF and Cantons of Bern and Geneva, Switzerland; MOST, Taiwan; TAEK, Turkey; STFC, United Kingdom; DOE and NSF, United States of America. In addition, individual groups and members have received support from BCKDF, CANARIE, Compute Canada, CRC and IVADO, Canada; Beijing Municipal Science \& Technology Commission, China; COST, ERC, ERDF, Horizon 2020 and Marie Sk{\l}odowska-Curie Actions, European Union; Investissements d'Avenir Labex, Investissements d'Avenir Idex and ANR, France; DFG and AvH Foundation, Germany; Herakleitos, Thales and Aristeia programmes co-financed by EU-ESF and the Greek NSRF, Greece; BSF-NSF and GIF, Israel; La Caixa Banking Foundation, CERCA Programme Generalitat de Catalunya and PROMETEO and GenT Programmes Generalitat Valenciana, Spain; G\"{o}ran Gustafssons Stiftelse, Sweden; The Royal Society and Leverhulme Trust, United Kingdom.
 
The crucial computing support from all WLCG partners is acknowledged gratefully, in particular from CERN, the ATLAS Tier-1 facilities at TRIUMF (Canada), NDGF (Denmark, Norway, Sweden), CC-IN2P3 (France), KIT/GridKA (Germany), INFN-CNAF (Italy), NL-T1 (Netherlands), PIC (Spain), ASGC (Taiwan), RAL (UK) and BNL (USA), the Tier-2 facilities worldwide and large non-WLCG resource providers. Major contributors of computing resources are listed in Ref.~\cite{ATL-SOFT-PUB-2020-001}.
 

\printbibliography

\clearpage 
 
\begin{flushleft}
\hypersetup{urlcolor=black}
{\Large The ATLAS Collaboration}

\bigskip

\AtlasOrcid[0000-0002-6665-4934]{G.~Aad}$^\textrm{\scriptsize 102}$,    
\AtlasOrcid[0000-0002-5888-2734]{B.~Abbott}$^\textrm{\scriptsize 128}$,    
\AtlasOrcid{D.C.~Abbott}$^\textrm{\scriptsize 103}$,    
\AtlasOrcid[0000-0002-2788-3822]{A.~Abed~Abud}$^\textrm{\scriptsize 36}$,    
\AtlasOrcid[0000-0002-1002-1652]{K.~Abeling}$^\textrm{\scriptsize 53}$,    
\AtlasOrcid[0000-0002-2987-4006]{D.K.~Abhayasinghe}$^\textrm{\scriptsize 94}$,    
\AtlasOrcid[0000-0002-8496-9294]{S.H.~Abidi}$^\textrm{\scriptsize 167}$,    
\AtlasOrcid[0000-0002-8279-9324]{O.S.~AbouZeid}$^\textrm{\scriptsize 40}$,    
\AtlasOrcid{N.L.~Abraham}$^\textrm{\scriptsize 156}$,    
\AtlasOrcid[0000-0001-5329-6640]{H.~Abramowicz}$^\textrm{\scriptsize 161}$
% ...
\bigskip
\\

% ...
$^{36}$CERN, Geneva; Switzerland.\\
% ...
$^{40}$Niels Bohr Institute, University of Copenhagen, Copenhagen; Denmark.\\
% ...
$^{53}$II. Physikalisches Institut, Georg-August-Universit\"{a}t G\"ottingen, G\"ottingen; Germany.\\
% ...
$^{94}$Department of Physics, Royal Holloway University of London, Egham; United Kingdom.\\
% ...
$^{102}$CPPM, Aix-Marseille Universit\'e, CNRS/IN2P3, Marseille; France.\\
$^{103}$Department of Physics, University of Massachusetts, Amherst MA; United States of America.\\
% ...
$^{128}$Homer L. Dodge Department of Physics and Astronomy, University of Oklahoma, Norman OK; United States of America.\\
% ...
$^{156}$Department of Physics and Astronomy, University of Sussex, Brighton; United Kingdom.\\
% ...
$^{161}$Raymond and Beverly Sackler School of Physics and Astronomy, Tel Aviv University, Tel Aviv; Israel.\\
% ...
$^{167}$Department of Physics, University of Toronto, Toronto ON; Canada.\\
% ...

\end{flushleft}


\end{document}